\documentclass[preprint,trackchanges]{aastex62}

\usepackage{amssymb,amsmath}
\usepackage{bm}
\usepackage{CJK}
\usepackage{graphicx}
\usepackage[utf8]{inputenc}
\DeclareUnicodeCharacter{2009}{~}
\usepackage{natbib}
\usepackage[caption=false]{subfig}

\newcommand{\xm}{$X_{\mathrm{max}}$}
\newcommand{\mxm}{$\left< X_{\mathrm{max}} \right>$}
\newcommand{\nm}{$N_{\mathrm{max}}$}
\newcommand{\sxm}{$\sigma(X_{\mathrm{max}})$}

\submitjournal{ApJ}

\watermark{Preprint}  

\begin{document}

\title{Depth of Ultra High Energy Cosmic Ray Induced Air Shower Maxima
  Measured by the Telescope Array Black Rock and Long Ridge FADC
  Fluorescence Detectors and Surface Array in Hybrid Mode}

\correspondingauthor{William Hanlon}
\email{whanlon@cosmic.utah.edu, ikeda@icrr.u-tokyo.ac.jp}

\author[0000-0001-6141-4205]{R.U. Abbasi}
\affiliation{High Energy Astrophysics Institute and Department of Physics
  and Astronomy, University of Utah, Salt Lake City, Utah, USA}

\author{M. Abe}
\affiliation{The Graduate School of Science and Engineering, Saitama University,
  Saitama, Saitama, Japan}

\author{T. Abu-Zayyad}
\affiliation{High Energy Astrophysics Institute and Department of Physics
  and Astronomy, University of Utah, Salt Lake City, Utah, USA}

\author{M. Allen}
\affiliation{High Energy Astrophysics Institute and Department of Physics
  and Astronomy, University of Utah, Salt Lake City, Utah, USA}

\author{R. Azuma}
\affiliation{Graduate School of Science and Engineering, Tokyo Institute of
  Technology, Meguro, Tokyo, Japan}

\author{E. Barcikowski}
\affiliation{High Energy Astrophysics Institute and Department of Physics
  and Astronomy, University of Utah, Salt Lake City, Utah, USA}

\author{J.W. Belz}
\affiliation{High Energy Astrophysics Institute and Department of Physics
  and Astronomy, University of Utah, Salt Lake City, Utah, USA}

\author{D.R. Bergman}
\affiliation{High Energy Astrophysics Institute and Department of Physics
  and Astronomy, University of Utah, Salt Lake City, Utah, USA}

\author{S.A. Blake}
\affiliation{High Energy Astrophysics Institute and Department of Physics
  and Astronomy, University of Utah, Salt Lake City, Utah, USA}

\author{R. Cady}
\affiliation{High Energy Astrophysics Institute and Department of Physics
  and Astronomy, University of Utah, Salt Lake City, Utah, USA}

% ...

\collaboration{(Telescope Array Collaboration)}

\begin{abstract}
  The Telescope Array observatory utilizes fluorescence detectors and
  surface detectors to observe air showers produced by ultra high
  energy cosmic rays in the Earth's atmosphere. Cosmic ray events
  observed in this way are termed hybrid data. The depth of air shower
  maximum is related to the mass of the primary particle that
  generates the shower. This paper reports on shower maxima data
  collected over 8.5 years using the Black Rock Mesa and Long Ridge
  fluorescence detectors in conjunction with the array of surface
  detectors. We compare the means and standard deviations of the
  observed \xm{} distributions with Monte Carlo \xm{} distributions of
  unmixed protons, helium, nitrogen, and iron, all generated using the
  QGSJet~II-04 hadronic model. We also perform an unbinned maximum
  likelihood test of the observed data, which is subjected to variable
  systematic shifting of the data \xm{} distributions to allow us to
  test the full distributions, and compare them to the Monte Carlo to
  see which elements are not compatible with the observed data. For
  all energy bins, QGSJet~II-04 protons are found to be compatible
  with Telescope Array hybrid data at the 95\% confidence level after
  some systematic \xm{} shifting of the data. Three other QGSJet~II-04
  elements are found to be compatible using the same test procedure in
  an energy range limited to the highest energies where data
  statistics are sparse.
\end{abstract}


\section{\label{sec:introduction}Introduction}
\begin{figure}
  \centering
  \includegraphics[width=\textwidth]{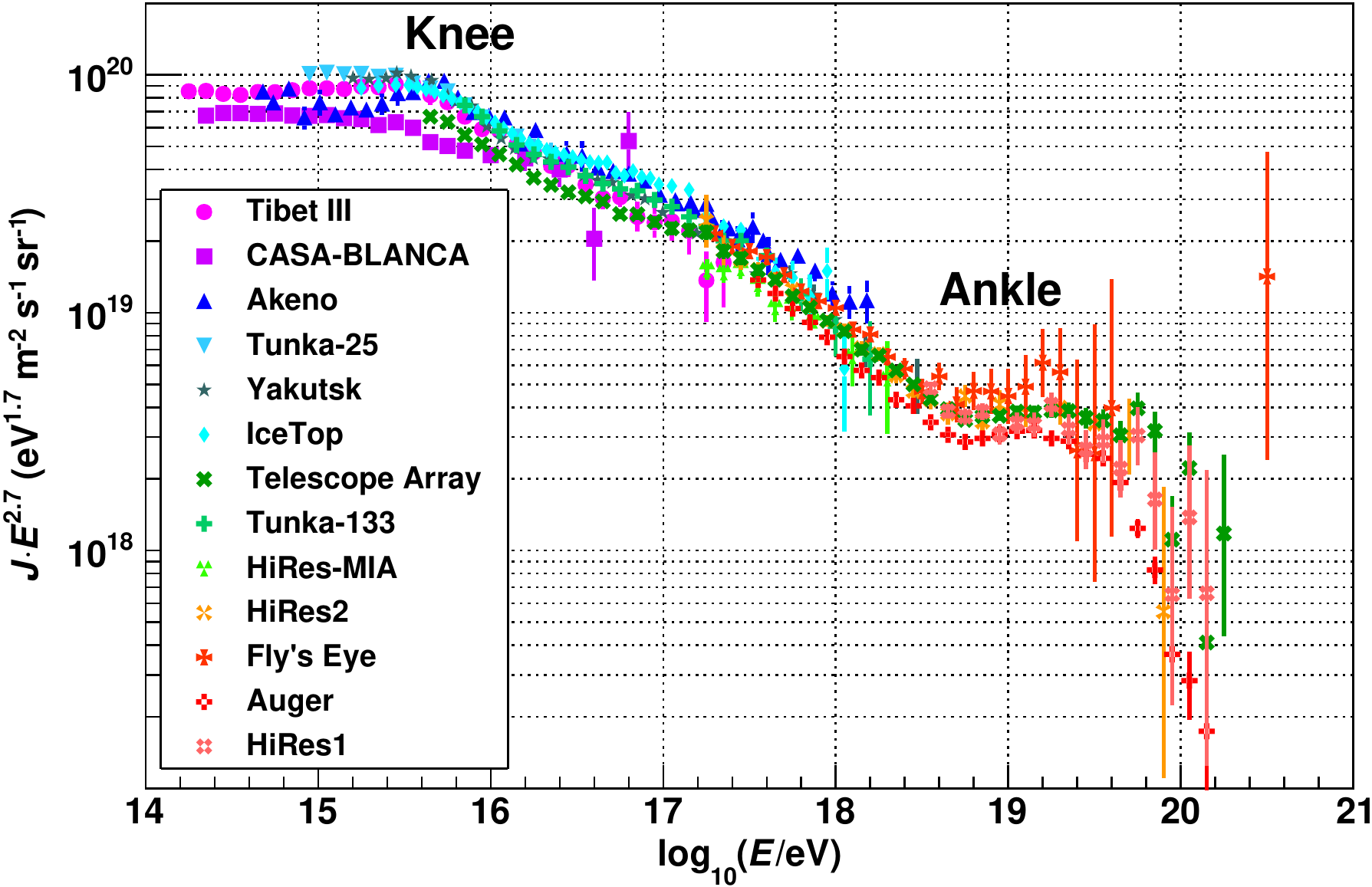}
  \caption{The cosmic ray spectrum observed by recent experiments. In
    this figure the flux is scaled by $E^{2.7}$. Below the ankle
    energy this is approximately the power law followed by the flux,
    and therefore appears flat in this figure. Near the energy of the
    ankle, the spectrum steepens, until flattening again near the
    ankle. Data from \citet{Aartsen:2013wda}, \citet{Abbasi:2007sv},
    \citet{AbuZayyad:2000ay}, \citet{Amenomori:2008sw},
    \citet{Bird:1994wp}, \citet{Fenu:2017hlc}, \citet{Fowler:2000si},
    \citet{Ivanov:2015pqx}, \citet{Knurenko:2015uoa},
    \citet{Nagano:1991jz}, \citet{Prosin:2015voa}.}
  \label{fig:crspec}
\end{figure}

Ultra high energy cosmic ray (UHECR) sources remain a mystery over a
century since they were first observed by Hess in
1912 \citep{HESS2013351}. Outstanding questions regarding the sources,
acceleration mechanisms, propagation, and chemical composition of
UHECRs have been studied now for over five decades, with the first of
the large air shower arrays exceeding an area of 1~km$^{2}$ reporting
results in 1961 \citep{Linsley:1961kt}. The UHECR spectrum, as shown in
Figure~\ref{fig:crspec}, exhibits structure that hints at correlated
changes in chemical composition and energy that can help us resolve
these long standing questions. Of particular interest are the energy
regions of UHECR flux dubbed the ``knee'' ($E \approx 10^{15.5}$~eV),
the ``ankle'' ($E \approx 10^{18.7}$~eV), and the ``GZK cutoff'' (or
suppression) ($E \approx 10^{19.8}$~eV). Prior to the knee, the cosmic
ray flux is remarkably stable for six decades of energy, decreasing
with energy as a power law, $E^{-\gamma}$, with $\gamma \approx
2.7$. The flux then steepens above the knee ($\Delta \gamma \approx
+0.4$), falling more rapidly. Around the energy of the ankle the flux
begins to flatten ($\Delta \gamma \approx -0.6$), until very rapidly
dropping off nearly completely above the GZK cutoff. Models that wish
to describe these changes in flux need to account for the maximum
injection energy of astrophysical sources, acceleration either at the
source or through other means such as shock waves, interactions with
the interstellar medium, and chemical composition of cosmic rays observed in
the Earth's atmosphere.

Composition provides a strong constraint on models describing UHECR
sources and is therefore a fundamental parameter in modelling
them. For instance, models that theorize about the origin of the knee
come in two flavors: astrophysical and interaction
models. Astrophysical models explain the knee as an intrinsic feature
of the energy spectra of individual chemical species, resulting from
magnetic rigidity dependence ($E_{\mathrm{max}} \propto Z$). Different
proposed acceleration mechanisms in magnetic field regions of galactic
supernovae remnants are theorized to boost the energy of particles to
PeV and higher energies, limited by some maximum energy. The
acceleration efficiency of higher $Z$ nuclei allows those heavier
elements to be boosted to higher energies than light elements. The
spectra of individual nuclei shows a series of cascading cutoffs as
energy increases. The all-particle cosmic ray spectrum around the knee
under this model theorizes increasing particle mass with energy. Other
versions of astrophysical models attempting to explain the origin of
the knee, use a leaky box model in which light galactic nuclei can not
be contained in the galaxy as energy increases due to their large
gyroradii. Heavier nuclei can not escape the galaxy as easily and
contribute to the observed flux as a larger proportion of elements at
higher energies. Interaction models of the knee propose that new
physics is in play as the air shower interacts in the atmosphere
producing, for example, undetected supersymmetric particles or other
exotic particles not yet observed in nature. See
\citet{Hoerandel:2004gv} for a review of many different models used to
explain the UHECR knee feature.

In the energy region of the ankle, the cosmic ray flux flattens,
indicating a slight rise in the flux compared to energies below
it. The ankle is traditionally thought of as the energy region where
cosmic rays of extragalactic origin begin to dominate the
spectrum. This is because there are few, if any, known sources in the
galaxy able to accelerate nuclei to $E \gtrapprox 10^{19}$~eV while
allowing the nuclei to remain contained in the galactic disk. A
signature of galactic sources of UHECRs at the energy of the ankle
would be anisotropy of arrival directions in the galactic plane,
something which is not observed. Historically, the flattening of the
spectrum at the ankle was described as the intersection of a steeply
falling galactic spectrum ($\gamma \approx 3.1$) and a flatter
extragalactic spectrum ($\gamma \approx 2-2.3$) at the energy of the
ankle. This model is known simply as the ankle model. More recent
models fitted to data from large cosmic ray experiments use the
signatures of propagation through the photon field of cosmic microwave
background (CMB) radiation in intergalactic space. These propagation
effects result in the suppression of flux above $10^{20}$~eV due to
photopion production with CMB photons (the GZK
mechanism \citep{Greisen:1966jv,Zatsepin:1966jv}), a bump due to pile up
in the flux for primaries with energies below the photopion production
energy, and a dip due to pair production during interaction with CMB
photons. This is called the dip model. The dip model predicts that the
galactic component of UHECRs disappears at a lower energy compared to
predictions of the ankle model. The dip model is relatively
insensitive to model parameters such as size and inhomogeneity of the
source distributions, cosmological evolution of sources, and maximum
acceleration energy. However, the dip model is sensitive to
composition of heavy nuclei in the spectrum. Heavier nuclei interact
and photospallate readily due to larger cross section than protons,
resulting in several primaries of lower total energy and a change in
the shape of the dip. See, for example, \citet{Aloisio:2006wv} for
further discussion of the dip model and the transition from galactic
to extragalactic cosmic rays. This model therefore makes testable
predictions based upon composition as well.

UHECR composition can be measured directly up to about $10^{14}$~eV
because the event rate here is about 1/m$^2$/h, sufficient for
balloon-borne or satellite experiments and their associated equipment
to measure particle mass. Above this energy the particle cascades
created by the cosmic ray primary inelastically colliding with an air
molecule must be observed if one wishes to collect sufficient data. To
do this, large ground based experiments are required to observe the
parts of the shower that survive to ground level or using fluorescence
detectors to observe the light produced by the air shower. Neither
method directly measures the mass of the primary, and a single
observation of muons on the ground or light generated from an air
shower can not reveal the mass of an individual primary. Therefore to
measure the mass composition of ultra high energy cosmic rays we must
resort to understanding the physics of extensive air showers,
identifying those observables that can be related to the primary mass,
and collecting large data sets to build statistical samples of
sufficient size to reliably measure the average mass in some energy
range. This method requires good hadronic models of high energy
interactions in matter to energy ranges not yet measured in the
lab. Measurements from the LHC reaches up to about $10^{17}$~eV in the
lab frame, whereas UHECR primary particle energies above $10^{20}$~eV
have been measured. Hadronic models which predict particle elasticity,
multiplicity, and interaction cross section currently require
extrapolation over a few decades of energy for the energy region below
the ankle and above.

UHECR composition measurements are performed best by fluorescence
detectors which observe the depth in the atmosphere where the
electromagnetic component of a cosmic ray induced shower reaches a
maximum. This atmospheric depth is called \xm{} and is measured in
g/cm$^{2}$. A toy model first developed by
Heitler \citep{Carlson:1937zz} demonstrates how \xm{} is related to the
primary particle energy and mass using a simple branching model of
electromagnetic (EM) showers in which a high energy electron primary
of energy $E_0$ collides inelastically with a target particle. The EM
shower is created and grows in size through the repeated processes of
pair production and bremsstrahlung. In this model, making the
simplistic assumption of a fixed interaction length, $\lambda$,
between interactions, two new particles are generated and added to the
shower for each existing particle. After $n$ interactions, the total
depth the shower has traveled is $n\lambda = X$, and the size of the
shower is $N(X) = 2^{n} = 2^{X/\lambda}$ The average energy of each
particle at depth $X$ is $E(X) = E_{0}/N(X) =
E_{0}/2^{X/\lambda}$. This process of particle generation at each
interaction length continues until the average energy per particle
decreases below some critical energy, $E_{\mathrm{c}}$, defined as the
energy at which particle energy lost due to collisions exceeds
radiative energy losses. When the average particle energy is equal to
$E_{\mathrm{c}}$, the shower reaches its maximum size, called \nm, and
the depth is \xm. Using the definition of $N(X)$, we find
$N_{\mathrm{max}} = E_{0}/E_{\mathrm{c}}$ and $X_{\mathrm{max}}
\propto \ln(E_{0}/E_{\mathrm{c}})$; the number of particles generated
at shower maximum is proportional to the primary particle energy, and
\xm{} is proportional to the log of that energy.

Showers initiated by a very high energy hadronic primary particle
exhibit similar relationships. If the primary particle has mass $A$
and energy $E_{0}$, we use the superposition principle to treat the
particle as $A$ independent nucleons each with an average initial
energy of $E_{0}/A$.  Using the Heitler model under the assumption of
the superposition principle we find $X_{\mathrm{max}} \propto
\ln(E_{0}/A)$. In reality hadronic showers are more complicated, since
for each hadronic interaction, on average 2/3 of the particles
produced are charged particles such as $\pi^{\pm}$ and 1/3 are
$\pi^0$. The $\pi^0$ rapidly decay into two photons which contribute
to the electromagnetic part of the shower. The relation between
particle mass and \xm{} predicted by the Heitler model is still
valid. The property of shower universality tells us that for showers
created by a hadronic primary particle of any mass the electromagnetic
component evolves in the same way, parameterized by the shower age,
$s$ \citep{Giller:2003xf}. Using this property we can use the same
method of observing \xm{} of a shower to determine the mass of the
primary particle, even if that particle is very light, such as a
proton, or much heavier, such as an iron nucleus. For details about
the treatment of hadronic showers as related to cosmic ray composition
refer to \citet{Engel:2011zzb} and \citet{Kampert:2012mx}.

\begin{figure}
  \centering
  \includegraphics[width=\textwidth]{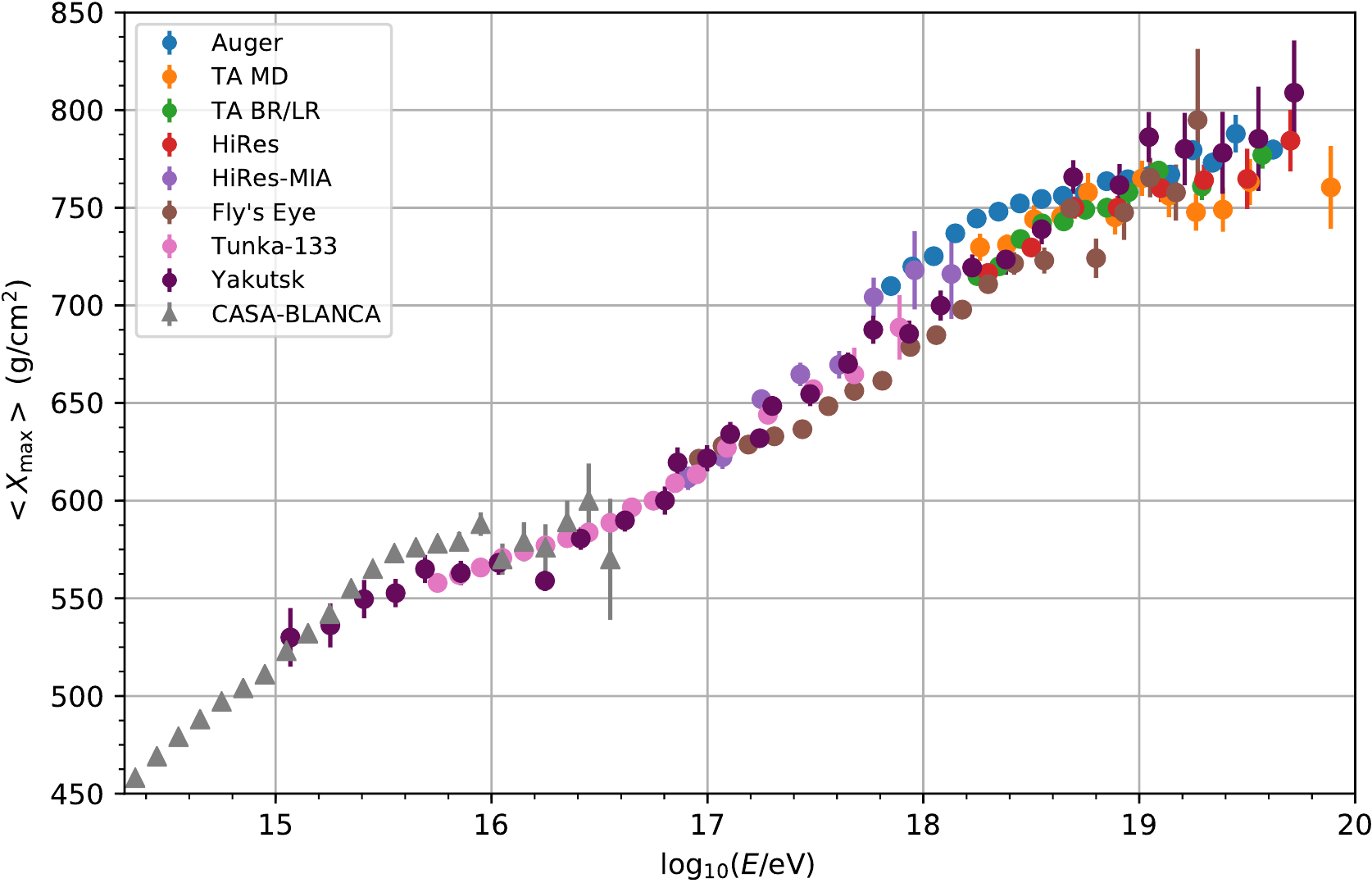}
  \caption{\mxm{} observed by recent experiments. The dependence of
    mean depth with $\log(E)$ over many decades of energy is readily
    apparent. Data from \citet{Aab:2014kda}, \citet{Abbasi:2009nf},
    \citet{Abbasi:2014sfa}, \citet{AbuZayyad:2000ay},
    \citet{Bird:1994wp}, \citet{Fowler:2000si}, \citet{Knurenko:2015apa},
    \citet{Prosin:2015voa}, and this work (TA BR/LR).  }
  \label{fig:crxmax}
\end{figure}

Heavier primary particles are therefore expected to reach shower
maximum at shallower depths in the atmosphere than light primaries. To
experimentally measure UHECR composition, one can use fluorescence
detectors to record energy and \xm{} for many showers. For a given
energy range \mxm{} for lighter primaries will be larger than for
heavy primaries. In addition because of the superposition principle,
the fluctuations in \xm{} are expected to be smaller for heavy
primaries. This data can be compared to models of individual primary
species or mixtures of elements to determine the composition of UHECRs
observed. Figure~\ref{fig:crxmax} shows \mxm{} measured over by many
experiments over the past 30 years.

Recent measurements of \xm{} over the past decade have vastly improved
statistics in the important region of the ankle and above, with the
three largest fluorescence based measurements of HiRes, Telescope
Array, and Auger. HiRes reported on composition measured by events
observed in stereo using simultaneous observation of two fluorescence
detectors \citep{Abbasi:2009nf} and Telescope Array has presented
results using hybrid reconstruction \citep{Abbasi:2014sfa}. Both
results found indications of composition consisting of light primaries
resembling mostly protons up to $10^{19.8}$~eV, by comparing \mxm{}
and \sxm{} of data to models. These results are consistent with the
view (up until that time), that UHECRs with energies at the ankle and
higher are most likely extragalactic protons. The existence of a
suppression in the flux at the energy predicted by the GZK mechanism,
first observed by HiRes \citep{Abbasi:2007sv}, and later confirmed by
Auger \citep{Abraham:2008ru}, fit in well with this scenario. However,
Auger's recent measurement of composition challenges this view. Auger
has shown an energy evolution in \mxm{} and \sxm{} to heavier
primaries above $10^{18.3}$~eV \citep{Aab:2014kda}. UHECR flux with
increasing mass above the ankle leads to unexpected models that have
been deemed ``disappointing'' for the
field \citep{Aloisio:2011fv}. Implications of such models are a lack of
photopion production on CMB photons (and therefore very few ultra high
energy cosmogenic neutrinos), no anisotropy of nearby sources due to
strong deflection in magnetic fields, and no cutoff in the spectrum
due to the GZK mechanism since the maximum energy of astrophysical
accelerators is too low \citep{Allard:2008gj,Aloisio:2011fv}. The
tension between these experimental results, and the implications for
particle astrophysics, provide the impetus for further, high precision
studies of UHECR composition such as this one.

This work presents an analysis of \xm{} data collected by the
Telescope Array experiment over an 8.5 year period. The \xm{}
distributions are collected in energy bins and \mxm{} and \sxm{} are
computed for each bin. Four sets of Monte Carlo, each representing a
single chemical element, are generated and then reconstructed in the
same manner as the data. The \xm{} distributions, \mxm{}, and \sxm{}
of the Monte Carlo are compared to the observed data. Statistical
tests are used to compare the compatibility of the Monte Carlo to the
data. Section~\ref{sec:apparatus} describes the design and operation
of the Telescope Array experiment. Section~\ref{sec:measurement}
describes the hybrid method of observation and how data is
reconstructed to measure \xm{} for air
showers. Section~\ref{sec:data_analysis} examines the data collected,
analysis cuts applied to the data, resolution and bias of observables
important to good \xm{} reconstruction, compares data to Monte Carlo,
and discusses the importance of understanding the different types of
biases in measuring \xm. Section~\ref{sec:hypothesis_tests} details
the statistical tests used to measure compatibility of the different
Monte Carlo sets to the data and the results of these
tests. Conclusions of this analysis are presented in
Section~\ref{sec:conclusions}.

\section{\label{sec:apparatus}Apparatus}
Telescope Array (TA) is one of the few detectors in the world able to
shed light on the composition of UHECRs. TA is the successor
experiment of the AGASA \citep{Ohoka:1996ww} and
HiRes \citep{AbuZayyad:2000uu,Boyer:2002zz} experiments. Expertise
using surface arrays from the AGASA experiment and fluorescence
detectors from the HiRes experiment is combined into a single cosmic
ray observatory able to observe ultra high energy cosmic ray flux over
four decades of energy.

\begin{figure}
  \centering
  \includegraphics[clip,width=\textwidth]{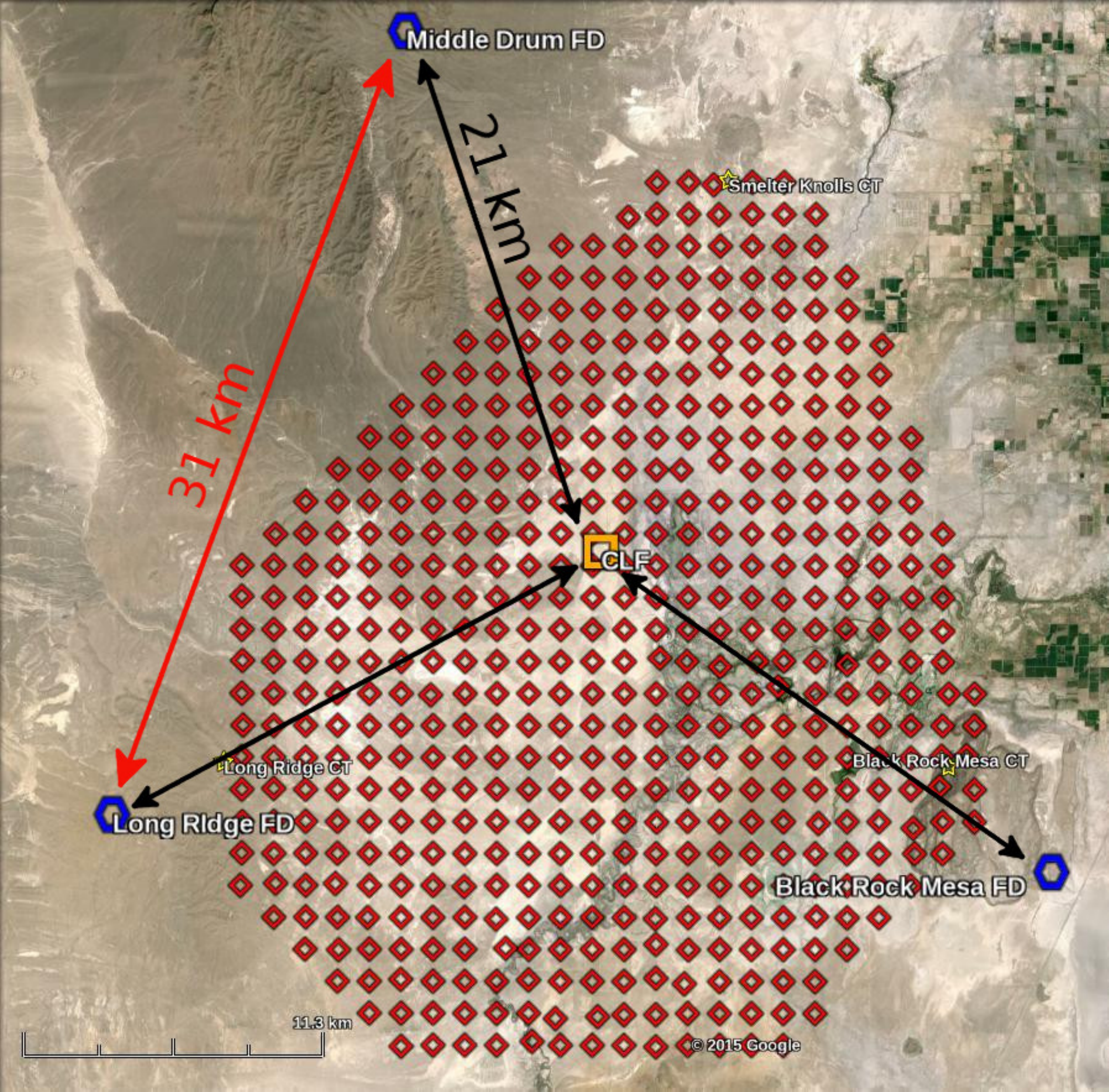}
  \caption{Location of Telescope Array surface detectors and
    fluorescence detector stations. Each red diamond indicates the
    location of one of the 507 surface detectors. The blue hexagons
    show the locations of the fluorescence detectors which look inward
    over the SD array. Each FD is about 30~km distant from its
    neighboring FDs. All FDs are placed 21~km from a central laser
    facility}
  \label{fig:ta_map}
\end{figure}

TA is located in Millard County Utah ($39.3^\circ$N and
$112.9^\circ$W, 1400~m above sea level), consisting of 507
scintillation surface detectors (SDs) sensitive to muons and
electrons, and 48 fluorescence telescopes located in three
fluorescence detector (FD) stations overlooking the counters. The
spacing of the counters in the SD array is 1.2~km and they are placed
over an area of approximately 700~km$^{2}$. Figure~\ref{fig:ta_map}
shows the physical locations of the SDs and FDs.

Each surface detector is made up of two layers of plastic scintillator
3~m$^{2}$ by 1.2~cm thick. Grooves running parallel along the length
of each layer are etched into each scintillator layer and 104
wavelength shifting fiber optic cables are embedded in them, for a
total length of 5~m of fiber in each layer. When a charged particle
passes through the scintillator and light is produced, the light is
transmitted to a photomultiplier tube (PMT) via the fiber. Each
scintillator layer has a dedicated PMT that is optically coupled to
the fiber bundle to detect the passage of charged particles. The
analog PMT signal is digitized via 12 bit FADC electronics operating
at 50~MHz sampling rate with signals stored in a local buffer. Each
FADC bin is 20~ns wide and a waveform consists of 128 FADC bins,
providing a waveform buffer 2.56~$\mu$s wide. Each SD electronics
suite also has a FPGA which continuously monitors the FADC waveforms
to monitor pedestals, and to determine if the event trigger condition
is met. When an SD measures a signal above threshold, it can announce
it to a remote DAQ via radio communications. An SD can record two
types of low level triggers: a level 0 trigger, in which an integrated
signal exceeding 15 FADC counts above pedestal is measured, and a
level 1 trigger in which an integrated signal exceeding 150 FADC
(equivalent to 3 MIPs or minimum ionizing particles) counts above
pedestal is measured.  These remote DAQ locations are referred to as
communication towers (CTs), as they monitor and receive data from many
SDs and make the decision about high level triggers based upon the low
level trigger logic of all SDs that is communicates with. If three or
more adjacent SDs announce level 1 triggers within an 8~$\mu$s window,
this constitutes a level 2 event trigger and the CT directs all SDs
that observed a level 0 trigger within $\pm 32 \mu$s of the event to
send the waveform data to the CT for storage. Each SD has an onboard
GPS unit to timestamp event triggers, so the time of particle passage
is also recorded by each SD and included as part of the event
information.

SD event reconstruction is done by examining all level 2 triggers and
finding SDs that have sufficient signal to noise ratio, and also are
connected in a small space-time window. Using the positions of the SDs
and their relative trigger times, the shower core and track direction
can be determined. Shower energy is measured by relating the SD signal
size 800~m from the shower axis (called S800) to a function which maps
S800 and the shower zenith angle to primary particle energy. This
mapping is determined by Monte Carlo simulation using CORSIKA and is
therefore model dependent. The final shower energy is found by scaling
this energy mapping via a scale factor by using real events observed
by both FD and SD, and correcting the energy to that determined by the
FD. Detailed information about the operation of the SD array can be
found in \citet{AbuZayyad:2012kk,Ivanov:2012tex}.

The three FD stations are placed on the periphery of the SD array and
look towards its center. Each is located the same distance from a
central laser facility (CLF) about 21~km away, which contains a
calibration laser that is fired throughout the night to monitor each
FD's response and to monitor atmospheric quality. The Middle Drum (MD)
FD station is located on the northern border of the SD array. It has
14 fluorescence telescopes that view $112^\circ$ in azimuth arranged
in two rings of elevation angle coverage. Ring 1 telescopes observe
from $3^\circ$ to $17^\circ$ in elevation and ring 2 telescopes
observe from $17^\circ$ to $31^\circ$. Seven telescopes are in each
ring. Each telescope consists of a 5.1~m$^2$ mirror which reflects
light from the sky onto a cluster of 256 PMTs arranged in a $16 \times
16$ array, with each PMT monitoring approximately 1 millisteradian
solid angle. Middle Drum is built using the same sample and hold
electronics and hardware that was used at the HiRes1 FD in the HiRes
experiment \citep{AbuZayyad:2000uu}. The Middle Drum site is also the
location of the TALE FD station, which is has ten telescopes and is
designed to observe low energy cosmic rays near the energy of the
ankle.

On the southeast and southwest borders of the SD array are the Black
Rock Mesa (BR) and Long Ridge (LR) fluorescence detector
stations. Each of these are made up of 12 telescopes in a two ring
configuration similar to Middle Drum. The electronics and hardware of
these stations were newly built for the Telescope Array experiment and
utilize FADC electronics. Each telescope has 256 PMTs focused onto a
6.8~m$^2$ mirror for light collection. Each PMT's analog signals are
digitized by FADC electronics which employ 12 bit digitizers operating
at 40~MHz. Before storage to the DAQ, four digital samples are summed
to provide an equivalent 14 bit, 10~MHz sampling rate providing a time
resolution of 100~ns. Each telescope employs a track finder module
which applies the trigger logic to incoming waveforms searching for
spatial patterns which indicate a track caused by an extensive air
shower. When an event trigger occurs a central computer orders the
readout of telescope electronics to store the data for later offline
analysis. More detailed information describing the construction and
operation of the BR and LR fluorescence detectors can be found in
\citet{Tameda:2009zza,Tokuno:2012mi}.

When an UHECR primary particle interacts in the atmosphere, an
extensive air shower results, producing copious amounts of electrons
and positrons among many other particles types. This electromagnetic
component of the shower interacts with atmospheric N$_2$ producing
fluorescence light, which is emitted isotropically and observed by FD
telescopes on the ground. Because of the typical large distances to
the showers and the relatively large size of each PMT's field of view,
the shower appears as a downward traveling line source. Each tube
observes the shower at different altitudes and therefore different
atmospheric depth. Tube signals will vary depending upon the amount of
light produced in the sky at that depth, the atmospheric clarity, and
distance to the shower. The goal of FD shower reconstruction is to
convert the signals and times measured by the passage of the shower
along with the fixed, known geometry of the PMTs, into the location
and direction of a distant shower track, as well as the energy of the
primary particle.

To accurately measure \xm{}, fluorescence detectors must be used. For
those events that also have their arrival time simultaneously measured
by the surface detector (SD) array though, the geometry of an
individual shower can be very well measured. These \textit{hybrid}
events are very valuable for use in high quality measurements of
\xm. Telescope Array began hybrid data collection in May 2008 and has
now analyzed over eight and a half years of data using this
method. This paper is the first to report on \xm{} measurements using
the BR and LR stations.

The results of \xm{} measurement using the Middle Drum FD and the SD
array are reported in \citet{Abbasi:2014sfa}. This analysis uses only
the combined data of both Black Rock and Long Ridge hybrid
events. Black Rock and Long Ridge employ identical electronics and
hardware design, which are different from that found at Middle
Drum. For example Black Rock and Long Ridge use larger mirrors than
Middle Drum, which affects triggering and acceptance of
tracks. Because hybrid reconstruction uses both the FDs and SDs the
locations of the FD stations relative the SD array border is also an
important consideration. Black Rock and Long Ridge, located 3 and 4~km
away from the SD border respectively, are closer to the SD array than
Middle Drum, which is 8~km away. This affects the
acceptance of low energy hybrid events. Black Rock and Long Ridge are
more efficient below $10^{18.5}$~eV than Middle Drum is. Because of
these differences Middle Drum reconstructed events are not considered
for this analysis.

\section{\label{sec:measurement}Measurement}
\subsection{\label{sec:hybrid_analysis_method}Hybrid Analysis Methodology}
Hybrid reconstruction combines the separate SD and FD data streams by
searching for time coincident events. The kinematic properties of the
shower, such as the charged particle depth profile, primary energy,
and \xm{}, are determined using the standard profile fitting procedure
applied to FD data. The time coincident SD data is used only to
improve the shower track geometry, because it can very
accurately measure the time of arrival and position of the shower
core. Hybrid reconstruction strives to use the same routines as those
used in the standard SD-only and FD-only analyses. 

Hybrid reconstruction of data consists of the following steps:
\begin{enumerate}
\item Tagging events which trigger both the SD and FD.
\item FD and SD geometry measurement.
\item Hybrid fitting.
\item FD Profile fitting.
\end{enumerate}

\subsubsection{\label{sec:hybrid_event_tag}Tagging Hybrid Events}
Tagging hybrid events is done by searching for hybrid event candidates
that independently trigger the SD array and either of the BR or LR FD
stations. For real data the FD data stream is searched for
downward-going events, and the SD data stream is searched for all
level 2 triggers. Level 2 triggers are events in which three or more
adjacent SD counters individually within 2400~m of each other, detect
a three minimum ionizing particle (MIP) signal within
8~$\mu$s \citep{AbuZayyad:2012kk}. FD downward-going events satisfy the
requirements of the FD event trigger and are traveling from the
atmosphere towards the ground \citep{Tameda:2009zza}. Hybrid events are
found by time matching SD and FD events using a 500~$\mu$s window.

For Monte Carlo data a shower library of pregenerated CORSIKA events
is used. The hybrid MC CORSIKA shower library is a collection of air
showers generated at predetermined energies and zenith angles. For
energies above $10^{18}$~eV TA has 250 showers generated per 0.1
decade in energy. We generate showers of arbitrary energy by
measuring the elongation rates of the Gaisser Hillas profile
parameters, and correcting them from their generated values to the
values expected for the energy being thrown for the MC. In this way we
can throw a continuous distribution of shower profiles for any
randomly chosen energy. Each CORSIKA shower library element is
thrown and set at a fixed zenith angle. Shower zenith angles are
chosen according to a $\sin\theta \cos\theta$ distribution for
$0^\circ \leq \theta \leq 60^\circ$. To generate a random shower the
following procedure is performed:

\begin{enumerate}
\item A shower energy is randomly chosen according to the energy
  distribution of the combined HiRes1 and HiRes2 observed spectrum
  \citep{Abbasi:2007sv}.
\item A pregenerated shower element contained in the energy bin chosen from the
  previous step is randomly selected.
\item Shower azimuthal angle is chosen by a uniform distribution
  for $0^\circ \leq \phi < 360^\circ$.
\item Shower core is randomly selected within a circle of 25~km
  radius centered on the CLF.
\end{enumerate}
Monte Carlo generation also assumes a given primary particle type and
other choices that must be made when running CORSIKA
\citep{1998cmcc.book.....H}, such as the high and low energy hadronic models.

At the conclusion of this stage of the reconstruction each event will
contain data from the SD counters and at least one of the FD
stations. A hybrid event candidate then has $\mathrm{SD} \oplus
\mathrm{BR}$ data, or $\mathrm{SD} \oplus \mathrm{LR}$ data, or
$\mathrm{SD} \oplus \mathrm{BR} \oplus \mathrm{LR}$ data, for events
energetic enough to have been observed simultaneously by both FD
stations.

This step of hybrid analysis is the only step which differs between
reconstructing real data and Monte Carlo data. The real data and
simulated data are packed into the same data format so that all
analysis which follows uses the same software.

\subsubsection{\label{sec:fd_sd_geometry}FD and SD Geometry Measurement}
This step of reconstruction determines the geometry of the shower as
independently observed by the FD and by the SD. FD plane fitting is
performed using the standard routines described in
\citet{Abu-Zayyad:2013jra,Stratton:2012pfa}. This procedure determines
the geometry of the air shower track relative to the observing
FD. This includes measuring the shower-detector plane (SDP) normal
vector ($\bm{\hat{n}}$), shower zenith angle ($\theta$), shower
azimuth angle ($\phi$), SDP angle ($\psi$), shower impact parameter
($R_{\mathrm{p}}$), core location and arrival time.

\begin{figure}
  \centering
  \includegraphics[clip,width=\textwidth]{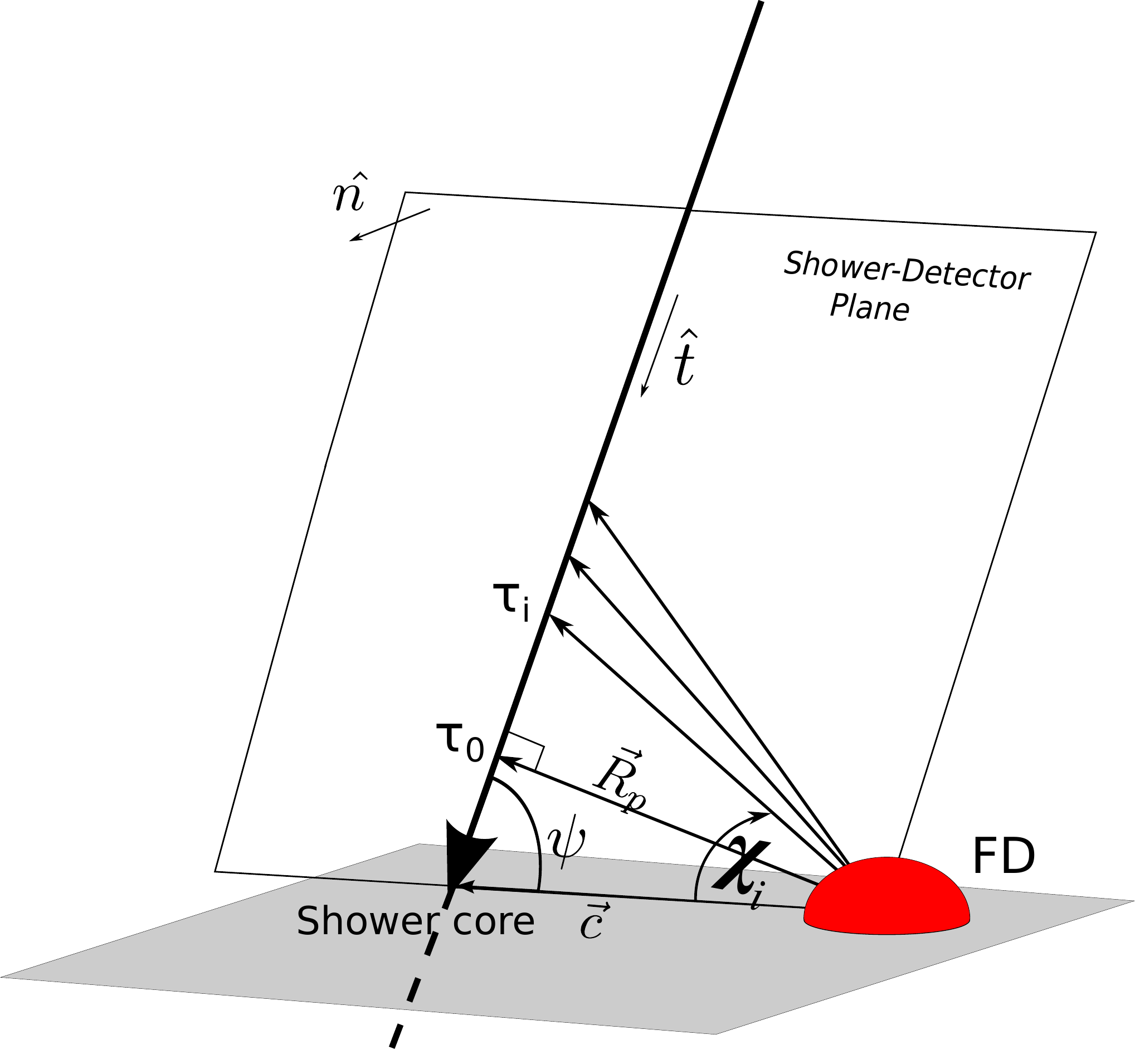}
  \caption{Geometry of an air shower track as viewed by a fluorescence
    detector. The track vector $\bm{\hat{t}}$ is a unit vector which
    points along the shower in the direction which the shower travels,
    $\bm{c}$ points from the FD to the point on the ground where the
    shower impacts (the core), $\bm{R}_{\mathrm{p}}$ is the impact
    parameter, and $\psi$ is the SDP angle. Each PMT that observes the
    shower has a viewing angle $\chi_i$ and triggers at time
    $t_i$. The time $\tau_i$ is the difference between the PMT trigger
    time and the light travel time from the shower axis. The SDP
    normal vector is $\bm{\hat{n}}$. All vectors and angles except
    $\bm{\hat{n}}$ are in the SDP.}
  \label{fig:fd_sdp}
\end{figure}

FD geometry reconstruction begins by selecting tubes that correlate
closely in time and space on the PMT cluster face. For any event
randomly triggered noise tubes are present as well. A filtering step
is performed to reject tubes that are not part of the shower
track. The SDP normal vector is found by finding those components of
the vector which minimize the following sum
\begin{equation} \label{eq:fd_plane_chi2}
\chi^2 = \sum_{i = 1}^{N} (\bm{\hat{n}} \cdot \bm{\hat{v}}_i)^2 N_{\mathrm{pe},i}
\end{equation}
where $N$ is the number of tubes along the shower track,
$\bm{\hat{v}}_i$ is the pointing direction of tube $i$ in the SDP, and
$N_{\mathrm{pe},i}$ is the tube signal measured as number of
photoelectrons. FDs can accurately measure the SDP normal because many
individual tubes are used to perform the fit. A typical shower
contains 50 tubes along the shower track. Figure~\ref{fig:fd_sdp}
illustrates the relationship among the different parts of the SDP.

Once the SDP is known, the shower impact parameter and the SDP angle
are calculated by the time vs. angle fit. The expected trigger time of
PMT $i$ is
\begin{equation} \label{eq:time_vs_angle}
\tau_i(\chi_i; R_{\mathrm{p}}, \psi, t_0) = t_0 +
\frac{R_{\mathrm{p}}}{c} \tan\left(\frac{\pi - \psi - \chi_i}{2}\right)
\end{equation}
$R_{\mathrm{p}}$, the shower impact parameter, also called the
pseudodistance, is the point of closest approach of the shower track
and the origin of the observing FD. $\psi$ is the SDP angle, or the
angle on the ground between the shower track and the core vector,
which points from the FD origin to the point of impact on the
ground. $\chi_i$ is the pointing direction of the PMT. $t_0$ is the
time when the shower front is located at the $R_{\mathrm{p}}$
point. $R_{\mathrm{p}}$, $\psi$, and $t_0$ are parameters to be
determined by fitting. A $\chi^2$ minimization is performed on the
function that measures the difference between the observed tube
trigger times and the expected trigger times, described by
Equation~\ref{eq:time_vs_angle}
\begin{equation} \label{eq:time_vs_angle_chi2}
\chi^{2} = \sum_{i=1}^{N} \frac{1}{\sigma^{2}_{i}} \left[ t_i - \left\{ t_0
  + \frac{R_{\mathrm{p}}}{c} \tan \left( \frac{\pi - \psi - \chi_i}{2}
  \right) \right\} \right]^2
\end{equation}
where $t_i$ is the observed trigger time and $\sigma_i$ is the timing
uncertainty for tube $i$.

SD reconstruction is performed as well to determine the core arrival
time and core location. A similar procedure for separating counters
which trigger during an event and those that are noise triggers is
employed by using time and space pattern recognition. Details about SD
reconstruction can be found in \citet{Ivanov:2012tex}.

At the end of this processing step, we have two independent
determinations of the shower's point of impact on the ground and its
time of arrival.

\subsubsection{\label{sec:hybrid_fitting}Hybrid Fitting}
\begin{figure}
  \centering
  \includegraphics[clip,width=\textwidth]{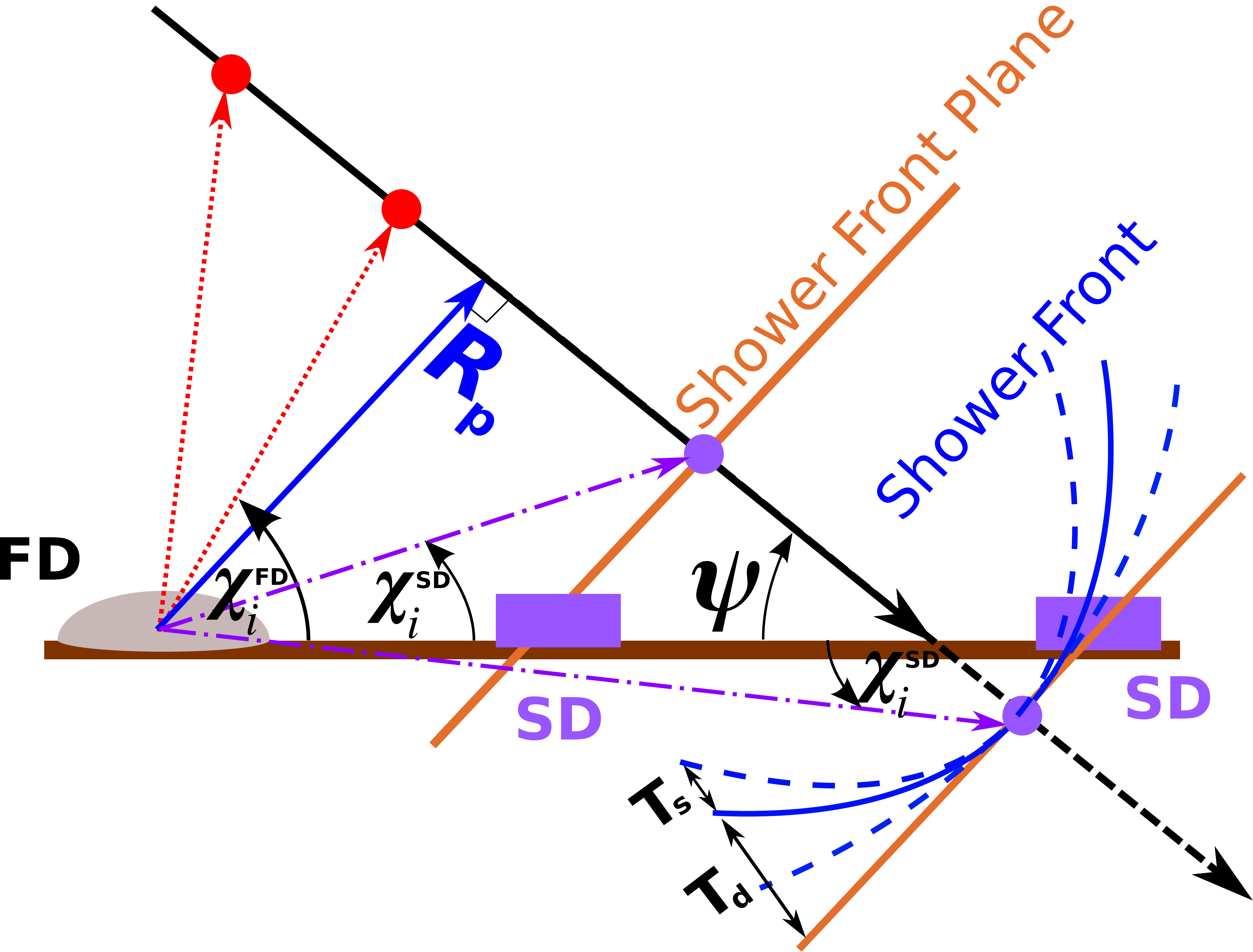}
  \caption{FD and SD components of shower geometry for a hybrid
    event. Each FD tube or SD counter can be considered a ``pixel''
    with known geometry and timing that marked the passing of the
    shower.  Timing of the individual SDs is corrected for shower
    front delay, where $T_{\mathrm{d}}$ is the mean delay from the
    shower front in particle arrival time at the SD, and
    $T_{\mathrm{s}}$ is the uncertainty on the delay time.}
  \label{fig:fd_sd_shower_geom}
\end{figure}

Hybrid fitting is done by casting the individual ``pixels'' of the FD
and SD that observe the passage of the shower into a common frame of
reference, then performing a four component $\chi^2$ minimization to
redetermine some parameters of the shower geometry. By combining the
independent FD and SD observations, the geometry fit is improved
compared to FD-only or SD-only observations. The hybrid function minimized
for this analysis is
\begin{equation} \label{eq:hybrid_chi2}
\chi^{2}_{\mathrm{hybrid}} = \chi^{2}_{\mathrm{FD}} +
\chi^{2}_{\mathrm{SD}} + \chi^{2}_{\mathrm{SDP}} + \chi^{2}_{\mathrm{COC}}
\end{equation}
and the parameters being minimized are shower core positon, $C_x$ and
$C_y$, zenith angle, $\theta$, azimuth angle, $\phi$, and shower core
arrival time, $t_c$. Figure~\ref{fig:fd_sd_shower_geom} shows how the
common geometry of the FD and SDs is constructed. The FD pixels are
treated as they typically are for normal FD reconstruction, each
assigned a pointing direction and trigger time to determine the
position and time of the shower point observed on the shower
axis. The geometry and timing of the SDs that observe a common event
are used to determine the pointing direction and times of those points
on the shower axis in the reference frame of the FD.

$\chi^{2}_{\mathrm{FD}}$ is the same as
equation~\ref{eq:time_vs_angle_chi2} using the FD data described in
Section~\ref{sec:fd_sd_geometry}. It is included in
$\chi^2_{\mathrm{hybrid}}$ because in this fitting procedure the
position of the shower core and the SDP are allowed to vary,
potentially changing the previously calculated parameters of
$R_{\mathrm{p}}$, $\psi$, and $t_0$. $\chi^{2}_{\mathrm{SDP}}$
(equation~\ref{eq:fd_plane_chi2}) is the same as described in
Section~\ref{sec:fd_sd_geometry} as well, allowing for varying shower
track geometry. Because we cast the SD positions and times into a
common reference frame as the FD tube data, $\chi^{2}_{\mathrm{SD}}$
uses equation~\ref{eq:fd_plane_chi2} to measure the best values for
$R_{\mathrm{p}}$, $\psi$, and $t_0$ of the shower
track. $\chi^{2}_{\mathrm{COC}}$ is a term used to directly relate the
determination of the shower core as observed by the SDs (the center of
charge) to the new shower core being fitted in
equation~\ref{eq:hybrid_chi2}. This term is calculated by
\begin{equation} \label{eq:chi2_coc}
\chi^2_{\mathrm{COC}} = \frac{(C_x - R_x)^2 + (C_y - R_y)^2}{\sigma_R^2}
\end{equation}
where $R_x$ and $R_y$ are the (fixed) components of the shower core as
observed by the SD array, and $\sigma_R$ is the SD uncertainty on the
core location.

\subsubsection{\label{sec:profile_fitting}Profile Fitting}
Once the shower geometry is improved by folding in the SD data, FD
profile fitting is done as it is normally done and described in
\citet{Abu-Zayyad:2013jra}. This means properties of the shower, such
as the primary particle energy and the depth of shower maximum, are
determined using only the light profile observed by the FD.

An inverse Monte Carlo method is used to determine the four parameters
of the Gaisser-Hillas equation \citep{1977ICRC....8..353G} as shown in
equation~\ref{eq:gh}, \nm{}, \xm{}, $X_0$, and $\lambda$, which, after
simulating the measured shower geometry, atmosphere, and detector
acceptance, best mimics the tube-by-tube response measured in the FD.
\begin{equation}\label{eq:gh}
N(x) = N_{\mathrm{max}} \left( \frac{x - X_0}{X_{\mathrm{max}} - X_0}
\right)^{\frac{X_{\mathrm{max}} - X_0}{\lambda}} \exp \left(
\frac{X_{\mathrm{max}} - x}{\lambda} \right)
\end{equation}
This analysis imposes constraints on the values of $\lambda$ and
$X_0$, fixing them to 70~g/cm$^2$ and -60~g/cm$^2$ respectively. It
has been observed that $\lambda$ and \nm{} are correlated, meaning
$\lambda$ is strictly not an independent parameter. Also, ground based
fluorescence detectors can not observe the depth of first interaction,
which $X_0$ is supposed to represent, but it is found to be physically
meaningless since it often takes negative
values \citep{Song:2004pk}. We find, for this experiment, that by
fixing these two parameters the bias in energy and \xm{} is reduced
for simulated proton induced showers over our energy range of
interest. Both data and Monte Carlo utilize databases of detector PMT
pedestals, PMT gains, and atmospheric conditions. These time-dependent
databases are generated from the actual data and describe accurately
the actual detector response recorded over each night's observations.

Once the best shower profile is determined, the energy of the primary
particle is found by integrating the shower profile and applying
corrections for energy transported by components of the shower not
observed by direct fluorescence detection, i.e., neutrinos and
muons. This correction is about 8\% to 5\% for proton induced showers
at $10^{18}$~eV and $10^{20}$~eV respectively. For iron induced
showers the missing energy correction is about 4\%
larger \citep{Barcikowski:2011xea}. This analysis applies the missing
energy correction calculated from simulated proton induced showers to
all data and all Monte Carlo. We therefore expect a relative energy
bias of around 4\% in the reconstruction of proton and iron Monte
Carlo data.

\begin{figure}
  \centering
  \subfloat[Surface detector display.]{%
    \includegraphics[clip,width=0.48\columnwidth]{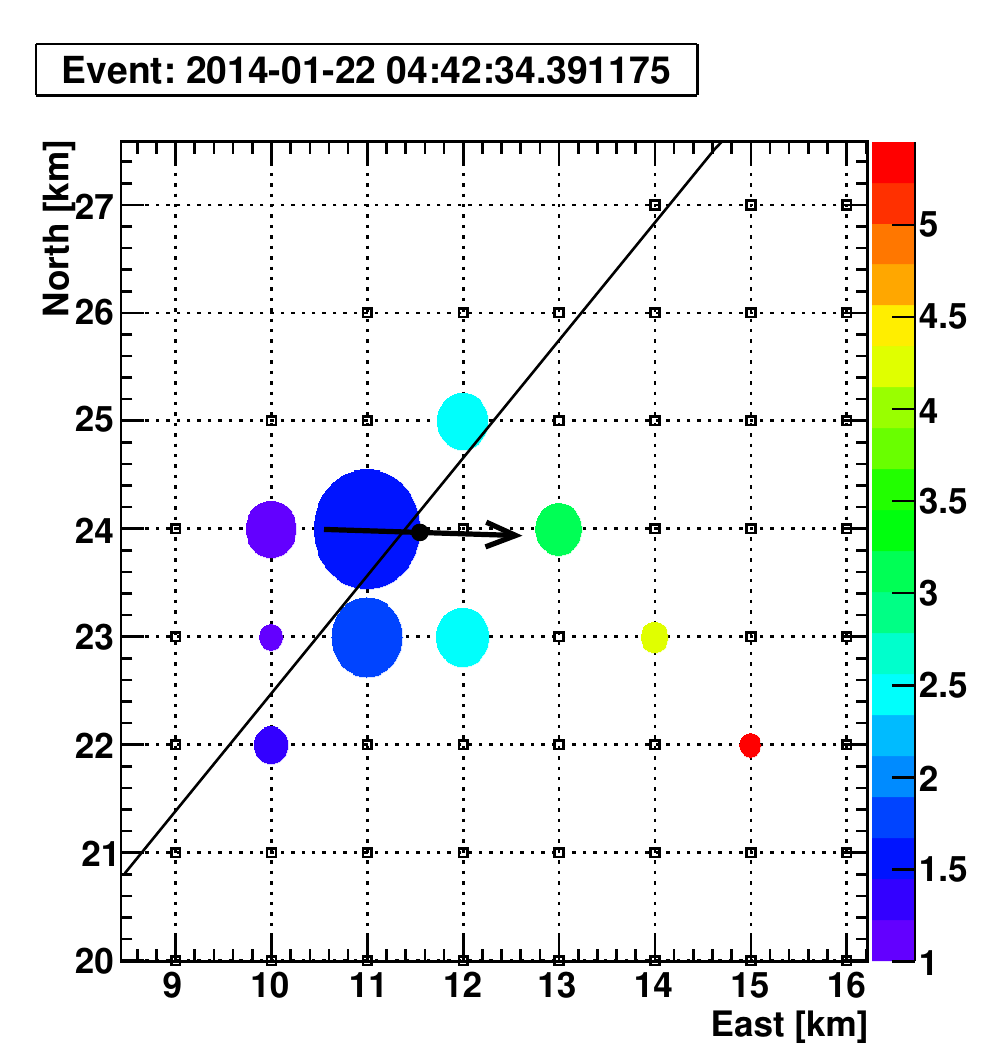}%
    \label{fig:hybrid_event_sd}%
  }
  \subfloat[Time vs angle fit.]{%
    \includegraphics[clip,width=0.48\columnwidth]{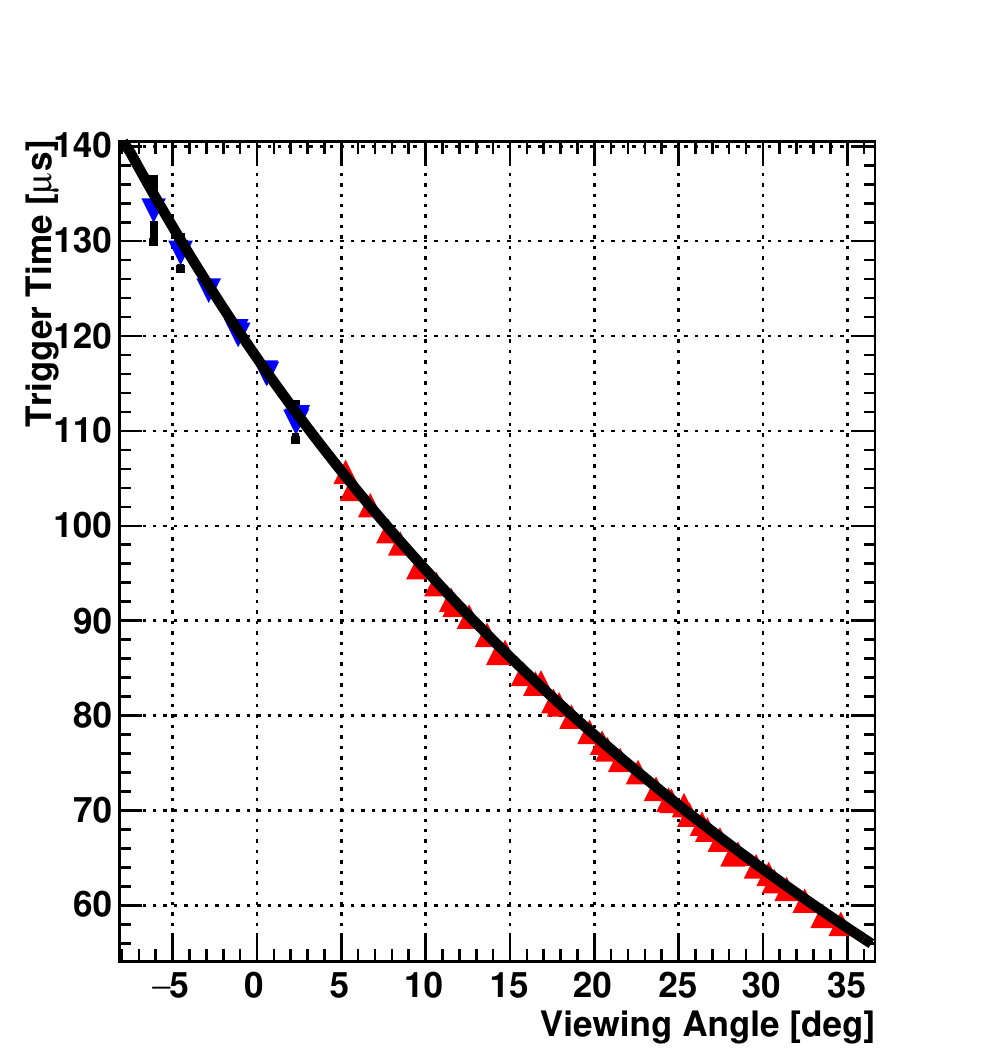}%
    \label{fig:hybrid_event_tvsa}%
  }

  \subfloat[Fluorescence detector display.]{%
    \includegraphics[clip,width=0.48\columnwidth]{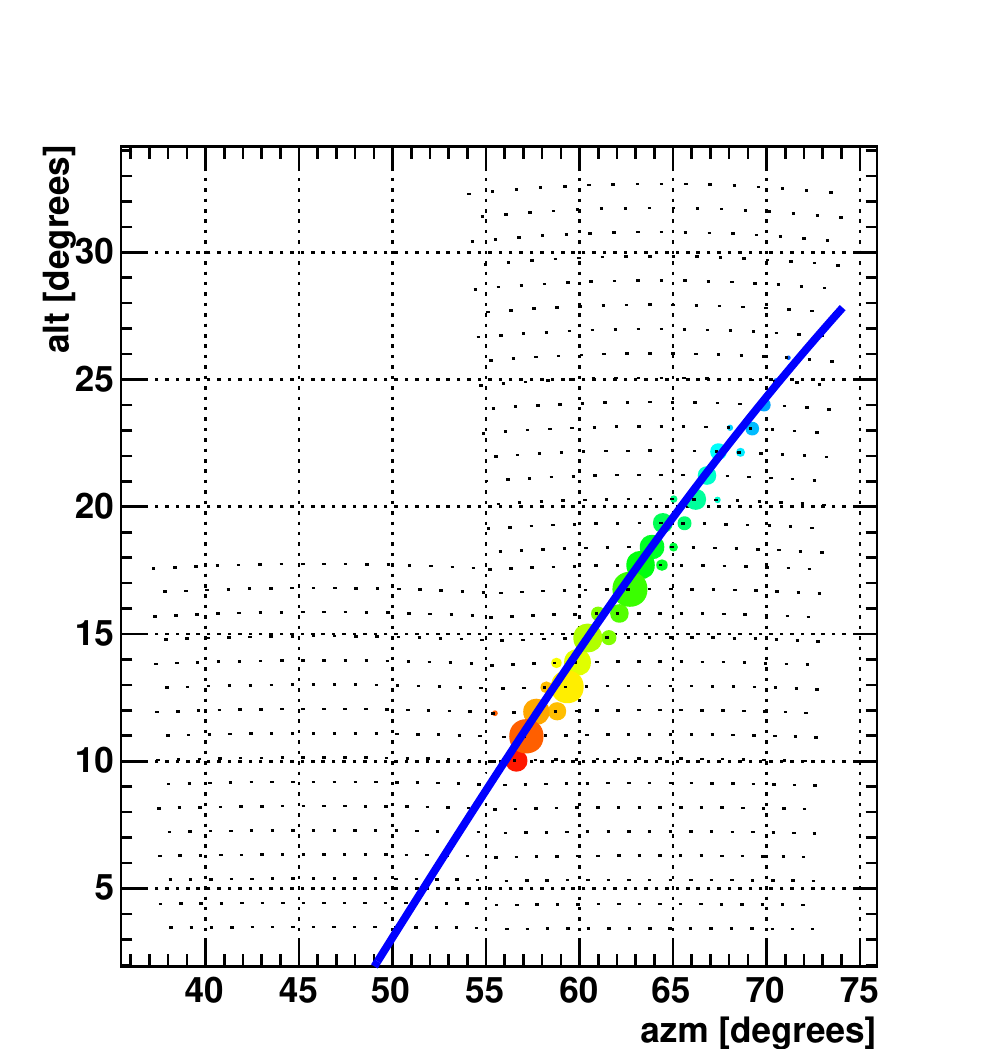}%
    \label{fig:hybrid_event_fd}%
  }
  ~
  \subfloat[Measured flux.]{%
    \includegraphics[clip,width=0.48\columnwidth]{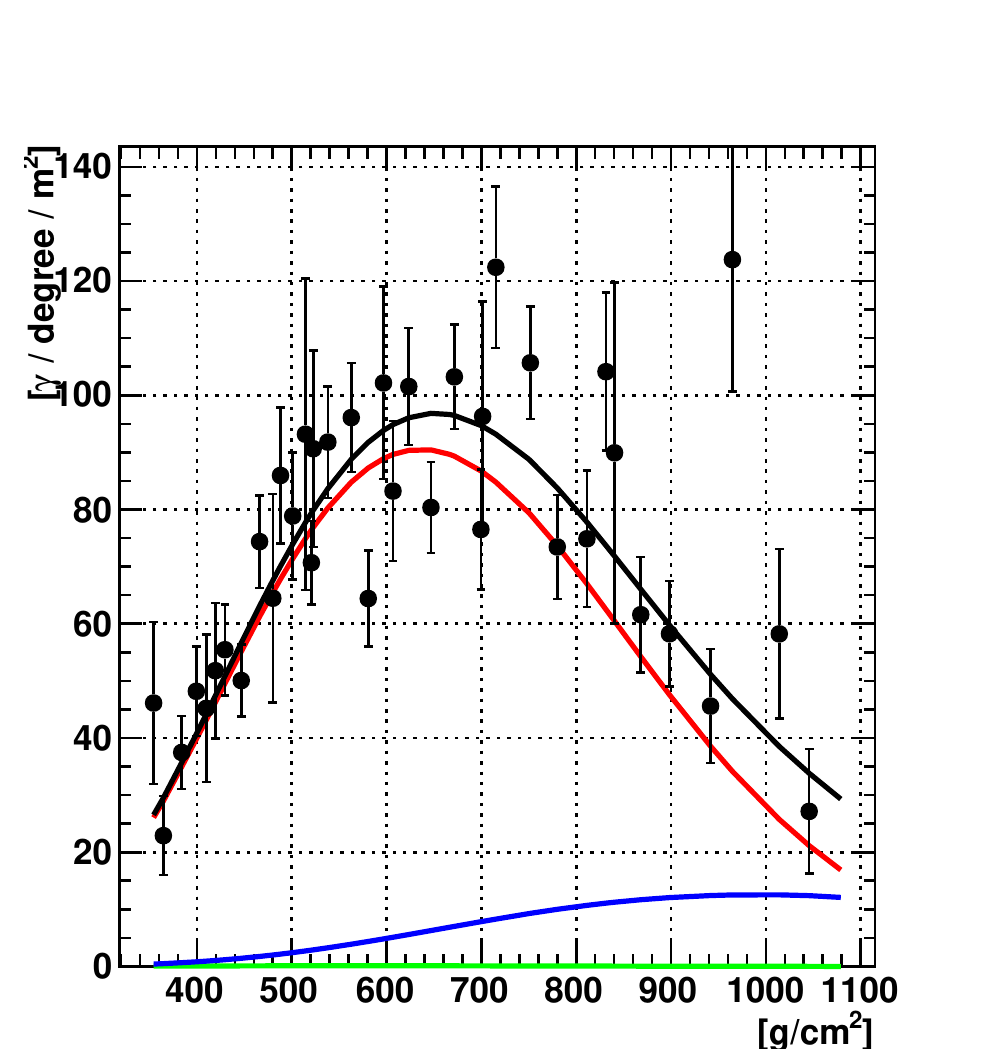}%
    \label{fig:hybrid_event_flux}%
  }
  \caption{Typical hybrid event seen by Telescope Array in hybrid mode.}
  \label{fig:hybrid_event}
\end{figure}

Figure~\ref{fig:hybrid_event} is an example of a typical hybrid event
observed by Telescope array. The surface detectors triggered in the
event are shown in Figure~\ref{fig:hybrid_event_sd}. Marker size
indicates signal measured by the SD and color indicates trigger. The
arrow shows the azimuthal angle of the shower track and the solid line
is the projection of the shower detector plane along the surface of
the earth along the line between the observing FD and the shower
core. Figure~\ref{fig:hybrid_event_tvsa} shows the hybrid time
vs. angle fit. Red triangles show the trigger time and viewing angle
of FD tubes the observed the passage of the shower. The blue triangles
show the same information for surface detectors. The sold line is the
fit to the combined SD and FD pixels using
equation~\ref{eq:time_vs_angle}. Figure~\ref{fig:hybrid_event_fd}
shows the FD tubes triggered in this event. Marker size indicates
signal size and color indicates trigger time. The solid line is the
shower detector plane found by fitting
equation~\ref{eq:fd_plane_chi2}.  The flux and uncertainty in
photons/degree/m$^2$ as a function of shower depth for each tube along
the shower detector plane are shown in
Figure~\ref{fig:hybrid_event_flux}. The solid lines show the simulated
flux generated for the best profile fit. Red is fluorescence flux,
blue is Rayleigh scattered flux, green is direct Cherenkov flux, and
black is the sum of all these flux components.

\section{\label{sec:data_analysis}Analysis of the Data}
The period of data collection covered in this work is 27 May 2008 to
29 November 2016, about 8.5 years. Because this analysis is done using
hybrid data, the collection period is limited by the approximately
10\% duty cycle of the fluorescence detectors. This analysis examines
1500 nights of data collected of these 8.5 years.

By performing time matching of SD and FD events as described in
Section~\ref{sec:hybrid_event_tag}, 17834 hybrid candidate events
(hybrid events that have not gone through full reconstruction and
analysis cuts) are found. The distribution of hybrid candidate events
by FD that observed them is

\medskip
\begin{tabular}{r l}
  17834 &hybrid candidate events \\
  10381 &BR events (mono \& stereo) \\
  8942  &LR events (mono \& stereo) \\
  8892  &BR events (mono) \\
  7453  &LR events (mono) \\
  1489  &BR + LR stereo events
\end{tabular}
\medskip

Monocular hybrid candidate events are hybrid events observed by only
one FD (either BR or LR for this analysis) and the SD array. Stereo
hybrid candidate events are events observed by both FDs and the SD
array. Even though an event is tagged as a stereo candidate event,
there is no guarantee that the event is able to be fully reconstructed
independently by each FD.  For example, a medium energy event located
much closer to one FD than the other, may not be of sufficient quality
to pass all cuts for the FD located farther away.

Reconstruction of the data proceeds as described in
Section~\ref{sec:hybrid_analysis_method}. Once profile fitting is done,
cuts are applied to the events to reject those that are poorly
reconstructed which may introduce \xm{} bias and degrade \xm{}
resolution. These cuts are chosen to maximize the number of events
collected without introducing large reconstruction biases. 

\begin{enumerate}
\item \textbf{Boundary cut}. This cut ensures the core of the event falls
  within the bounds of the SD array. Additionally, the core location
  must not be within 100~m of the boundary. This is to ensure the
  charged particle distribution on the ground is fully contained
  within the array, allowing accurate reconstruction by the SDs.
\item \textbf{Track length cut}. This cut ensures the shower track as
  observed by the FDs is sufficiently long enough to allow enough PMTs
  to observe it. Short shower tracks may be caused by distant showers
  or showers moving towards the detector, both instances of which
  indicate unfavorable geometry for accurate reconstruction. We
  require a shower track of $10^\circ$ or greater to accept the track.
\item \textbf{Good tube cut}. Good tubes are tubes that have sufficient
  signal-to-noise and are spatially and temporally part of the
  shower track as determined by the FD plane fitting routines. We
  require 11 or more good tubes to accept a track.
\item \textbf{SDP angle cut}. This cut rejects events with SDP angle
  ($\psi$) greater than $130^\circ$. Showers with $\psi$ greater than
  $90^\circ$ have some component of the track vector pointing towards
  the observing FD.  As this angle grows, the shower is seen more and
  more head on, increasing the contribution of direct Cherenkov light
  received. Showers with large direct Cherenkov signals are difficult
  to reconstruct accurately and are rejected.
\item \textbf{Time extent cut}. This cut rejects tracks with time
  profiles less than 7~$\mu$s. This is another cut to ensure short
  tracks, potentially approaching the observing FD directly, are
  removed from the data.
\item \textbf{Zenith angle cut}. Tracks with large zenith angle
  ($\theta$) are difficult to reconstruct by both the SD and
  FD. FD-only reconstruction can accurately reconstruct events with
  relatively large zenith angles ($\sim
  75^\circ$) \citep{Abu-Zayyad:2013jra}. The upper zenith angle limit
  for SD-only reconstruction is $\sim
  45^\circ$ \citep{AbuZayyad:2012ru,Ivanov:2012tex}. This cut removes
  tracks with zenith angles greater than $55^\circ$.
\item \textbf{Hybrid geometry $\chi^2$ cut}. This cut requires that
  the reduced $\chi^2$ of the hybrid fitting described in
  Section~\ref{sec:hybrid_fitting} must be less than 5 to accept the
  track. This ensures good geometry reconstruction when combining the
  SD and FD geometry information.
\item \textbf{Profile $\chi^2$ cut}. This cut requires that the
  reduced $\chi^2$ of the profile fit described in
  Section~\ref{sec:profile_fitting} be less than 10. This ensures the
  light profile of the track is well observed and the inverse Monte
  Carlo process sufficiently well simulated the shower as observed by
  the FD.
\item \textbf{\xm{} bracketing cut}. Once the shower profile is
  reconstructed, the atmospheric depth vs. angle for observed parts of
  the shower is known. Using this depth profile, the minimum depth
  observed, $X_{\mathrm{low}}$, and the maximum depth observed,
  $X_{\mathrm{high}}$, as well as \xm{} are calculated. The \xm{}
  bracketing cut requires that the fitted value of \xm{} be greater
  than $X_{\mathrm{low}}$ and less than $X_{\mathrm{high}}$ for the
  track to be accepted. This cut ensures that the turnover from rising
  shower size before \xm{} and the falling size after it are in the
  field of view of the observing FD. This is required to get a good
  profile fit. If \xm{} is not bracketed, the Gaisser-Hillas profile
  fit will often fail to accurately measure \xm{} and \nm{}, which
  also causes large uncertainty on the primary particle energy.
\item \textbf{Energy cut}. This cut is used to ensure the energy of
  the shower is not less than $10^{18.2}$~eV. Showers with energies
  below this cut are difficult to reconstruct by both the SD and FDs
  independently due to low signal to noise. The hybrid aperture falls
  steeply below $E = 10^{18.5}$~eV. This cut represents the design
  limit of our detectors operating in hybrid mode.
\item \textbf{Weather cut}. Events that occur during bad weather
  nights are rejected. Weather is monitored by operators in the field
  and logged hourly. Operators record the state of cloud coverage in
  the four cardinal directions as well as the amount of overhead
  coverage and cloud thickness as judged by eye. These weather codes
  are used to categorize nights into ``excellent'', ``good'', or
  ``bad'' weather nights. This analysis uses nights deemed
  ``excellent'' or ``good''. Nights that are recorded as bad are
  rejected from the data. To ensure consistency among similar
  analyses, the same weather cut criteria as used by the FD monocular
  spectrum analysis \citep{Abu-Zayyad:2013jra} is also used here.
\end{enumerate}

Hybrid monocular events which pass all of the analysis cuts are
accepted as part of the final event set. Stereo hybrid events must go
through one more selection step before final acceptance. A stereo
event is independently reconstructed using the data as observed by the
BR FD and as observed by the LR FD. The cuts above are applied
separately to the BR and LR reconstructed hybrid event information. If
only one of the two site's data passes cuts, then the hybrid event
data is accepted using that site's reconstruction parameters. If both
site's data passes the cuts, then the site's data with the profile
reduced $\chi^2$ closest to one is accepted. The same set of cuts and
the same selection procedures are used on data and Monte Carlo data.

After the cuts have been applied to the data 3330 events remain. The
distribution of fully reconstructed and accepted data events is

\medskip
\begin{tabular}{r l}
  3330  &hybrid accepted events \\
  1743  &BR events (mono \& stereo) \\
  1587  &LR events (mono \& stereo) \\
  1676  &BR events (mono) \\
  1504  &LR events (mono) \\
   150  &BR + LR stereo events
\end{tabular}
\medskip

\subsection{\label{sec:data_monte_carlo}Data/Monte Carlo Comparison}
\begin{figure}
  \centering
  \subfloat[Data/Monte Carlo track length comparison.]{%
    \includegraphics[clip,width=0.48\columnwidth]{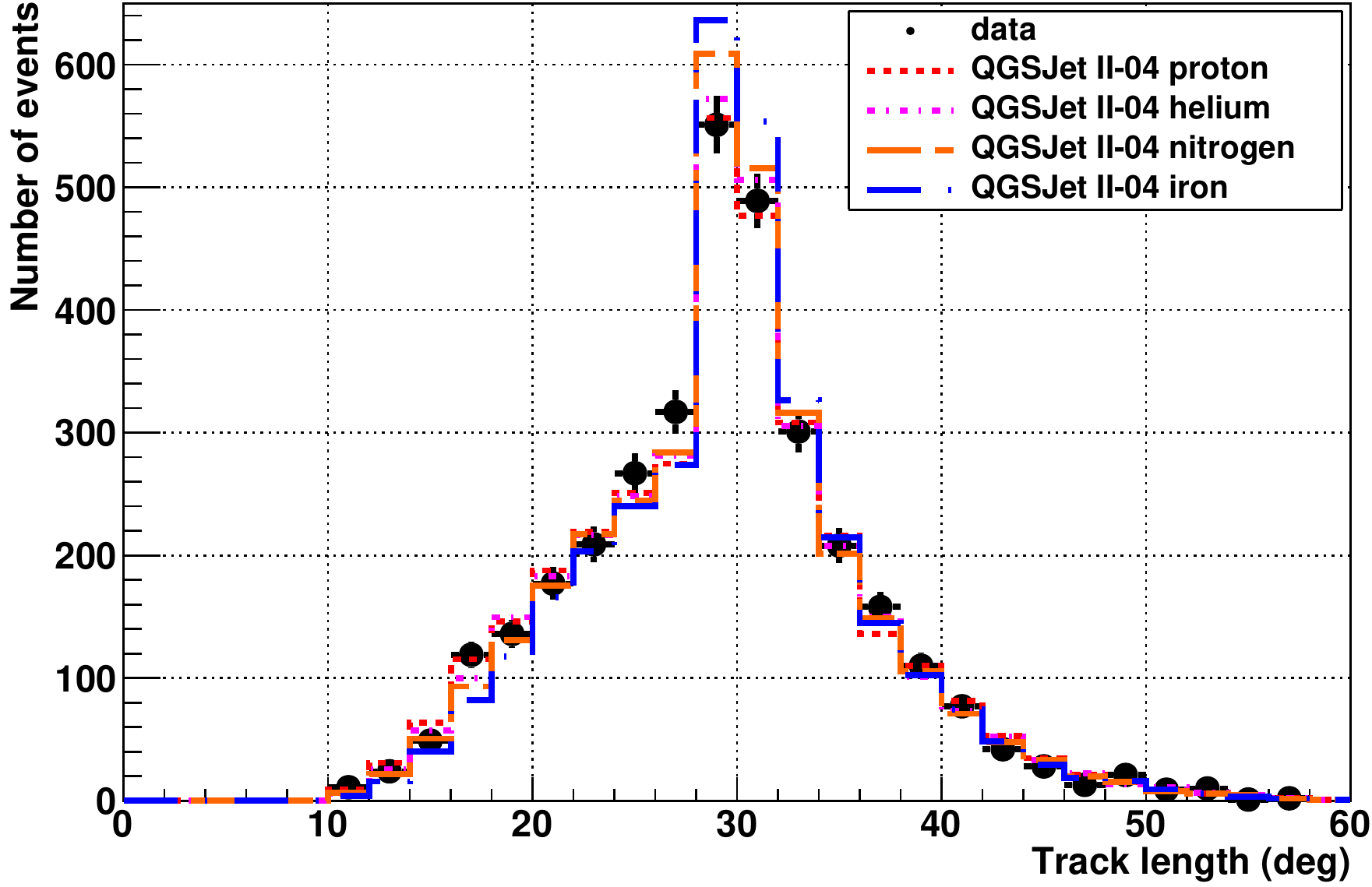}%
    \label{fig:dataMC_tracklen}%
  }
  ~
  \subfloat[Data/Monte Carlo number of good tubes comparison.]{%
    \includegraphics[clip,width=0.48\columnwidth]{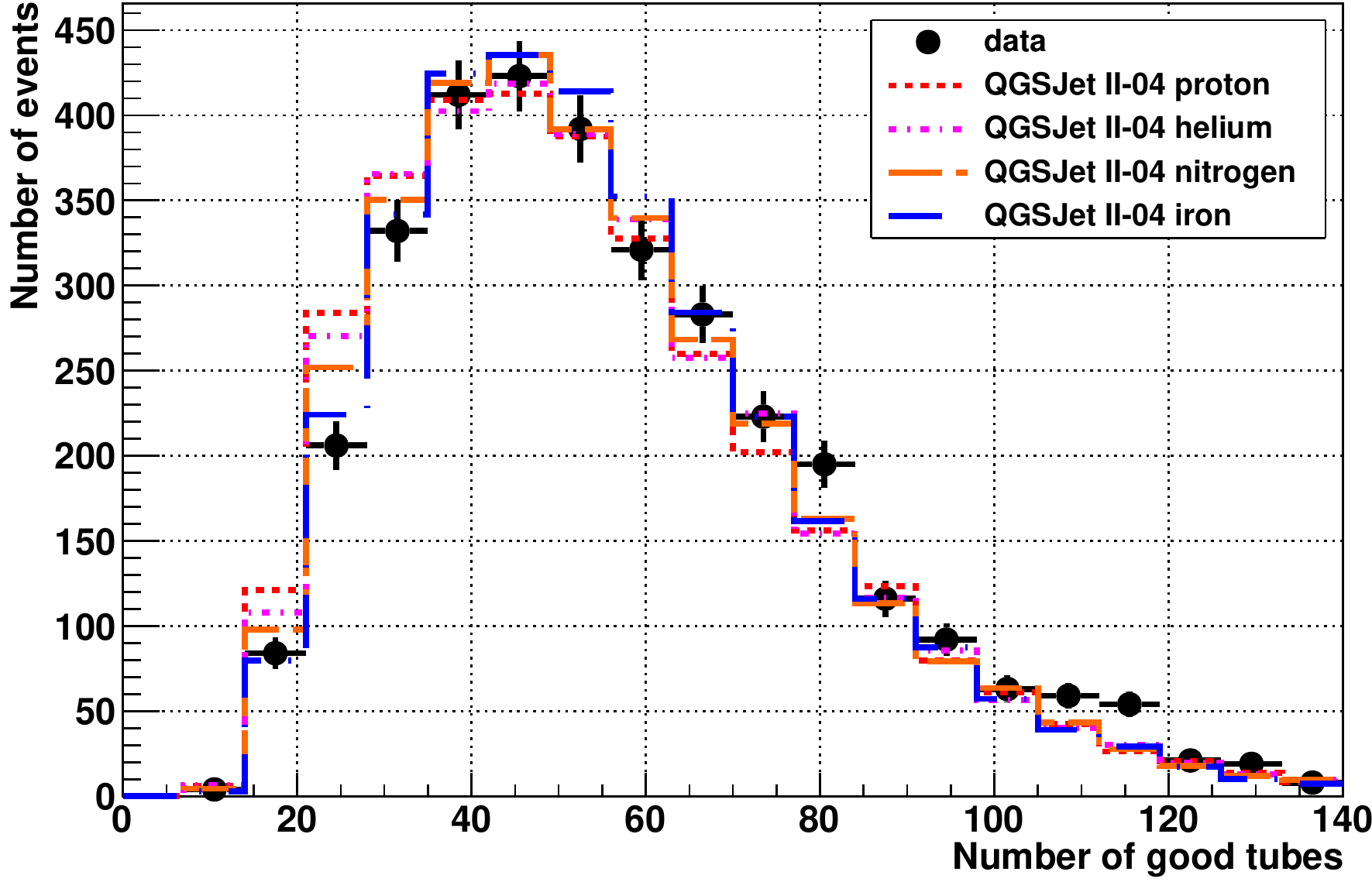}%
    \label{fig:dataMC_ngoodtube}%
  }
  
  \subfloat[Data/Monte Carlo $\psi$ angle comparison.]{%
    \includegraphics[clip,width=0.48\columnwidth]{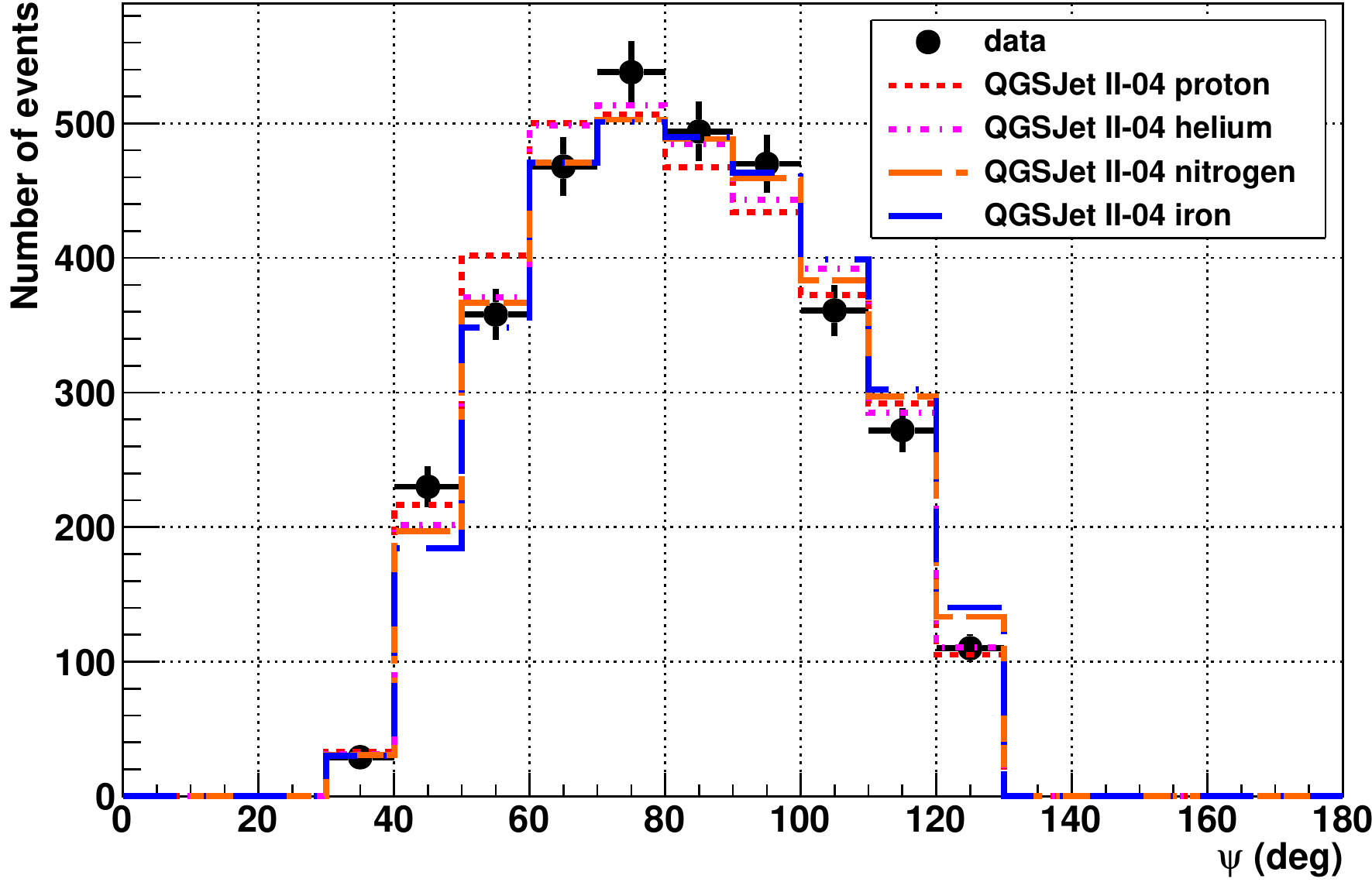}%
    \label{fig:dataMC_psi}%
  }
  ~
  \subfloat[Data/Monte Carlo track time extent comparison.]{%
    \includegraphics[clip,width=0.48\columnwidth]{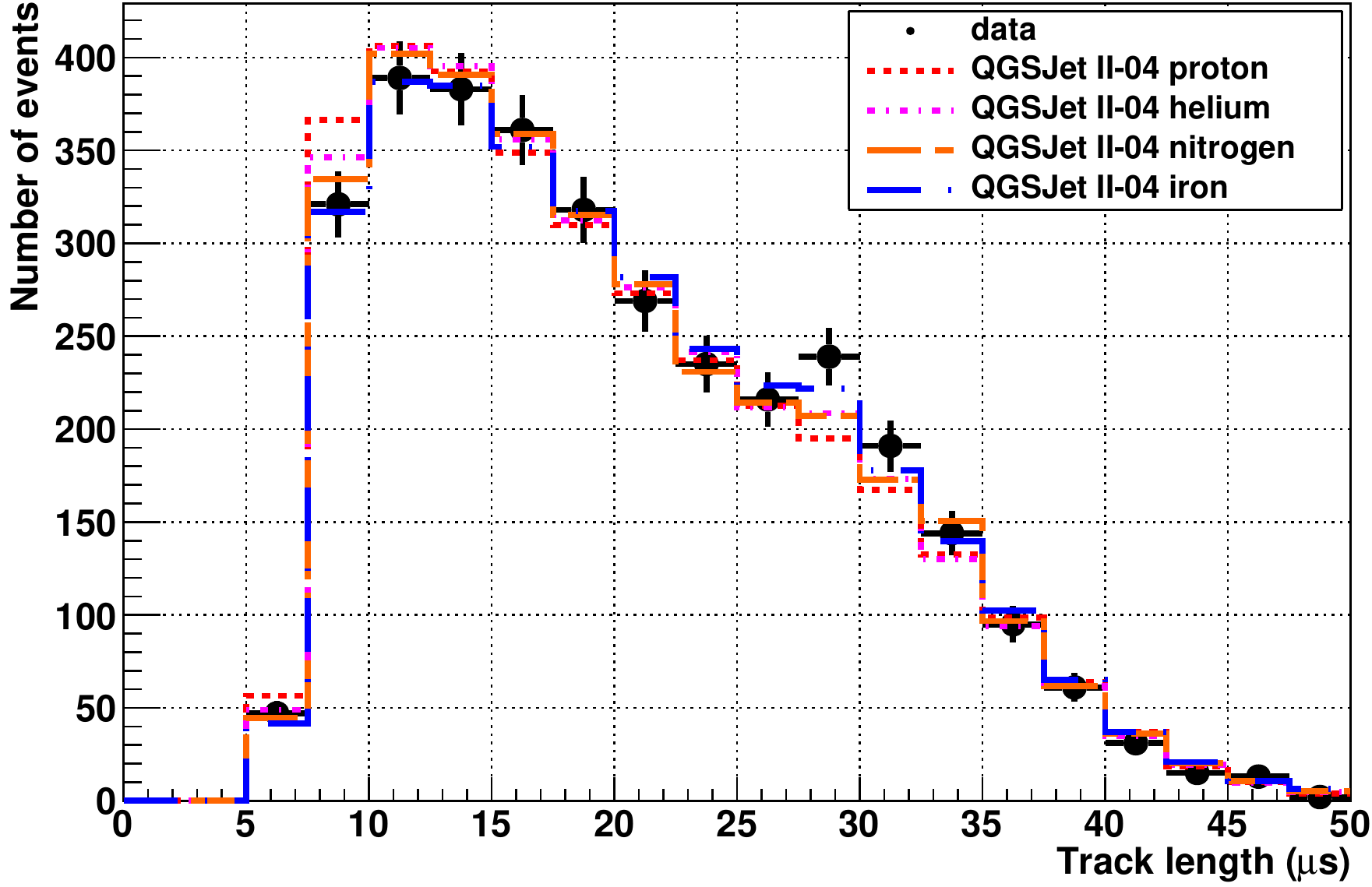}%
    \label{fig:dataMC_timeext}%
  }

  \subfloat[Data/Monte Carlo zenith angle comparison.]{%
    \includegraphics[clip,width=0.48\columnwidth]{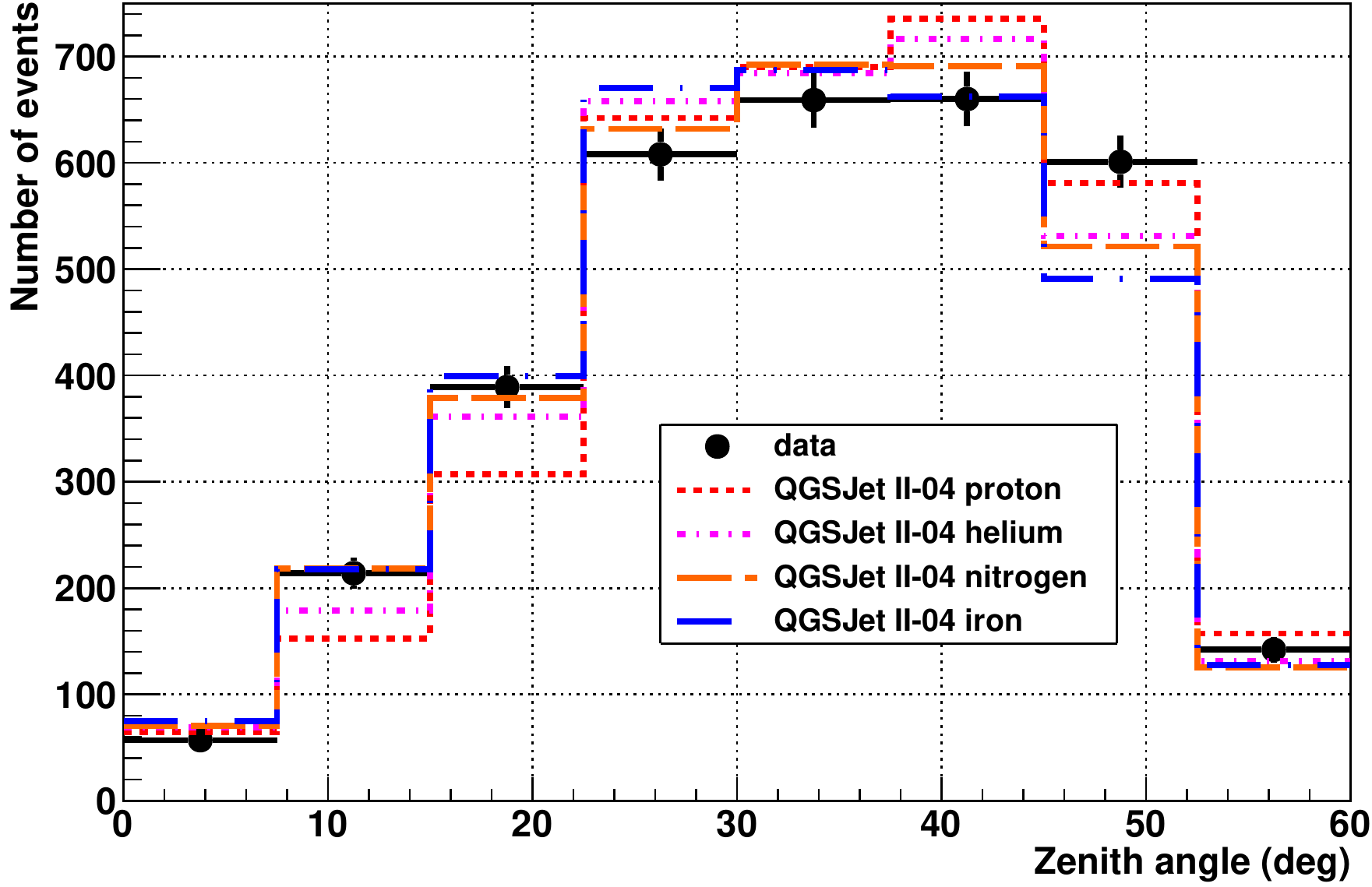}%
    \label{fig:dataMC_theta}%
  }
  ~
  \subfloat[Data/Monte Carlo geometry fit $\chi^2$ comparison.]{%
    \includegraphics[clip,width=0.48\columnwidth]{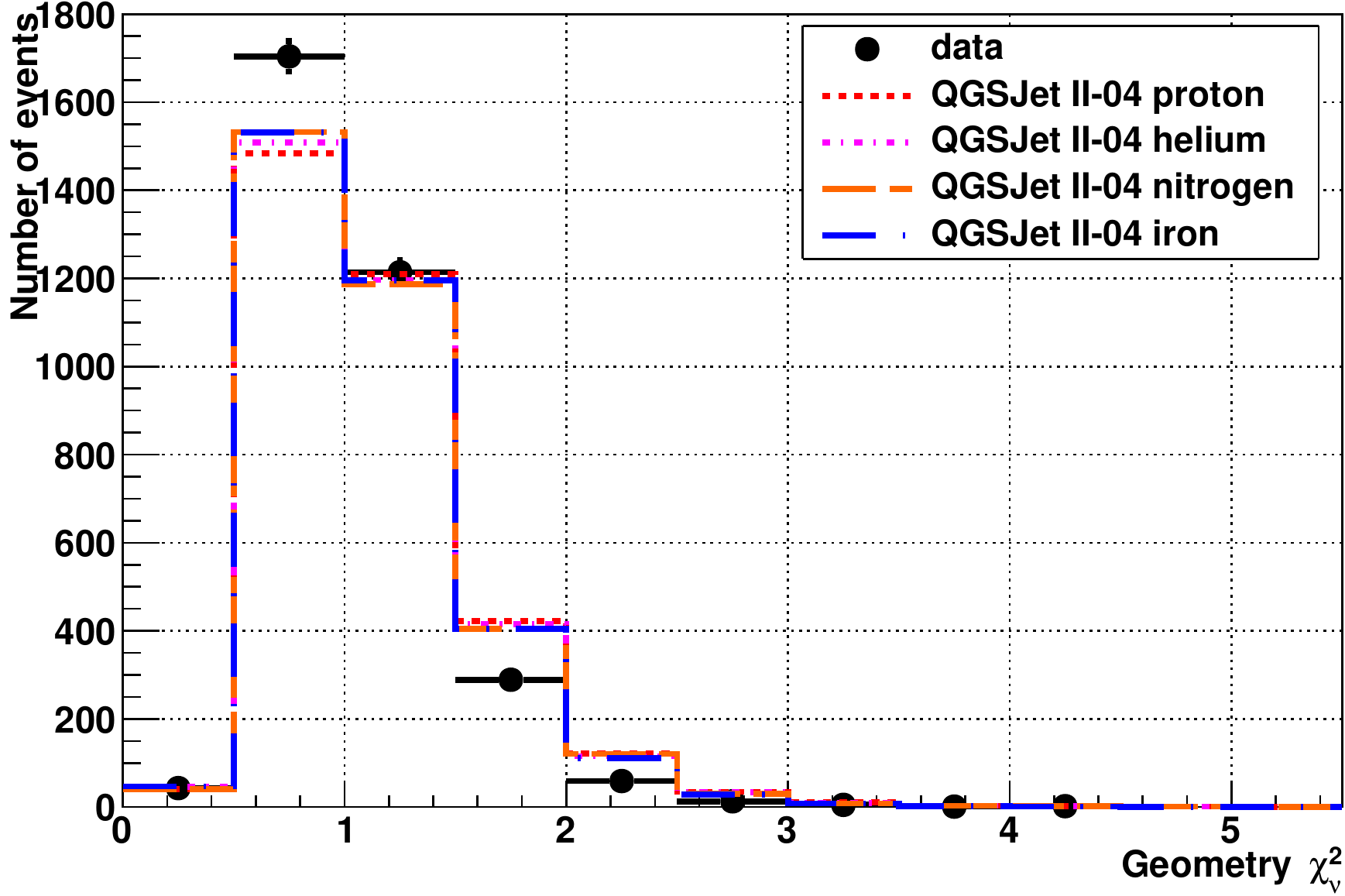}%
    \label{fig:dataMC_geomchi2}%
  }
  \caption{Data/Monte Carlo plots I.}
  \label{fig:dataMC_plots1}
\end{figure}

\begin{figure}
  \centering
  \subfloat[Data/Monte Carlo profile fit $\chi^2$ comparison.]{%
    \includegraphics[clip,width=0.48\columnwidth]{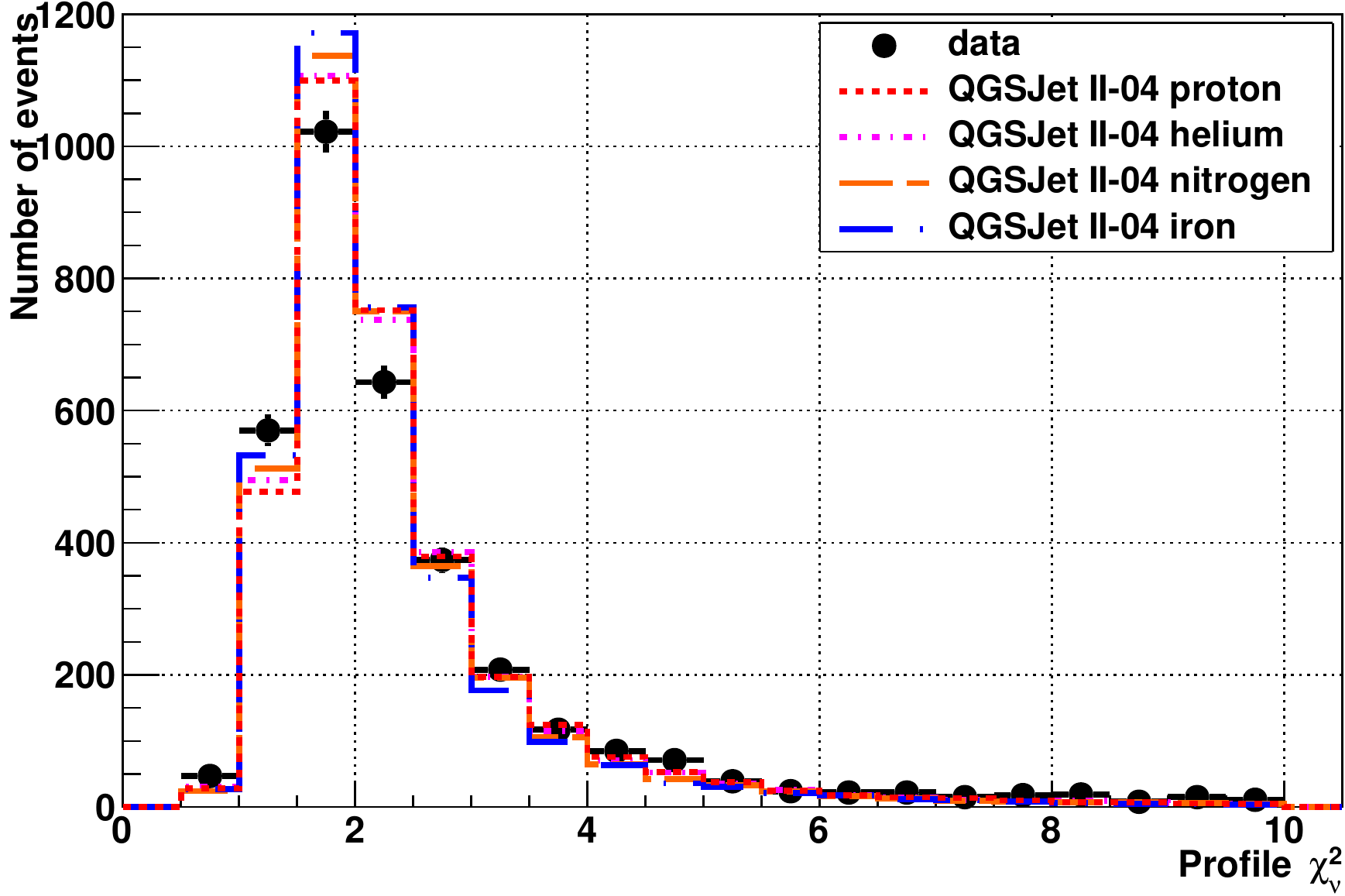}%
    \label{fig:dataMC_profchi2}%
  }
  ~
  \subfloat[Data/Monte Carlo energy comparison.]{%
    \includegraphics[clip,width=0.48\columnwidth]{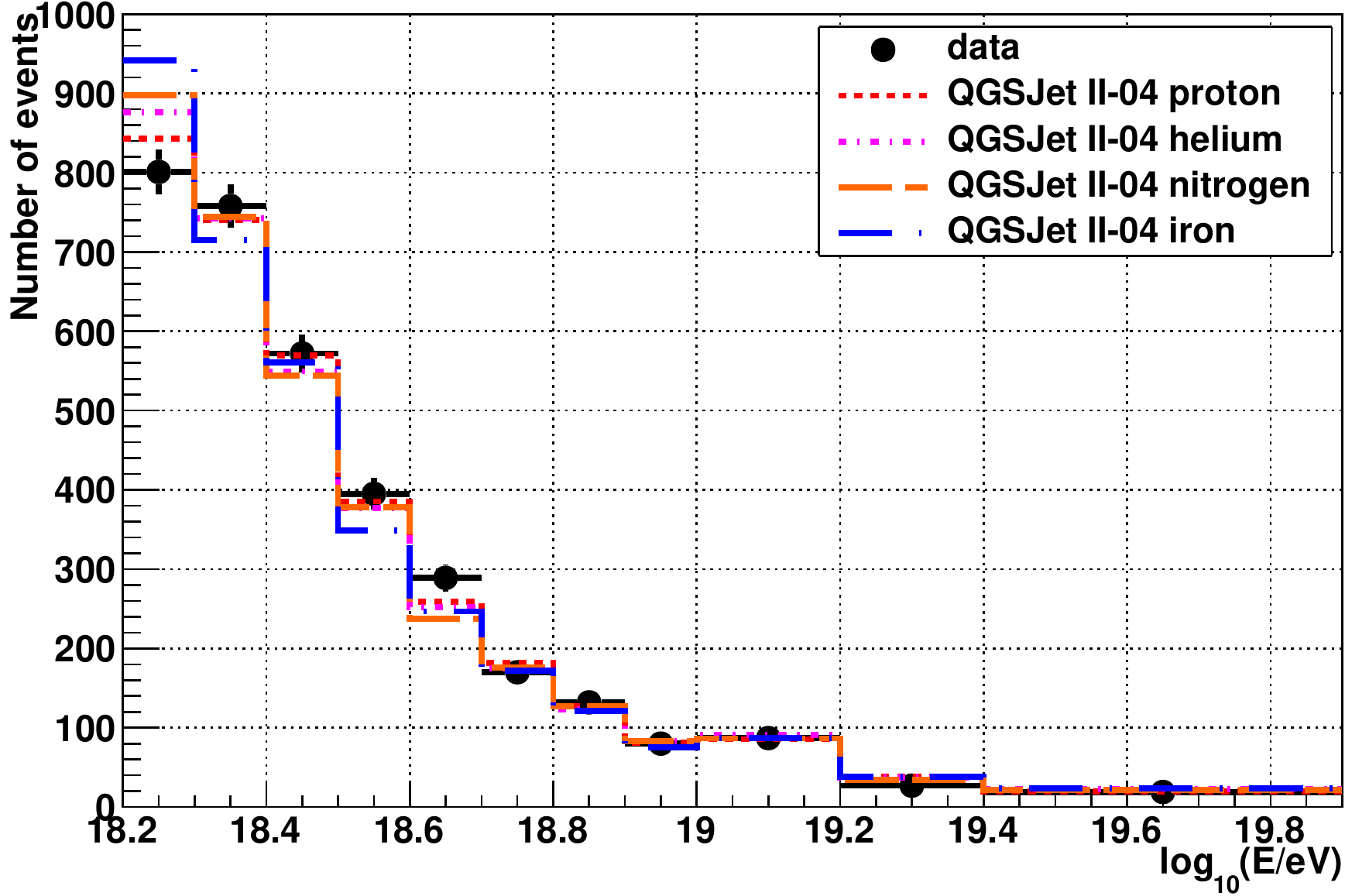}%
    \label{fig:dataMC_energy}%
  }
  
  \subfloat[Data/Monte Carlo $X_{\mathrm{low}}$ comparison.]{%
    \includegraphics[clip,width=0.48\columnwidth]{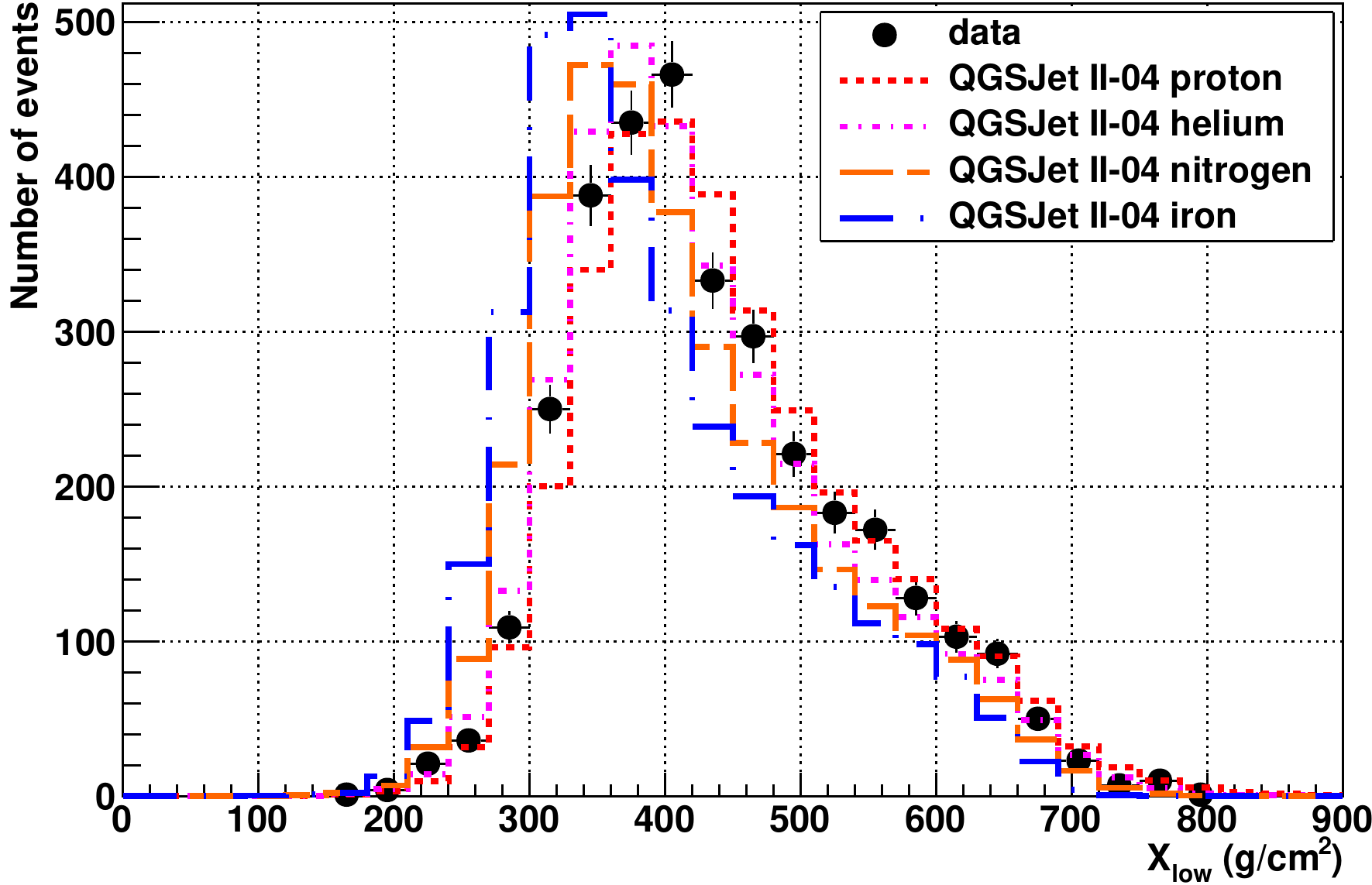}%
    \label{fig:dataMC_xlow}%
  }
  ~
  \subfloat[Data/Monte Carlo $X_{\mathrm{high}}$ comparison.]{%
    \includegraphics[clip,width=0.48\columnwidth]{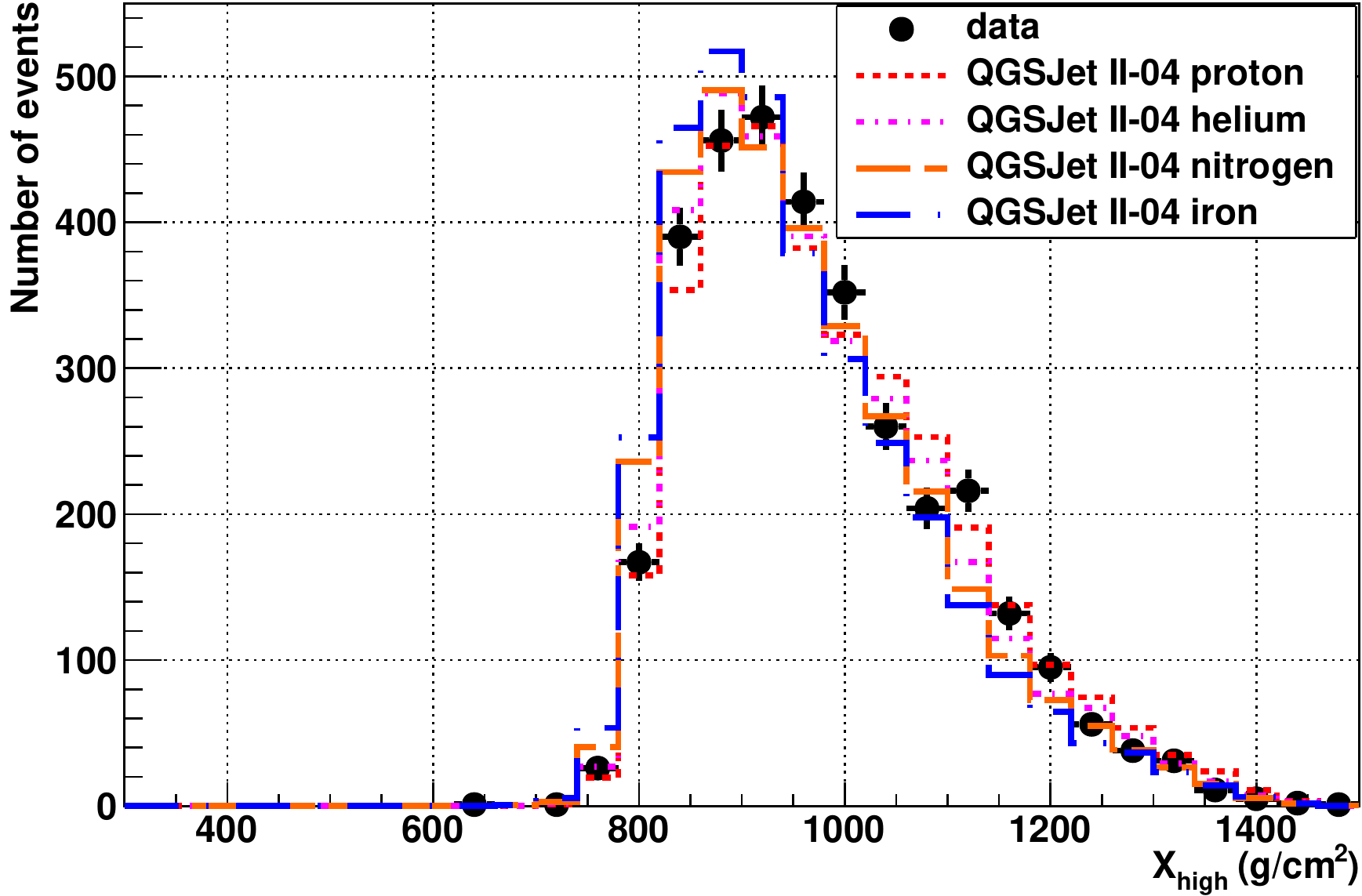}%
    \label{fig:dataMC_xhigh}%
  }
  \caption{Data/Monte Carlo plots II.}
  \label{fig:dataMC_plots2}
\end{figure}

Monte Carlo is stored in the same format and analyzed in the same
manner as data. This allows us to understand the acceptance of our
detector by detailed simulations and to perform data/Monte Carlo
comparisons to test the agreement of data with different composition
models and mixtures. Figures~\ref{fig:dataMC_plots1} and
\ref{fig:dataMC_plots2} compare data and Monte Carlo of the parameters
used by the analysis cuts for four different primary species: protons,
helium, nitrogen, and iron. All of the Monte Carlo data used in this
analysis is generated in CORSIKA using the QGSJet~II-04 hadronic
model. For all plots, the accepted energy range is $E \geq
10^{18.2}$~eV and all Monte Carlo simulated data histograms are
normalized by area to the area of the data histogram. Each Monte Carlo
data set represents six years of TA operations with about 10 times the
statistics collected in the data over that same period. The TA Monte
Carlo mimics the real operating conditions of the FDs and SDs, by
using time-dependent databases of the real operating conditions in the
field, such as pedestals, atmosphere, and tube gains. This information
is used in reconstruction as well as event simulation. As the figures
show, the simulation mimics the observed data well between the four
chosen models.

To measure the bias and resolution of our detector for a given
observable parameter $X$, for all reconstructed Monte Carlo events we
histogram the difference $X_{\mathrm{recon}} - X_{\mathrm{true}}$,
where $X_{\mathrm{recon}}$ is the reconstructed value of the parameter
and $X_{\mathrm{true}}$ is the true value of the parameter. The
parameter bias is the sample mean of this distribution and the
resolution is the sample standard
deviation. Table~\ref{tab:mc_bias_res} shows the measured bias and
resolution of this analysis for four primary species for all
reconstructed Monte Carlo events with $E_{\mathrm{true}} \geq
10^{18.2}$~eV. The table shows the reconstruction biases for \xm{} is
very small, about -1~g/cm$^2$ for protons and -4~g/cm$^2$ for iron,
both of which are much smaller than the \xm{} resolutions of
17~g/cm$^2$ and 13~g/cm$^2$ respectively. Energy bias is less than 2\%
for protons and -6.5\% for iron. We expect a larger energy bias for
iron because when the shower energy is computed the missing energy
correction assumes a proton primary (see
Section~\ref{sec:profile_fitting}). In all cases the energy resolution
is less than 6\% for the four primary species shown. Angular
resolution and bias for the geometric parameters are acceptably small in
all cases, less than a degree, which is expected for hybrid
reconstruction. Bias and resolution of the shower impact parameter is
of order 0.1\% of the average observed distance of $R_{\mathrm{p}}$.
The reconstruction accuracy of $X_{\mathrm{core}}$ and
$Y_{\mathrm{core}}$, the $x$ and $y$ components of the shower core
location on the ground are also very good.

\begin{table}
  \centering
\begin{tabular}{l rr rr rr rr}
  \hline
  \hline
  &\multicolumn{2}{c}{proton} &\multicolumn{2}{c}{helium}
  &\multicolumn{2}{c}{nitrogen} &\multicolumn{2}{c}{iron} \\
  \cline{2-9}
  &bias &res. &bias &res. &bias &res. &bias &res.\\
  \hline
  \xm (g/cm$^{2}$) &-1.1 &17.2 &-3.3 &15.7 &-3.8 &14.2 &-3.8 &13.2\\
  Energy (\%) &1.7 &5.7 &-1.1 &5.1 &-3.5 &4.4 &-6.5 &3.9\\
  $\theta$ (deg) &0.014 &0.377 &0.006 &0.364 &0.0005 &0.3553 &-0.003 &
  0.344\\
  $\phi$ (deg) &-0.020 &0.410 &-0.017 &0.399 &-0.015 &0.389 &-0.015
  &0.374\\
  $\psi$ (deg) &0.074 &0.397 &0.088 &0.385 &0.112 &0.375 &0.135
  &0.356\\
  $R_{\mathrm{p}}$ (m) &18.9 &39.8 &20.1 &39.0 &21.1 &38.9 &22.3
  &37.7\\
  $X_{\mathrm{core}}$ (m) &-3.6 &49.8 &-3.4 &49.9 &-3.3 &50.6 &-3.7
  &51.9\\
  $Y_{\mathrm{core}}$ (m) &8.7 &42.9 &8.1 &42.3 &8.3 &42.9 &8.4 &43.3\\
  \hline
\end{tabular}
\caption{Bias and resolution of BR/LR hybrid \xm{} analysis
  reconstruction. All primary species are generated using the
  QGSJet~II-04 hadronic model.}
\label{tab:mc_bias_res}
\end{table}

\subsection{\label{sec:xmax_bias}\texorpdfstring{\xm{}}{Xmax} Biases}
\xm{} bias in our simulation comes in two parts: bias due to detector
acceptance and bias due to reconstruction. Reconstruction bias is bias
that is affected by operating condition of the detector, selection of
cuts, composition and hadronic model dependence, and proper modeling
of the detector and air shower physics in the Monte Carlo. Acceptance
bias is affected by physical detector design and detector response,
such as choice of triggering algorithm.

\begin{figure}
  \centering
  \includegraphics[width=\textwidth]{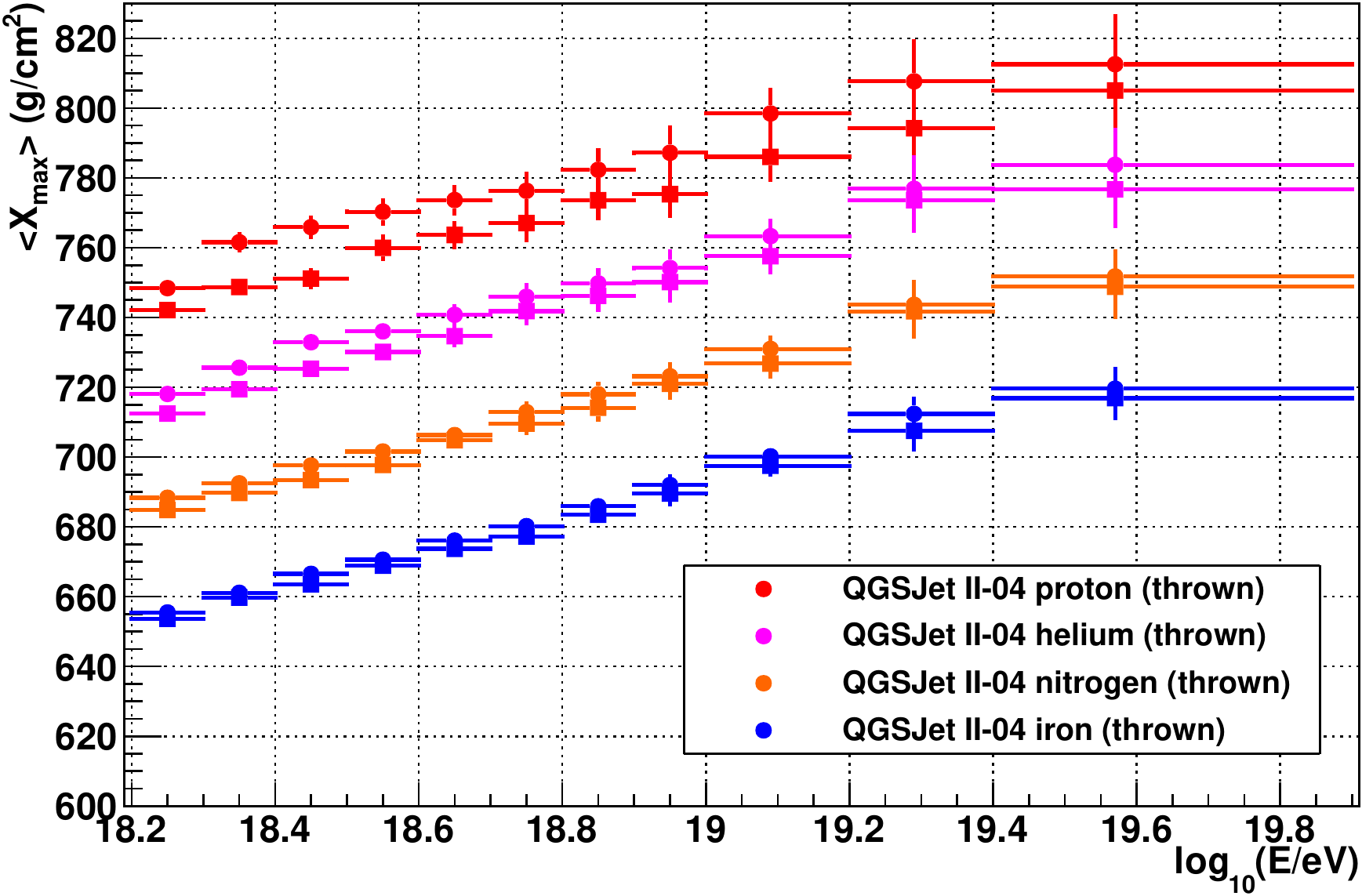}
  \caption{QGSJet~II-04 Monte Carlo \mxm{} used in this
    analysis. Circles represent \mxm{} of the true distributions prior
    to any reconstruction. Squares represent the \mxm{} after
    reconstruction. The difference in an energy bin between the thrown
    \mxm{} and the reconstructed \mxm{} (acceptance bias) is caused
    mainly by detector acceptance, with a small contribution from
    reconstruction biases as well. Light elements have larger
    acceptance bias because the primary effect of this type of bias is
    to cause the loss of very deeply penetrating events in the tails
    of the distributions. Uncertainties in the means are calculated
    using the equivalent exposure in the data.}
  \label{fig:all_species_thrown_recon_mean_xmax}
\end{figure}

Acceptance bias predominantly affects the deeply penetrating tail of
the \xm{} distribution. This is because there is an upper bound to the
maximum depth to which an air shower can be observed due to limited
atmospheric mass overburden, which is part of the detector design
(placement on the Earth's surface). For very deeply penetrating
primaries, the ability to reconstruct via fluorescence observation is
limited by the following scenarios: 1) the air shower track has small
zenith angle and \xm{} occurs at the ground level or below, or 2) the
air shower track achieves shower maximum in air, but therefore has a
very large zenith angle. The result is that as a function of energy,
\xm{} acceptance bias in TA is seen as a systematic shift of the mean
of the \xm{} distribution to smaller depths and narrowing in the width
(RMS) of the \xm{} distribution in each energy bin. This effect is
shown explicitly in the Monte Carlo distributions in
Figure~\ref{fig:all_species_thrown_recon_mean_xmax}.

This effect is dependent upon the mass and energy of the primary
particle; light particles penetrate more deeply and shower maximum
occurs at deeper depths on average with increasing energy and
decreasing primary mass. It is also dependent upon the physics of
UHECR hadronic interactions, which are not known for our energy range
of interest. This appears as model dependence through our choice of
hadronic generator in CORSIKA simulations. More recent hadronic models
tuned to LHC results, such as QSGJet~II-04 \citep{Ostapchenko:2010vb}
and EPOS LHC \citep{Pierog:2013ria}, generate events that penetrate
more deeply on average than older models, such as
QGSJet~II-03 \citep{Ostapchenko:2007qb}.

The sum of acceptance bias and reconstruction bias is called total
bias, and it is important for us to understand because it appears as a
systematic shift in the final reconstructed \xm{} distribution
compared to the true generated \xm{}. Given that there is some
combination of hadronic model and mixture of elements that represent
the true distribution of \xm{} that is impinging upon the Earth's
atmosphere, detectors with acceptance bias will never be able to fully
reproduce the true distribution in nature simply by plotting the
distribution of reconstructed events. Acceptance bias will distort the
observed distribution, because information about the distribution is
simply lost and it will typically appear as if it is the result of a
heavier mixture of elements. To correct for this type of bias, one can
attempt unfolding of the data, or resort to even more restrictive sets
of cuts such as done by the Auger experiment \citep{Aab:2014kda}.

An alternate method to understand measured composition is to simply
use Monte Carlo to simulate biases incurred due to detector acceptance
and compare the measured \xm{} distribution to the biased, simulated
one. This is the method chosen by TA. It is important to understand
how a measurement deals with this issue of acceptance bias before
attempting to compare composition results between different
experiments.

\subsection{\label{sec:ta_xmax_data}TA \texorpdfstring{\xm{}}{Xmax} data}
\begin{figure}
  \centering
  \subfloat[$18.2 \leq \log_{10}(E/\mathrm{eV}) < 18.3$]{%
    \includegraphics[clip,width=0.48\columnwidth]{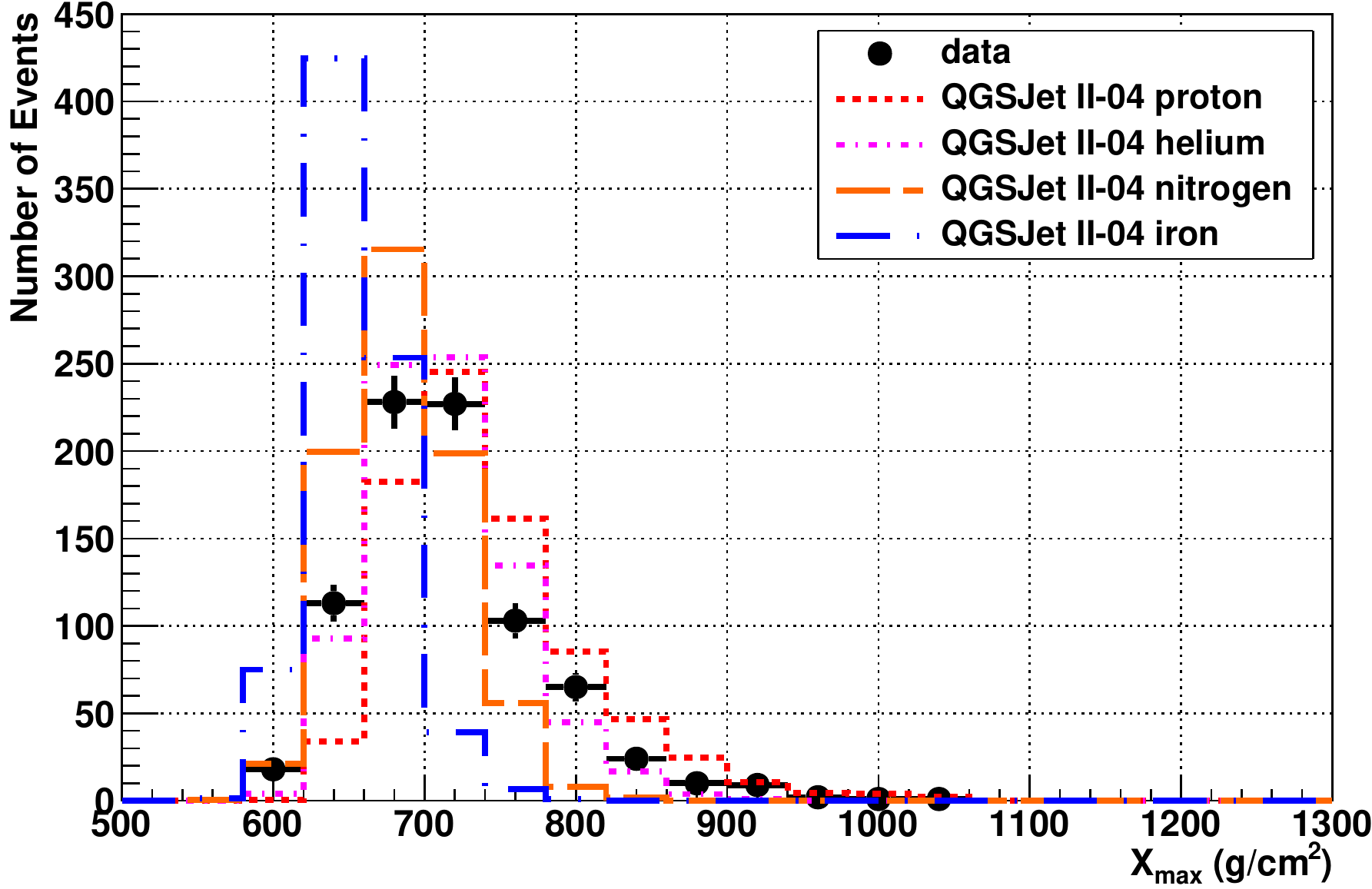}%
    \label{fig:dataMC_xmax_bin_00}%
  }
  ~
  \subfloat[$18.3 \leq \log_{10}(E/\mathrm{eV}) < 18.4$]{%
    \includegraphics[clip,width=0.48\columnwidth]{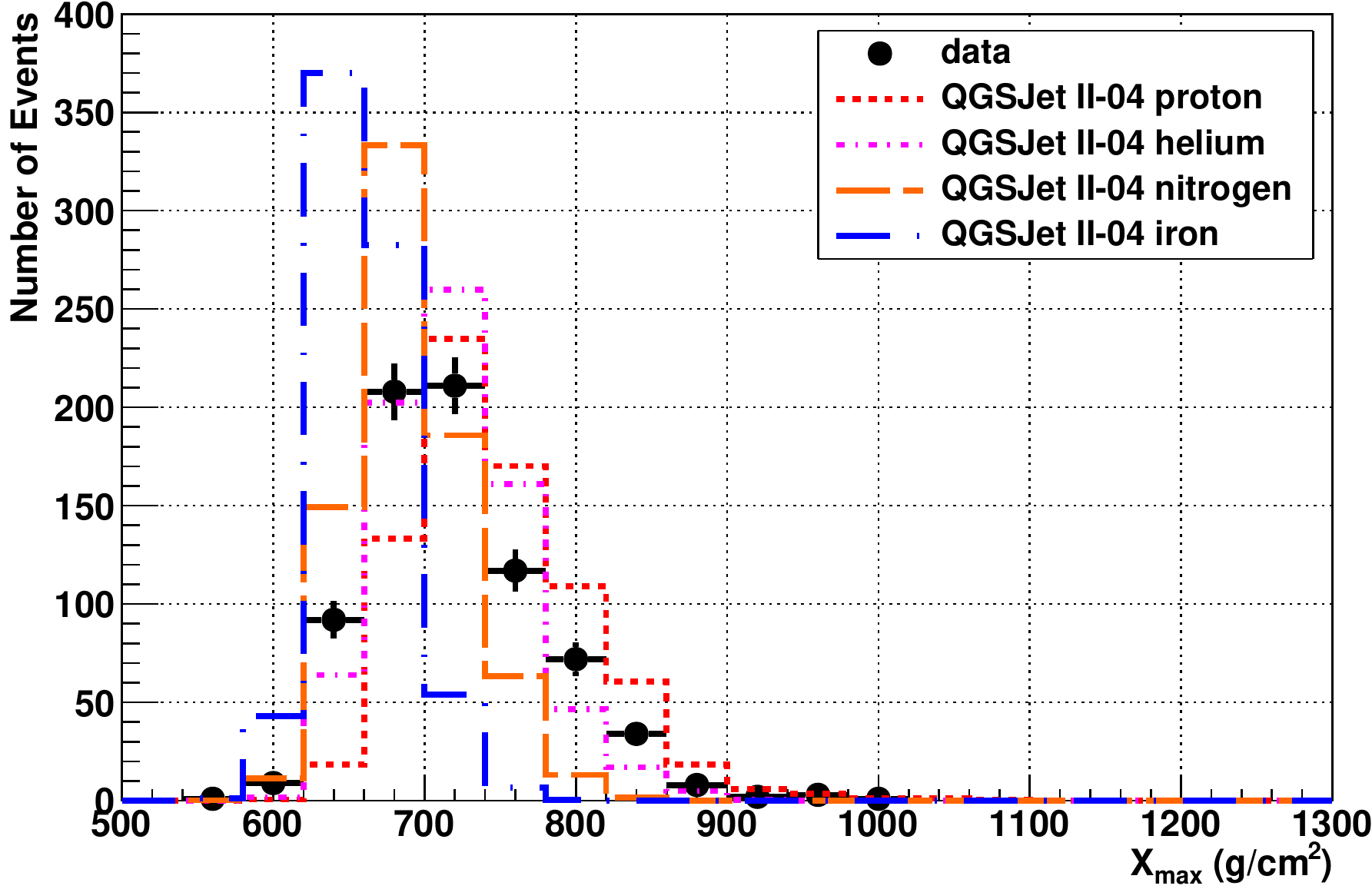}%
    \label{fig:dataMC_xmax_bin_01}%
  }

  \subfloat[$18.4 \leq \log_{10}(E/\mathrm{eV}) < 18.5$]{%
    \includegraphics[clip,width=0.48\columnwidth]{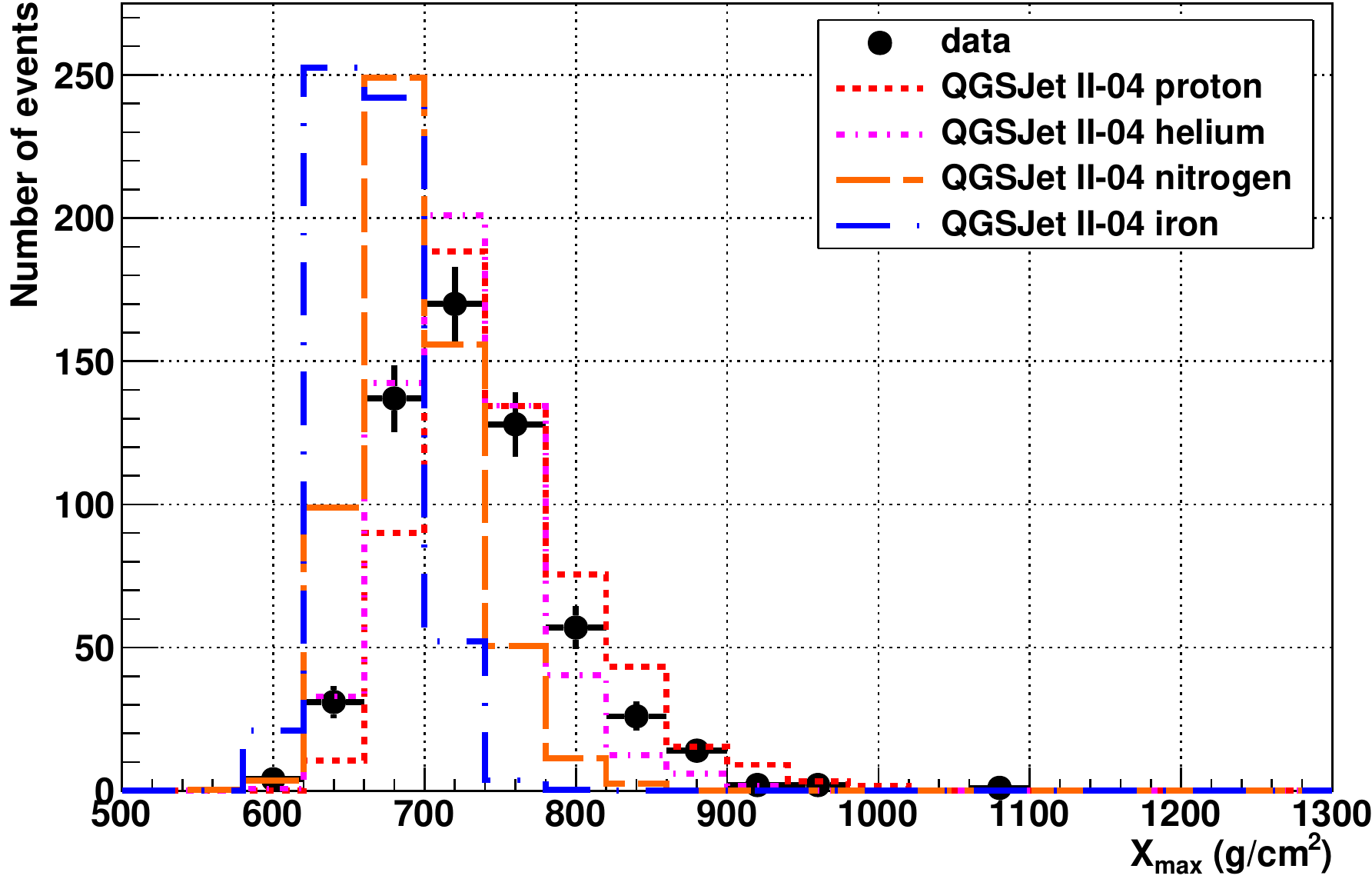}%
    \label{fig:dataMC_xmax_bin_02}%
  }
  ~
  \subfloat[$18.5 \leq \log_{10}(E/\mathrm{eV}) < 18.6$]{%
    \includegraphics[clip,width=0.48\columnwidth]{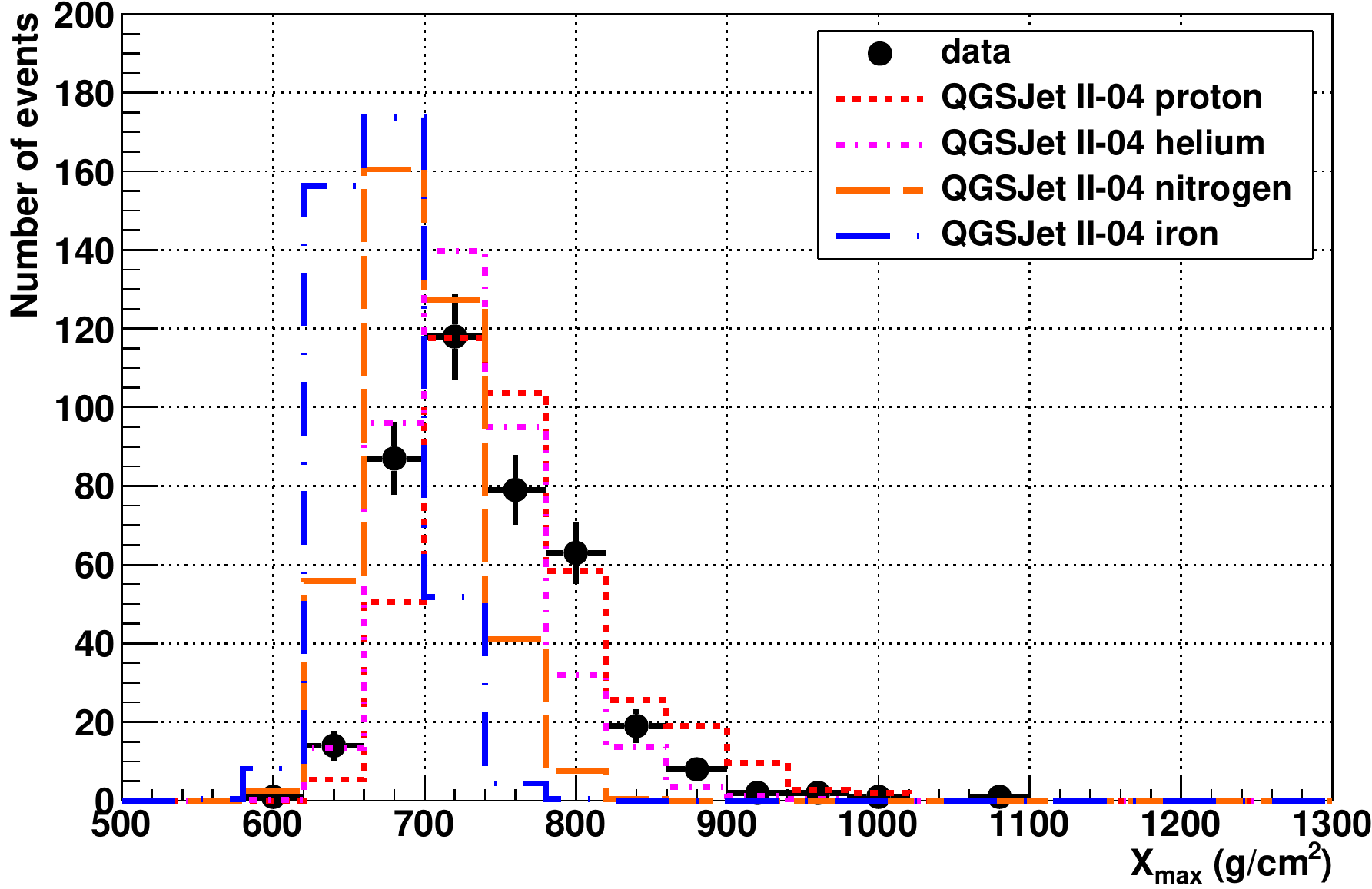}%
    \label{fig:dataMC_xmax_bin_03}%
  }

  \subfloat[$18.6 \leq \log_{10}(E/\mathrm{eV}) < 18.7$]{%
    \includegraphics[clip,width=0.48\columnwidth]{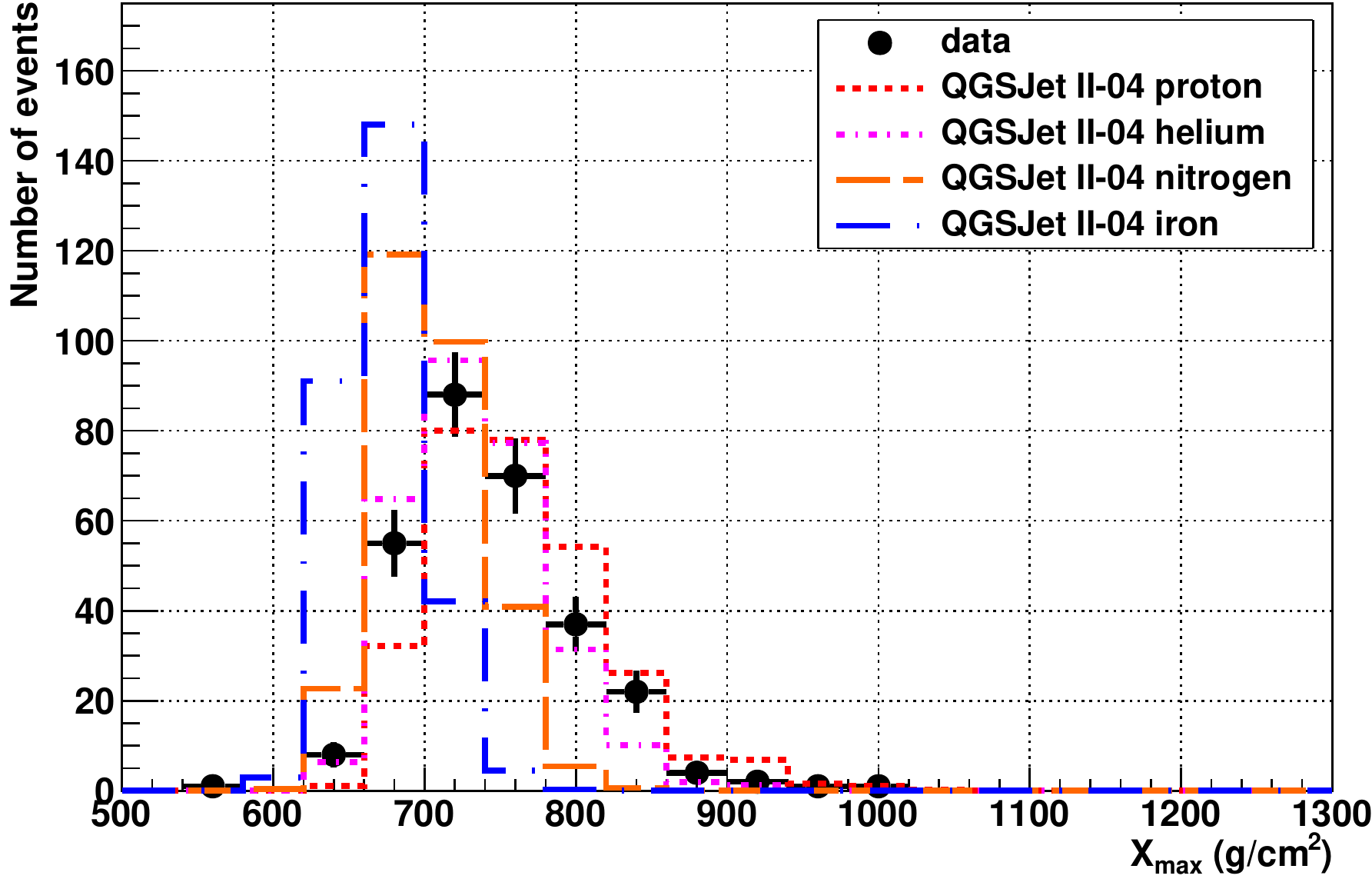}%
    \label{fig:dataMC_xmax_bin_04}%
  }
  ~
  \subfloat[$18.7 \leq \log_{10}(E/\mathrm{eV}) < 18.8$]{%
    \includegraphics[clip,width=0.48\columnwidth]{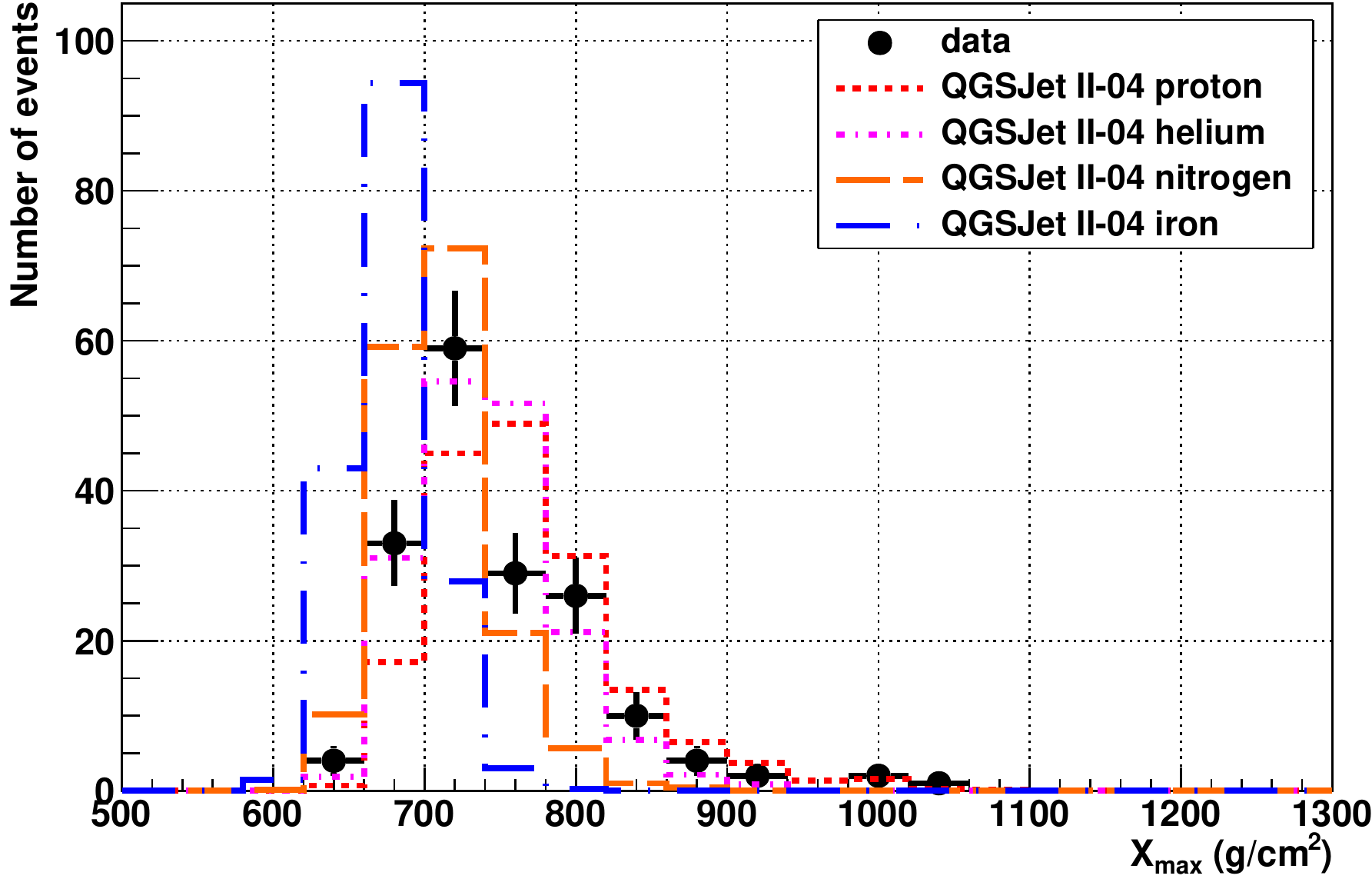}%
    \label{fig:dataMC_xmax_bin_05}%
  }
  \caption{\xm{} distributions in energy bins for $18.2 \leq
    \log_{10}(E/\mathrm{eV}) < 18.8$. The data is compared to Monte
    Carlo \xm{} distributions generated using the QGSJet~II-04
    hadronic model for four primary elements.}
  \label{fig:dataMC_xmax_01}
\end{figure}

\begin{figure}
  \centering
  \subfloat[$18.8 \leq \log_{10}(E/\mathrm{eV}) < 18.9$]{%
    \includegraphics[clip,width=0.48\columnwidth]{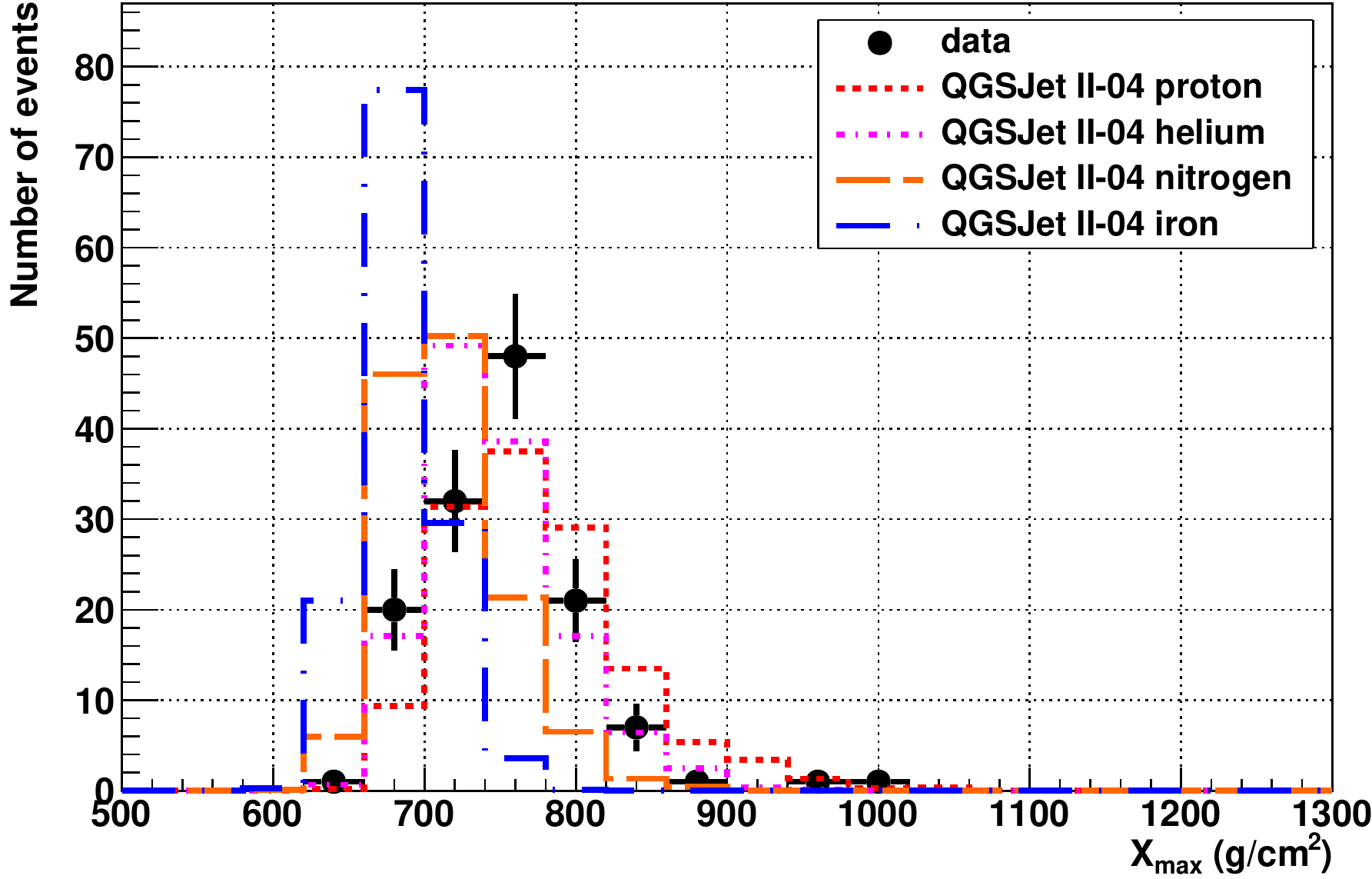}%
    \label{fig:dataMC_xmax_bin_06}%
  }
  ~
  \subfloat[$18.9 \leq \log_{10}(E/\mathrm{eV}) < 19.0$]{%
    \includegraphics[clip,width=0.48\columnwidth]{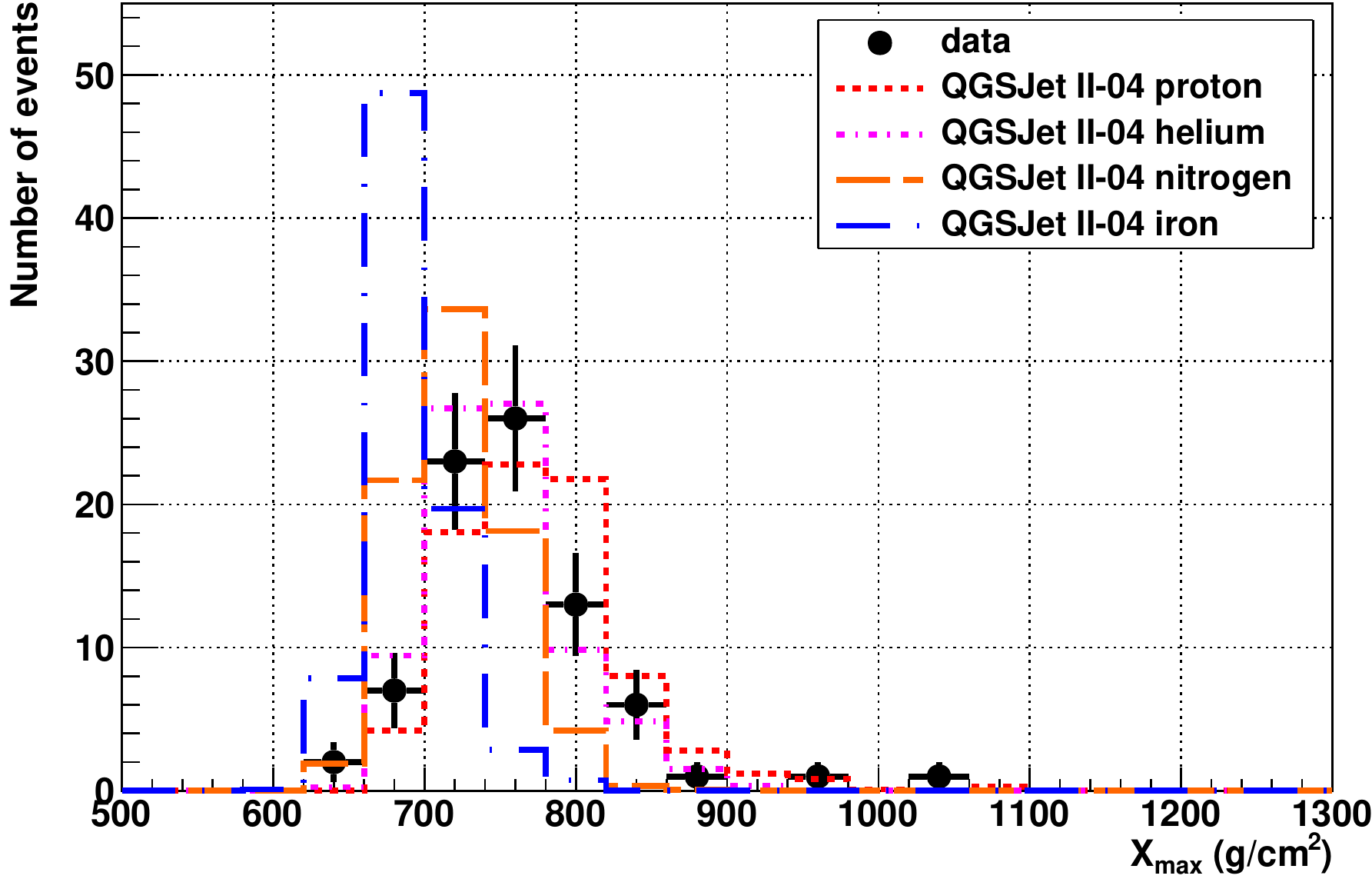}%
    \label{fig:dataMC_xmax_bin_07}%
  }

  \subfloat[$19.0 \leq \log_{10}(E/\mathrm{eV}) < 19.2$]{%
    \includegraphics[clip,width=0.48\columnwidth]{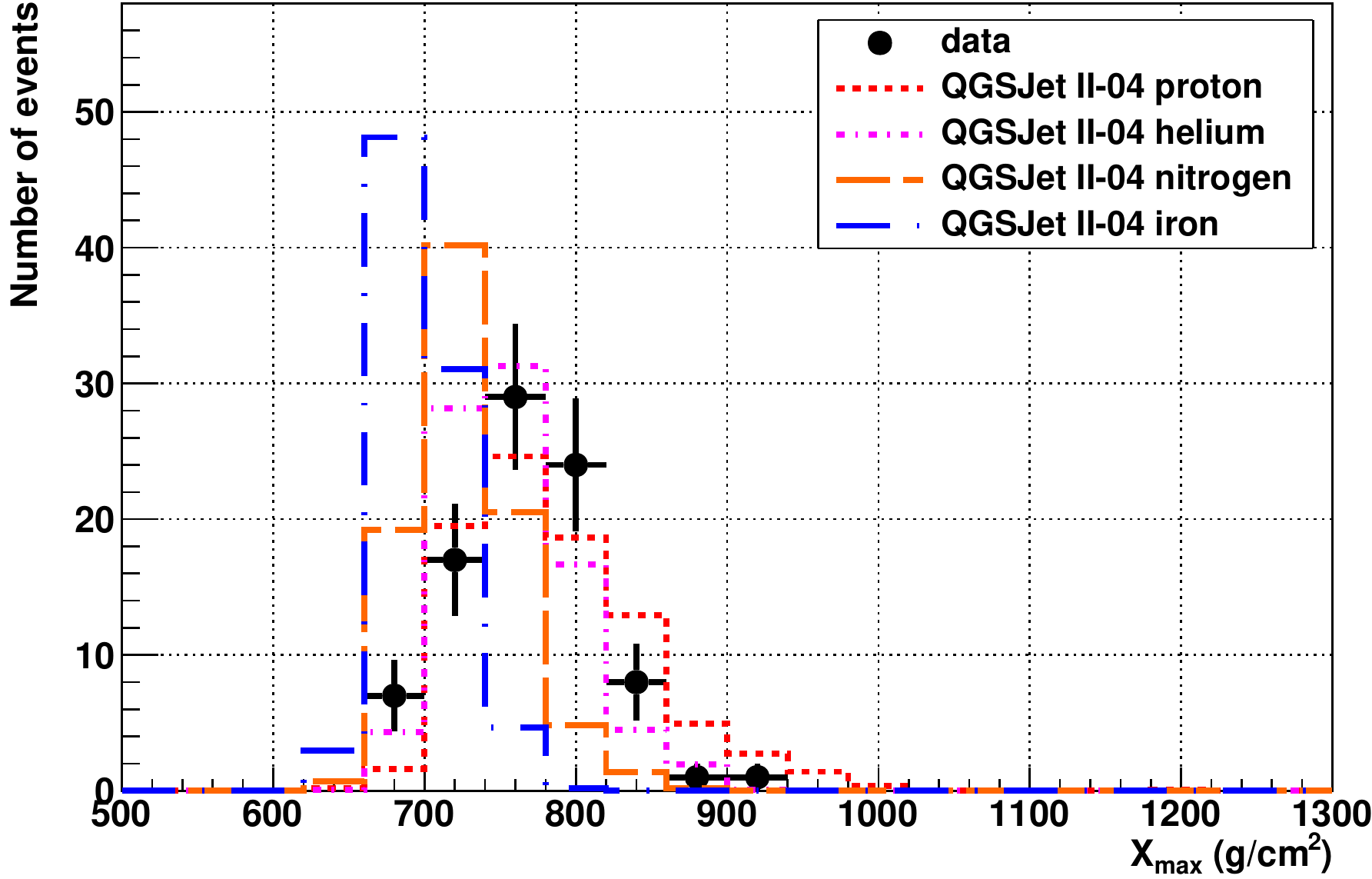}%
    \label{fig:dataMC_xmax_bin_08}%
  }
  ~
  \subfloat[$19.2 \leq \log_{10}(E/\mathrm{eV}) < 19.4$]{%
    \includegraphics[clip,width=0.48\columnwidth]{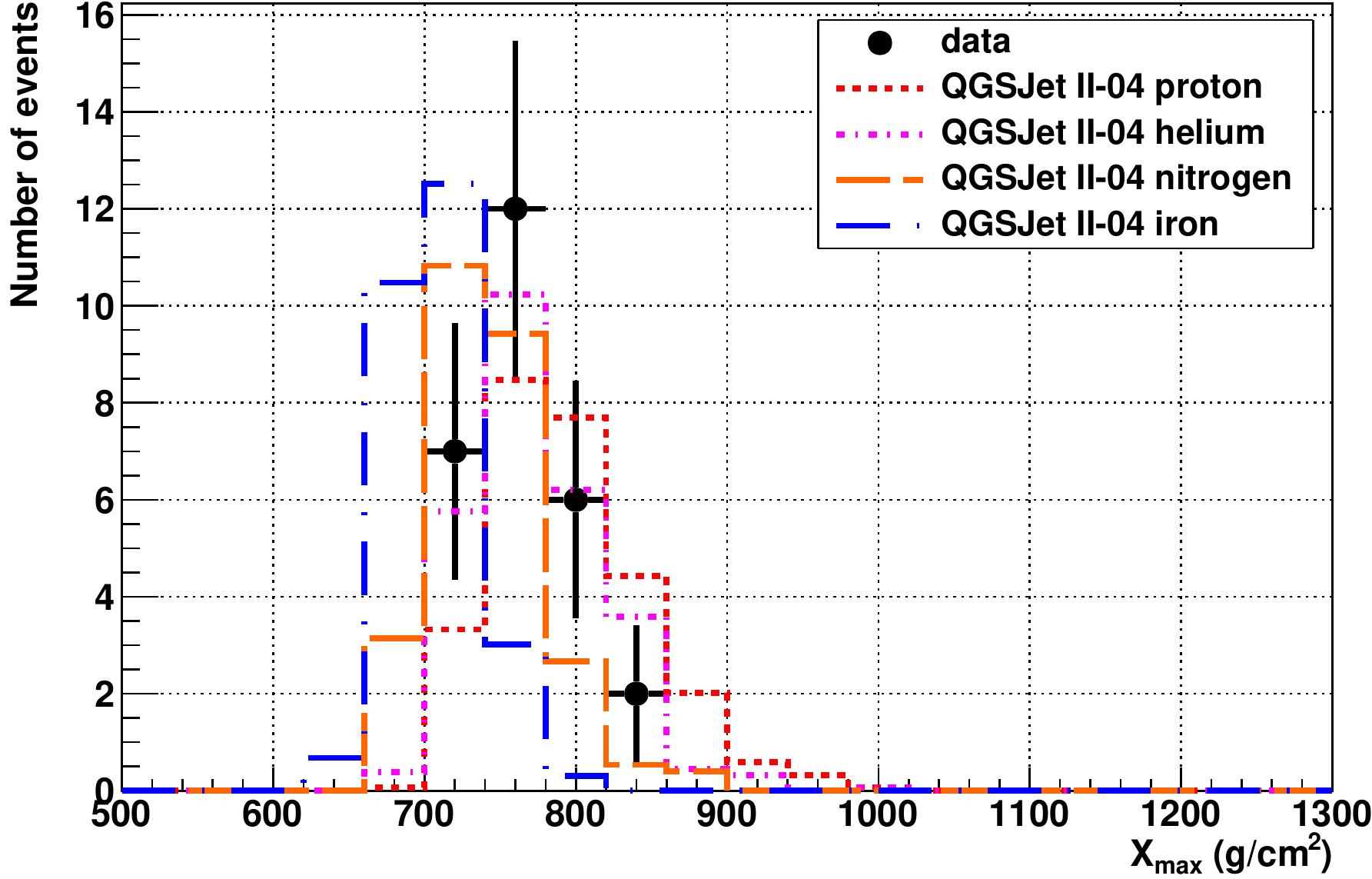}%
    \label{fig:dataMC_xmax_bin_09}%
  }

  \subfloat[$19.4 \leq \log_{10}(E/\mathrm{eV}) < 19.9$]{%
    \includegraphics[clip,width=0.48\columnwidth]{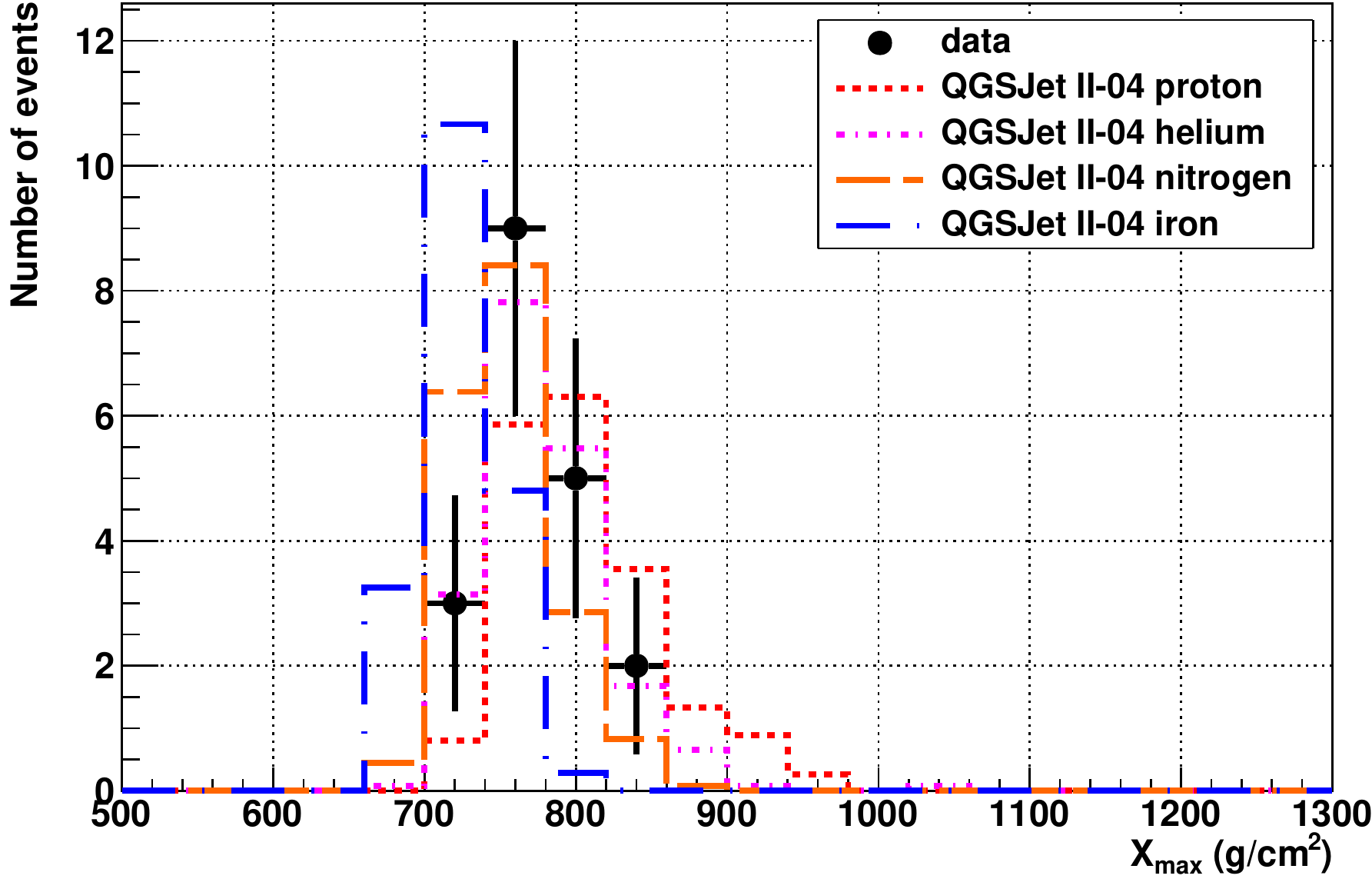}%
    \label{fig:dataMC_xmax_bin_10}%
  }~\hspace{0.5\textwidth}
  \caption{\xm{} distributions in energy bins for $18.8 \leq
    \log_{10}(E/\mathrm{eV}) < 19.9$. The data is compared to Monte
    Carlo \xm{} distributions generated using the QGSJet~II-04
    hadronic model for four primary elements.}
  \label{fig:dataMC_xmax_02}
\end{figure}

TA hybrid \xm{} data is binned by energy into eleven energy
bins. Below $10^{19}$~eV, there are sufficient statistics to use 0.1
decade wide energy bins, to provide $\gtrapprox 100$ events per
bin. Above $10^{19}$~eV the bins are widened to try and capture more
events. Figures~\ref{fig:dataMC_xmax_01} and \ref{fig:dataMC_xmax_02}
show the \xm{} distributions measured in this analysis. The
distributions for reconstructed QGSJet~II-04 Monte Carlo are shown as
well. For each energy bin, the histogram of each individual species is
normalized by area to the area of the data histogram. In a given
energy bin, lighter elements have larger \mxm{} as expected because of
the relationship $X_{\mathrm{max}} \propto \ln(E_{0}/A)$ discussed in
Section~\ref{sec:introduction}

Each figure shows that the means and standard deviations of the
distributions of the simulated elements decrease with increasing mass
as we expect. We can use them to compare to the data to determine
which pure element drawn from the QGSJet~II-04 model most resembles
the data and which elements may be excluded. Such a comparison does
not imply that we believe cosmic rays in nature to be composed of a
single chemical element in any given energy bin we've observed. In a
future paper, we will investigate the compatibility of TA data with
mixtures of elements. In this current work, we only compare TA data to
pure CORSIKA elements.

UHECR composition measurements typically utilize the first and second
moments of \xm{} distributions of data and Monte Carlo to compare
observed results to those expected for the models under
investigation. These individual quantities are too limited to fully
understand the details of \xm{} distributions, particularly for light
elements which exhibit prominent non-Gaussian tails. While examining
\mxm{} and \sxm{} as a function of energy is still useful, especially
to place an experiment in historical context with older measurements,
utilizing more powerful statistical techniques is more appropriate
given that more powerful computers now exist to make these
calculations much more practical. For this reason we will make our
primary visual comparisons of data and Monte Carlo by simultaneously
examining \mxm{} and \sxm{}, which is a more powerful way to
understand the relationship of the data and Monte Carlo.

\begin{figure}
  \centering
  \subfloat[$18.2 \leq \log_{10}(E/\mathrm{eV}) < 18.3$]{%
    \includegraphics[clip,width=0.48\columnwidth]{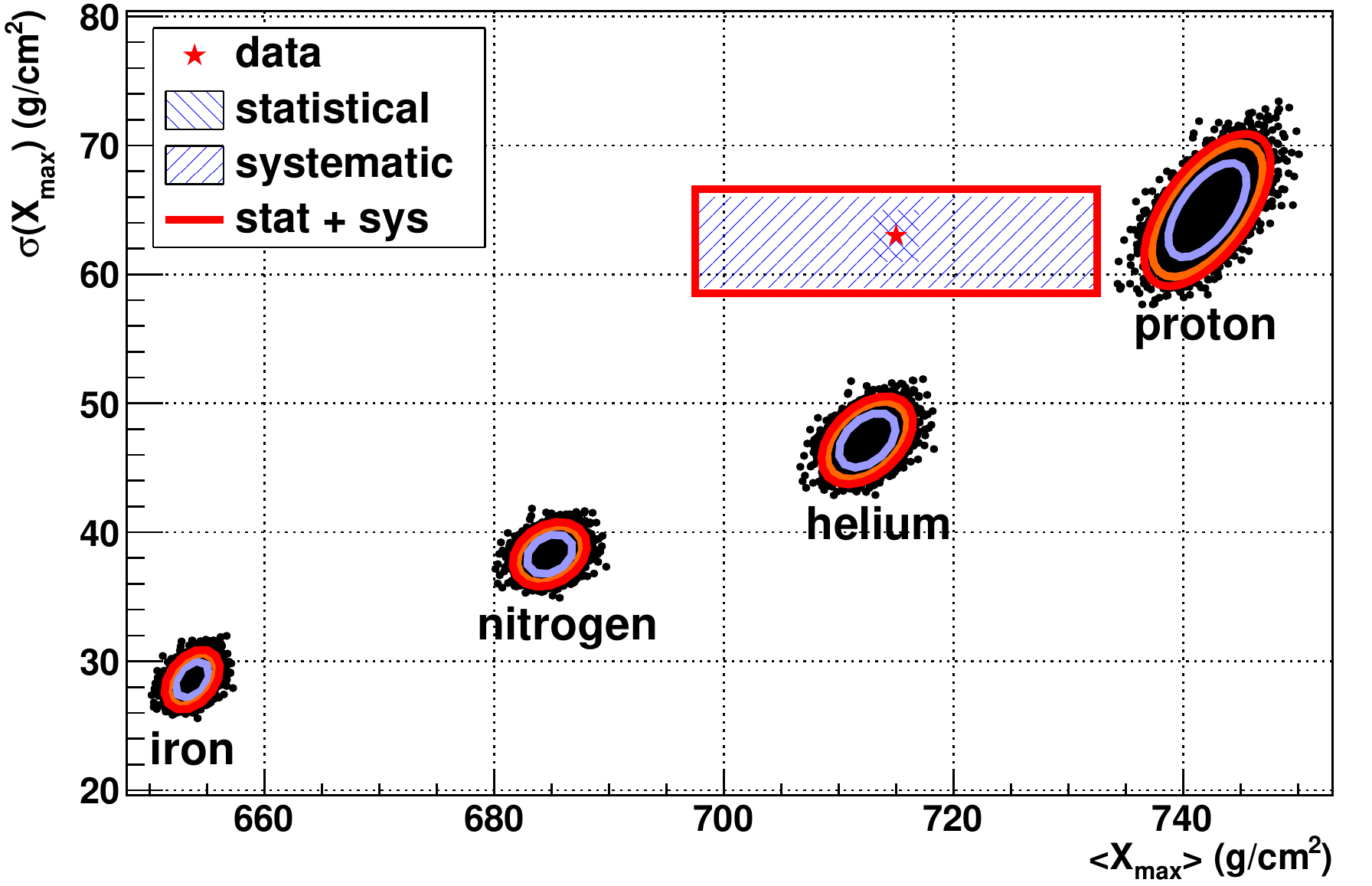}%
    \label{fig:mxm_sxm_bin_00}%
  }
  ~
  \subfloat[$18.3 \leq \log_{10}(E/\mathrm{eV}) < 18.4$]{%
    \includegraphics[clip,width=0.48\columnwidth]{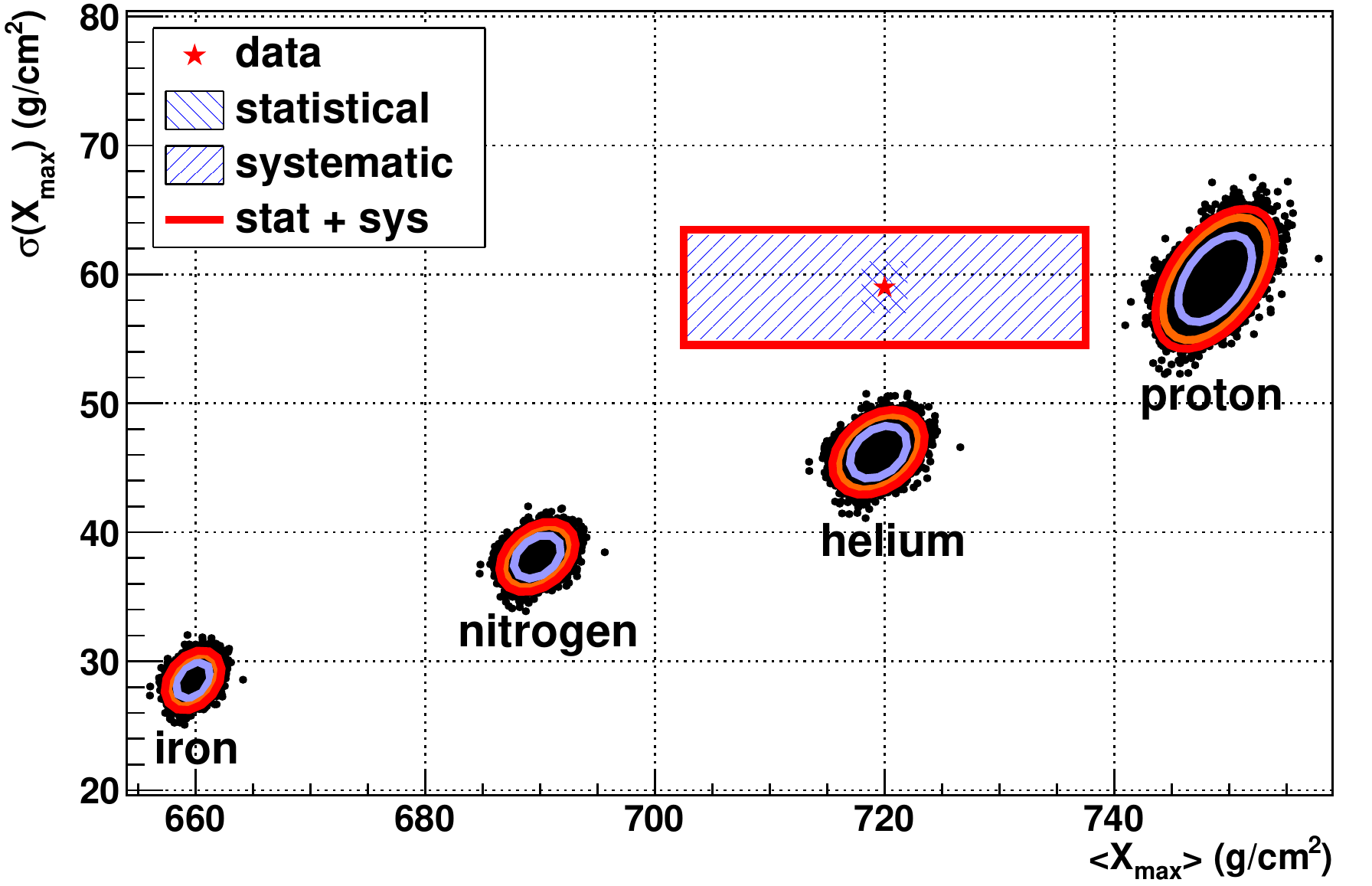}%
    \label{fig:mxm_sxm_bin_01}%
  }

  \subfloat[$18.4 \leq \log_{10}(E/\mathrm{eV}) < 18.5$]{%
    \includegraphics[clip,width=0.48\columnwidth]{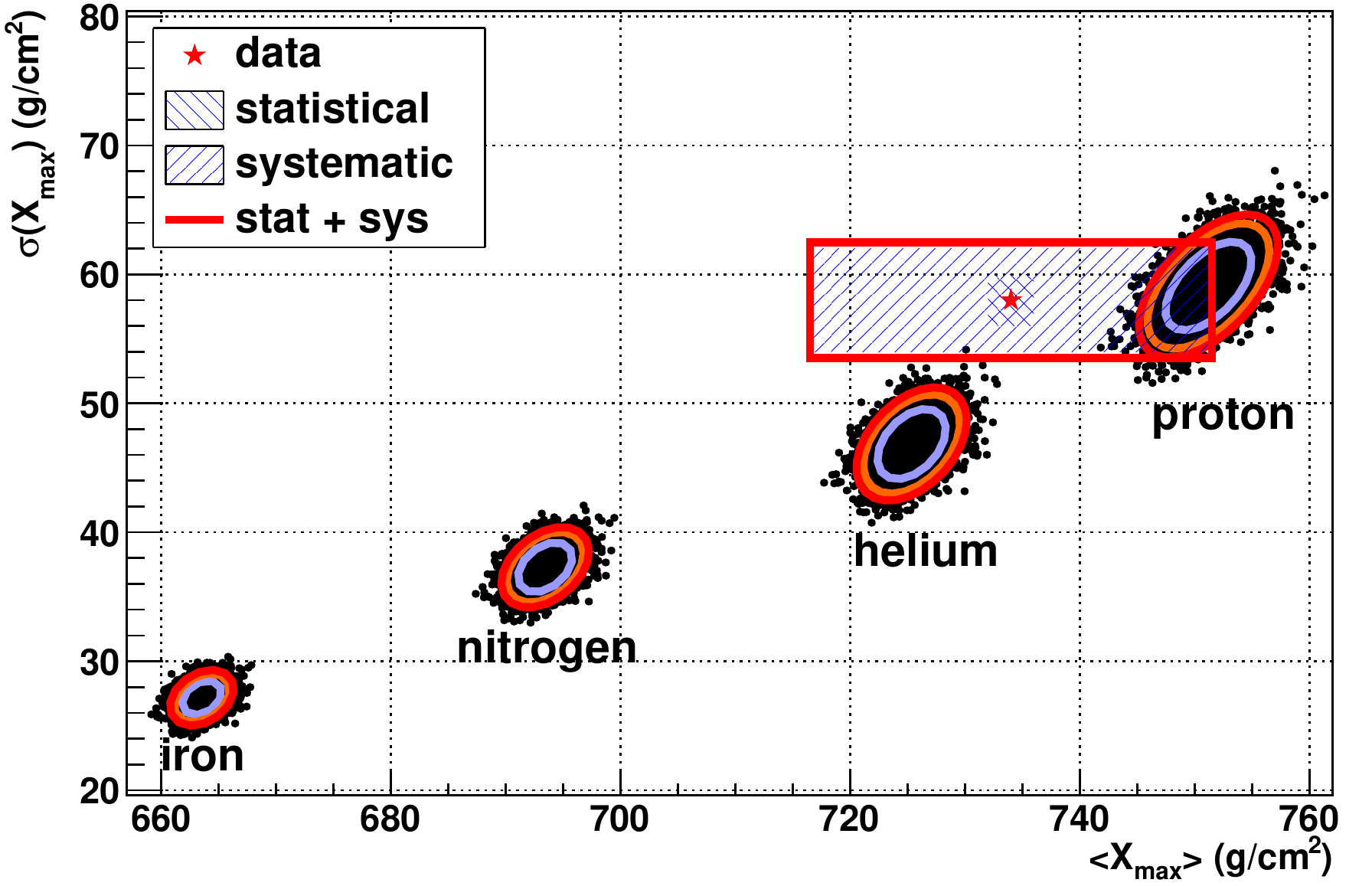}%
    \label{fig:mxm_sxm_bin_02}%
  }
  ~
  \subfloat[$18.5 \leq \log_{10}(E/\mathrm{eV}) < 18.6$]{%
    \includegraphics[clip,width=0.48\columnwidth]{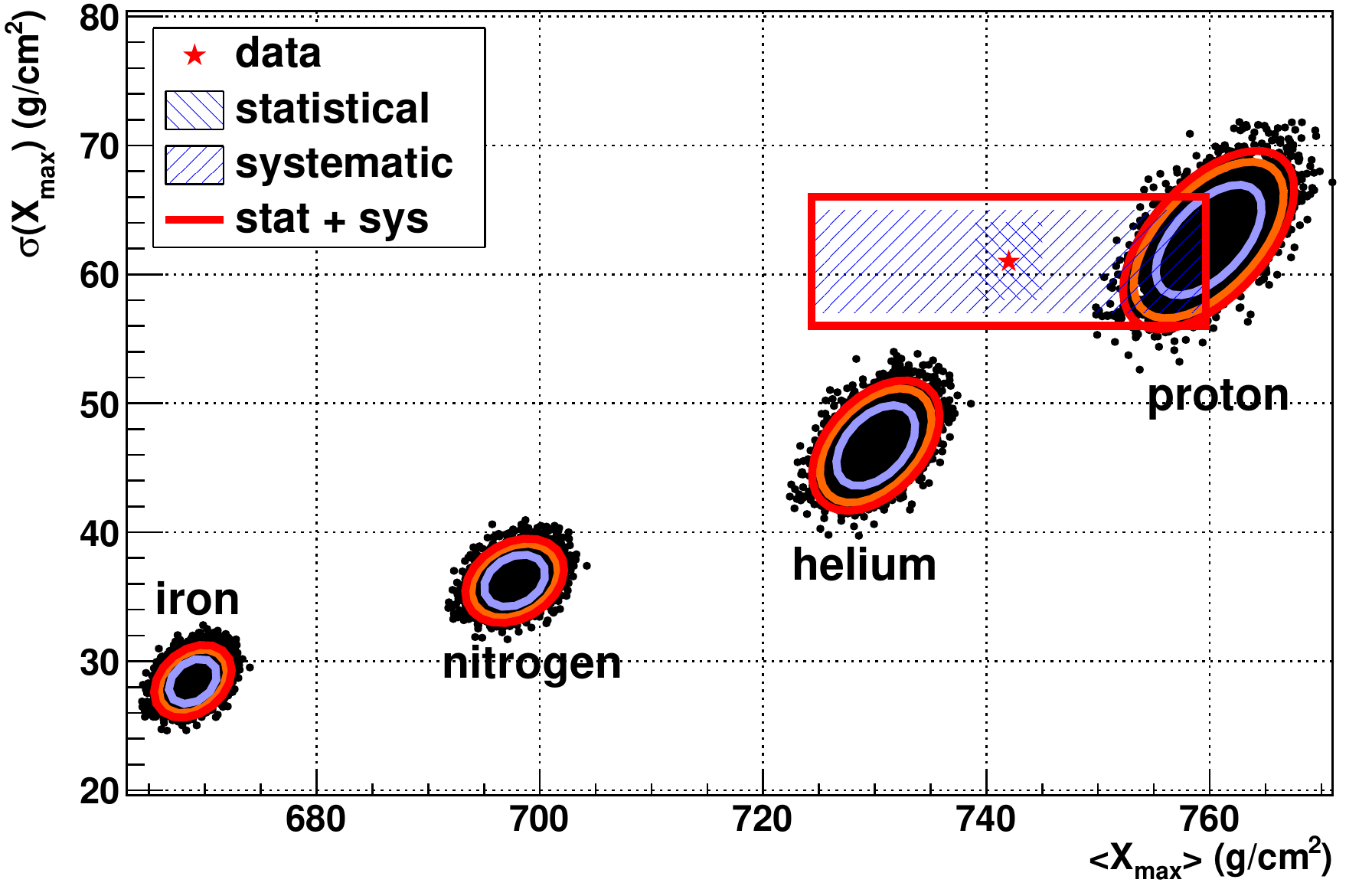}%
    \label{fig:mxm_sxm_bin_03}%
  }

  \subfloat[$18.6 \leq \log_{10}(E/\mathrm{eV}) < 18.7$]{%
    \includegraphics[clip,width=0.48\columnwidth]{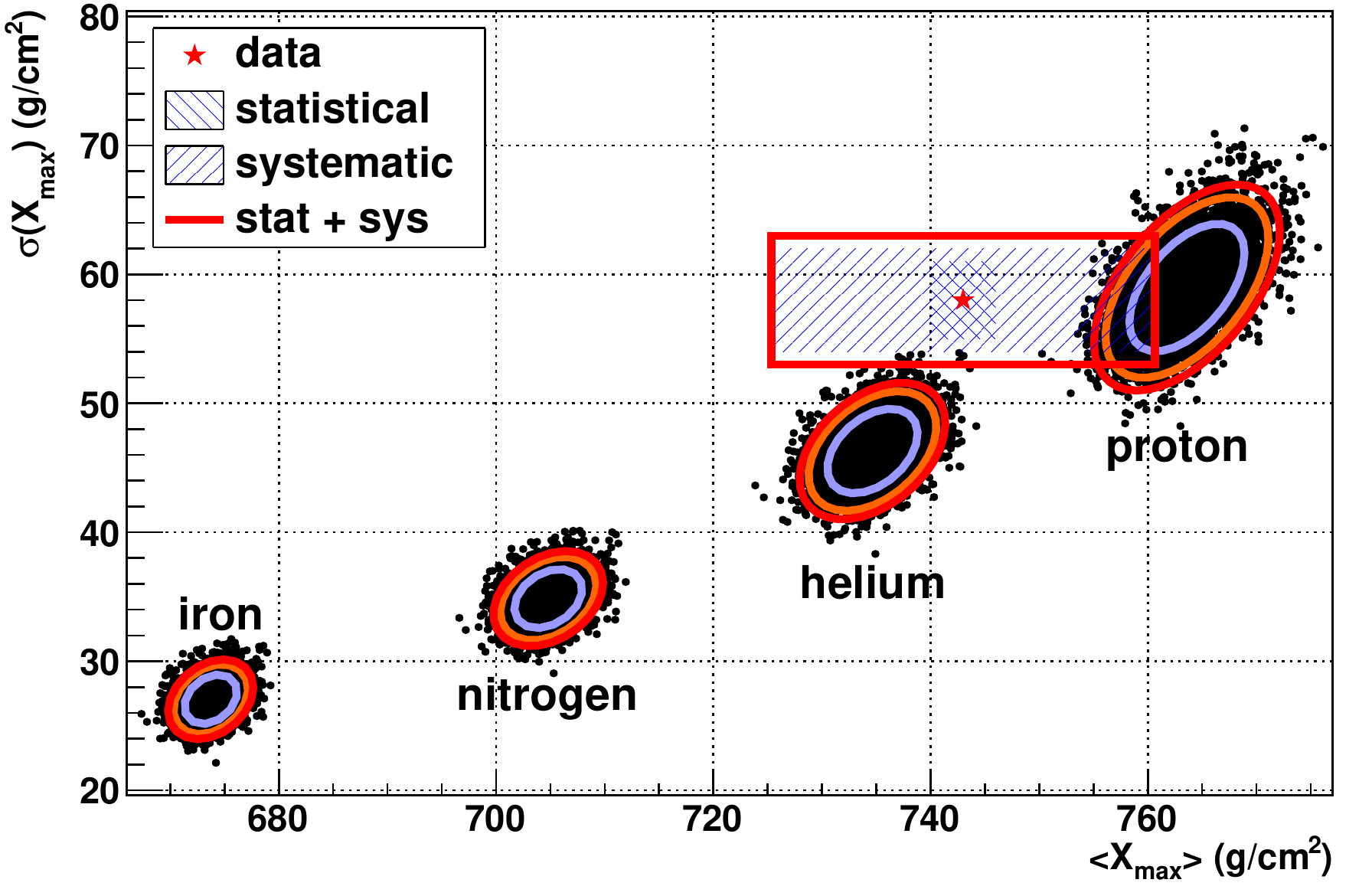}%
    \label{fig:mxm_sxm_bin_04}%
  }
  ~
  \subfloat[$18.7 \leq \log_{10}(E/\mathrm{eV}) < 18.8$]{%
    \includegraphics[clip,width=0.48\columnwidth]{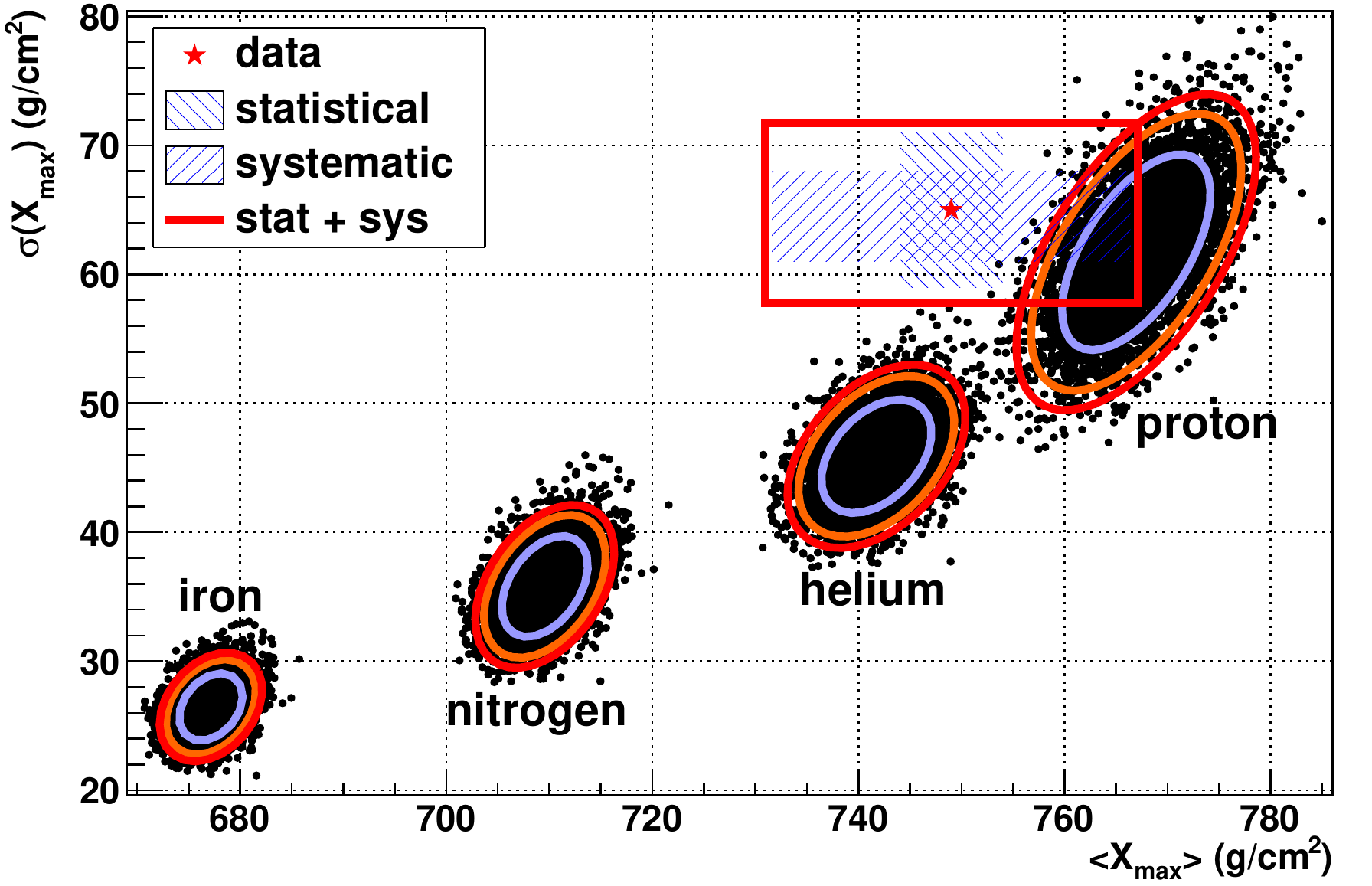}%
    \label{fig:mxm_sxm_bin_05}%
  }
  \caption{Measurements of data and QGSJet~II-04 Monte Carlo \mxm{}
    and \sxm{} in energy bins for $18.2 \leq \log_{10}(E/\mathrm{eV})
    < 18.8$. Each Monte Carlo chemical element shows the 68.3\% (blue
    ellipse), 90\% (orange ellipse), and 95\% (red ellipse) confidence
    intervals.} 
  \label{fig:mxm_sxm_01}
\end{figure}

\begin{figure}
  \centering
  \subfloat[$18.8 \leq \log_{10}(E/\mathrm{eV}) < 18.9$]{%
    \includegraphics[clip,width=0.48\columnwidth]{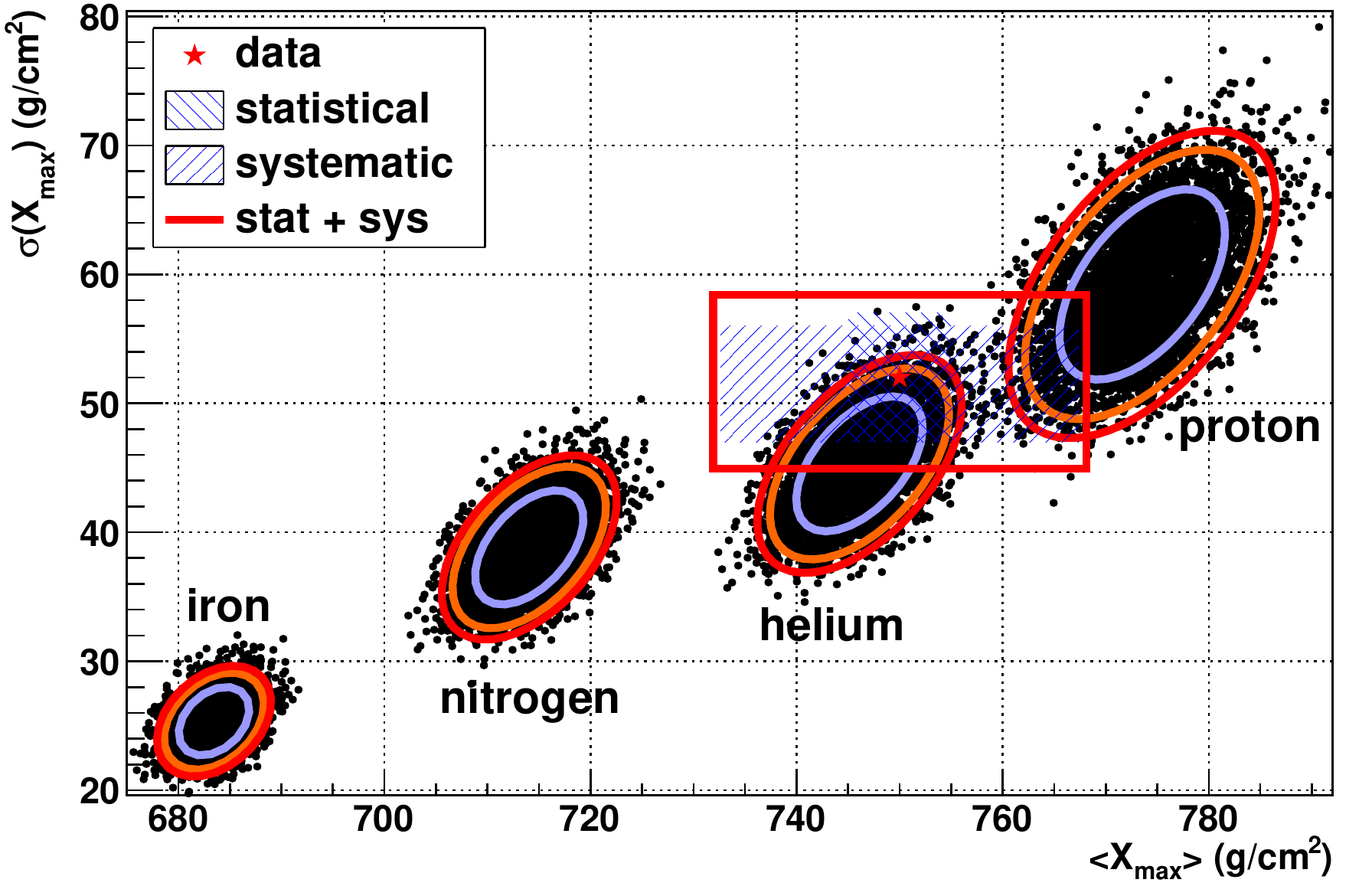}%
    \label{fig:mxm_sxm_bin_06}%
  }
  ~
  \subfloat[$18.9 \leq \log_{10}(E/\mathrm{eV}) < 19.0$]{%
    \includegraphics[clip,width=0.48\columnwidth]{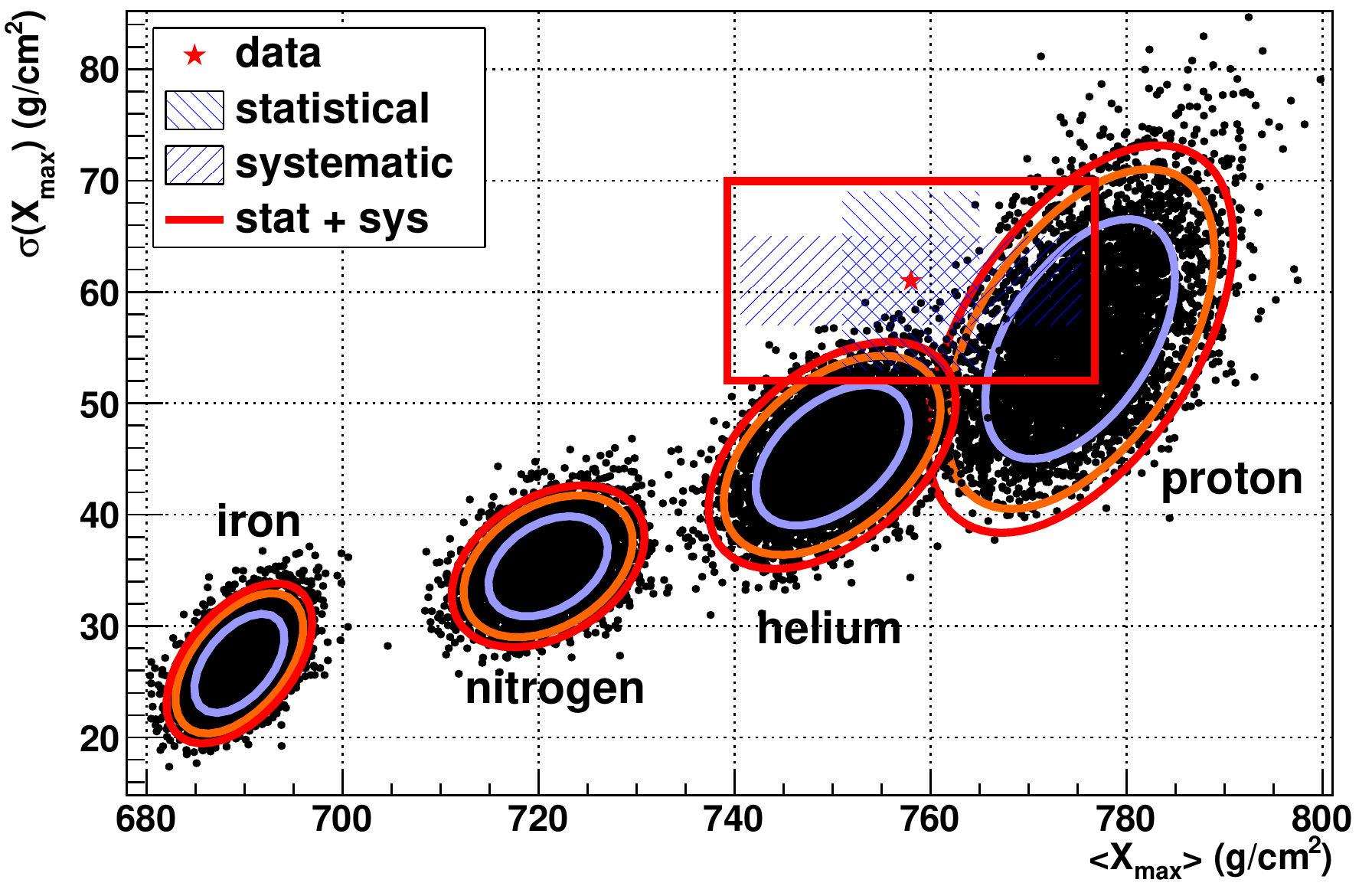}%
    \label{fig:mxm_sxm_bin_07}%
  }

  \subfloat[$19.0 \leq \log_{10}(E/\mathrm{eV}) < 19.2$]{%
    \includegraphics[clip,width=0.48\columnwidth]{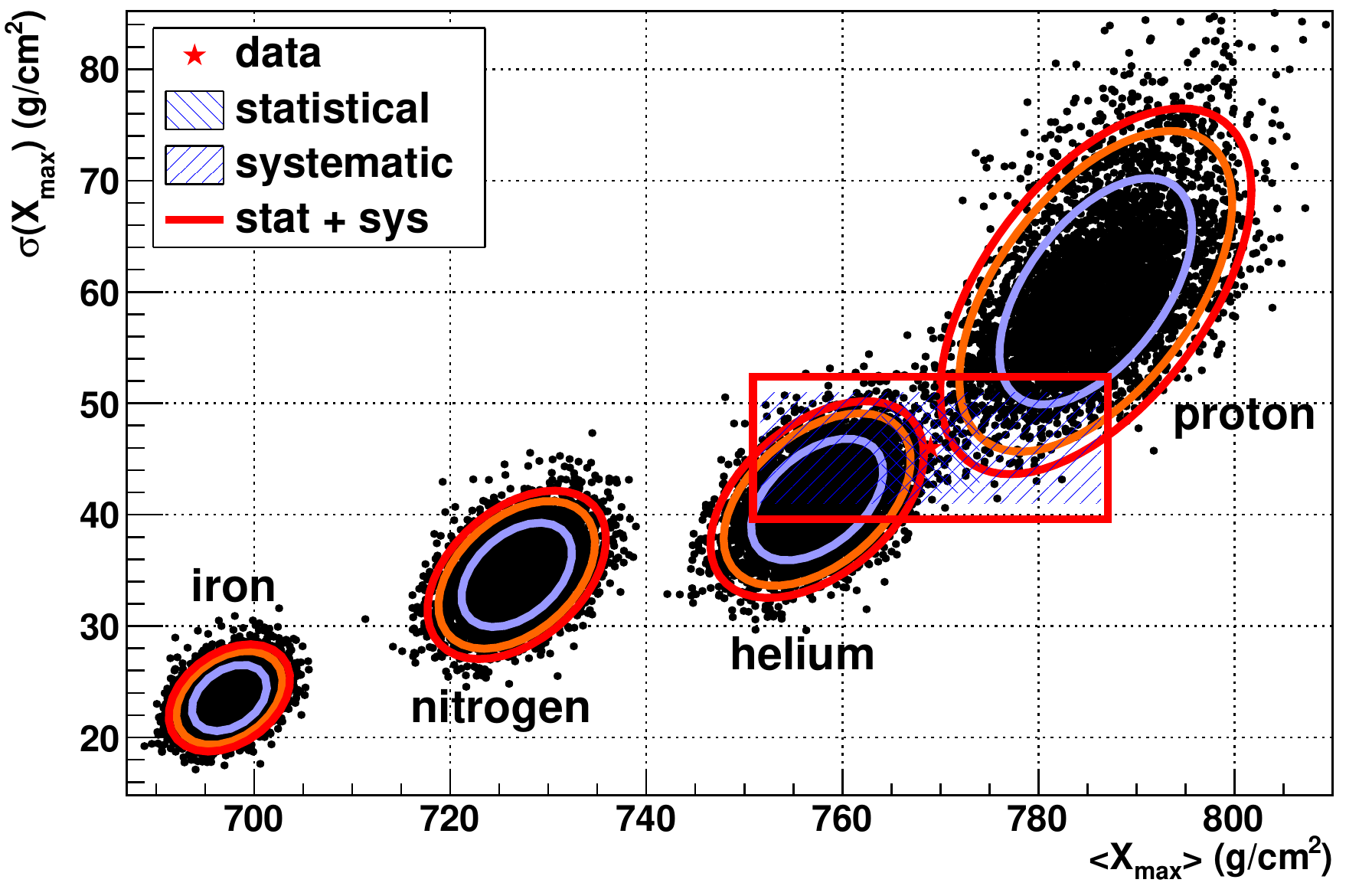}%
    \label{fig:mxm_sxm_bin_08}%
  }
  ~
  \subfloat[$19.2 \leq \log_{10}(E/\mathrm{eV}) < 19.4$]{%
    \includegraphics[clip,width=0.48\columnwidth]{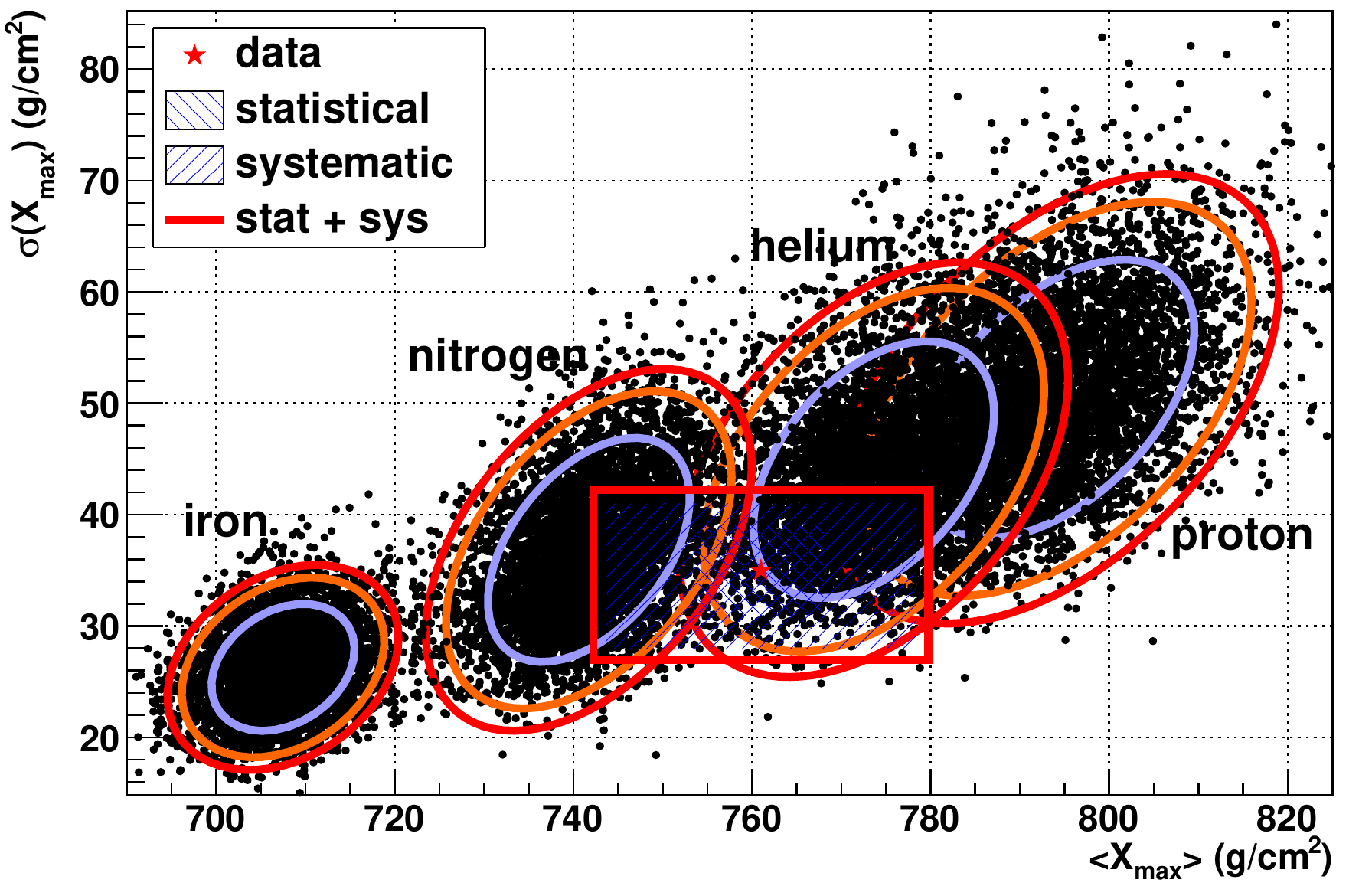}%
    \label{fig:mxm_sxm_bin_09}%
  }

  \subfloat[$19.4 \leq \log_{10}(E/\mathrm{eV}) < 19.9$]{%
    \includegraphics[clip,width=0.48\columnwidth]{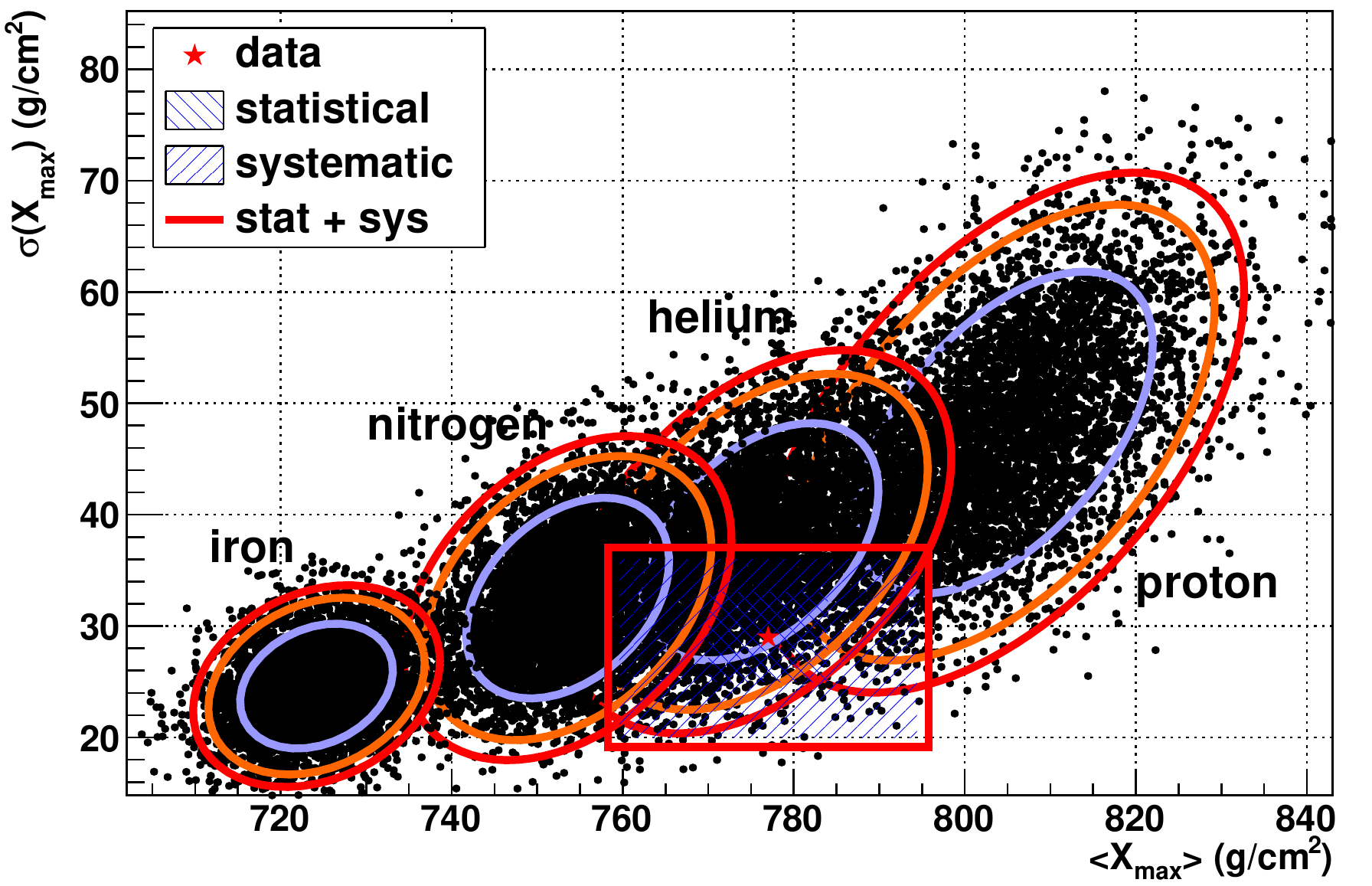}%
    \label{fig:mxm_sxm_bin_10}%
  }~\hspace{0.5\textwidth}
  \caption{Measurements of data and QGSJet~II-04 Monte Carlo \mxm{}
    and \sxm{} in energy bins for $18.8 \leq \log_{10}(E/\mathrm{eV})
    < 19.9$. Each Monte Carlo chemical element shows the 68.3\% (blue
    ellipse), 90\% (orange ellipse), and 95\% (red ellipse) confidence
    intervals.}
  \label{fig:mxm_sxm_02}
\end{figure}

Recalling the discussion of the relationship of mass and energy to the
mean and width of the \xm{} distribution from
Section~\ref{sec:introduction}, we can examine the signature of a
given element as observed by 8.5 years of exposure in TA in hybrid
mode by simultaneously measuring the distributions of \mxm{} and
\sxm{}. Light elements will have both larger \mxm{} and \sxm{}
distributions, because of the larger fluctuations in first the
interaction in the atmosphere and subsequent shower development. We
will also be able to see the affect of TA's acceptance on the
reconstructed distributions and compare them to the observed data. To
do this the reconstructed \xm{} distribution for a single element,
such as QGSJet~II-04 protons shown in Figures~\ref{fig:dataMC_xmax_01}
and \ref{fig:dataMC_xmax_02}, is sampled according to the same number
of events recorded in the data for a given energy bin. \mxm{} and
\sxm{} of the energy are calculated and recorded for this sample. This
procedure is then 5000 times. The distribution of \mxm{} and \sxm{} is
used to calculate the 68.3\%, 90\%, and 95\% confidence intervals. The
entire \mxm{} and \sxm{} calculated by this method is then plotted as
a 2-dimensional distribution along with the computed confidence
intervals. This procedure is repeated for the other three chemical
elements used in the analysis. Figures~\ref{fig:mxm_sxm_01} and
\ref{fig:mxm_sxm_02} show this measurement for all observed energy
bins. The \mxm{} and \sxm{} of the data observed in each energy bin is
also recorded as a single red star. Additionally, the statistical,
systematic, and combined statistical and systematic error bounds are
marked around the data.

Figures~\ref{fig:mxm_sxm_bin_00} and \ref{fig:mxm_sxm_bin_01},
corresponding to the energy range $10^{18.2} - 10^{18.4}$~eV, show
that each of the four modeled chemical elements have clear separation
and are individually resolvable by TA in those energy bins given our
acceptance and statistics in the data (801 and 758 events,
respectively). The \sxm{} of the data resembles QGSJet~II-04 protons,
but the \mxm{} of the data is lower by about our systematic
uncertainty in those energy bins. Note that we do not account for
systematic uncertainties in the QGSJet~II-04 model, which will be
discussed in
Section~\ref{sec:hypothesis_tests}. Figures~\ref{fig:mxm_sxm_bin_02} -
\ref{fig:mxm_sxm_bin_05}, corresponding to the energy range $10^{18.4}
- 10^{18.8}$~eV show that \sxm{} of the data continue to resemble
QGSJet~II-04 protons, and \mxm{} of the data falls within the 68.3\%
confidence interval of the proton distributions within the data's
systematic uncertainty. We also notice an effect of decreasing
statistics in the data, by observing the increase in the size of the
confidence intervals of the individual Monte Carlo
elements. Figure~\ref{fig:mxm_sxm_bin_06}, corresponding to the energy
range $10^{18.8} - 10^{18.9}$~eV, shows a relatively large downward
fluctuation in \sxm{} of the data. In this energy bin, the 68.3\%
confidence intervals of QGSJet~II-04 proton and helium both fall
within the bounds of the systematic uncertainty of the data. In
Figure~\ref{fig:mxm_sxm_bin_07}, corresponding to the energy range
$10^{18.9} - 10^{19.0}$~eV, \sxm{} of the data fluctuates up from the
previous energy bin and the systematic error bounds of the data falls
within the 68.3\% confidence interval of protons. In this energy,
because of the small statistics in the data (80 events), the largest
confidence intervals of proton and helium begin to overlap. This
indicates that given TA's current exposure in this energy bin, we are
losing our ability make precise statements about the signature of pure
light chemical elements. The ability to distinguish among nitrogen
from iron, or nitrogen from helium or protons remains. In
Figure~\ref{fig:mxm_sxm_bin_08}, corresponding the energy range
$10^{19.0} - 10^{19.2}$~eV, the 95\% confidence intervals of proton
and helium are once again separated, but here we have doubled the size
of the energy bin in an attempt to enable us to still make a
reasonably good measurement of \mxm{} and \sxm{} in the data. The
68.3\% confidence intervals of both proton and helium fall within the
bounds of the systematic uncertainty of the
data. Figures~\ref{fig:mxm_sxm_bin_09} and \ref{fig:mxm_sxm_bin_10},
corresponding to the energy range $10^{19.2} - 10^{19.9}$~eV, show
that TA's ability to resolve individual QGSJet~II-04 elements is
degraded due to the overlap of the confidence intervals. According to
these figures, when considering only the joint distributions of \mxm{}
and \sxm{}, within the data's systematic uncertainty the data may
resemble QGSJet~II-04 proton, helium, or nitrogen.

Figure~\ref{fig:xmax_elongation} shows only the means of the \xm{}
distributions presented in Figures~\ref{fig:dataMC_xmax_01},
\ref{fig:dataMC_xmax_02}, \ref{fig:mxm_sxm_01}, and
\ref{fig:mxm_sxm_02}, also called the elongation rate, of the observed
data ($d\left<X_{\mathrm{max}}\right>/d\log E$), as well as
reconstructed Monte Carlo for four primary species. The gray band
around the data points indicates the systematic uncertainty in \mxm{}
of 17.4~g/cm$^{2}$ estimated for this analysis.

\subsection{\label{sec:systematics}Systematic Uncertainties}
Systematic uncertainties in the \mxm{} and \sxm{} are evaluated for
four sources: detector modeling, atmosphere, fluorescence
yield, and the reconstruction algorithm.

The pointing accuracies of the phototubes ($\pm 0.05$~degrees) and the
relative timing between FD and SD (240ns) dominate the detector
effects. These effects give $\pm 3.3$~g/cm$^2$ and $\pm 3.8$~g/cm$^2$
uncertainty in \mxm, and $\pm 1.7$~g/cm$^2$ and $\pm 4.0$~g/cm$^2$ in
\sxm, respectively. Some events are detected by both FD
stations (BR and LR), and the \xm{} differences of such stereo
events can also be used to estimate the detector effect. We found
that the BR-LR difference is smaller than 10~g/cm$^2$ in \mxm{} for
events with energies greater than $10^{18.2}$~eV.

The atmospheric effect is dominated by the amount of aerosols.  We
have $\sim 15\%$ uncertainty of aerosols in terms of vertical aerosol
optical depth (VAOD), which gives a shift in \mxm{} of
3.4~g/cm$^2$. Variations in atmospheric aerosols potentially has a
large affect on \sxm. Aerosols are measured every 30 minutes by the
central laser facility (CLF) \citep{Tomida:2013wka}. If we compare
how \sxm{} varies using the VAOD data measured by the CLF, the effect on
\sxm{} is found to be 18.9~g/cm$^2$.  Another effect comes from the
atmospheric profile, i.e., the pressure and density of the atmosphere as
functions of height. When we reconstruct data using an atmospheric
database that uses NOAA National Weather Service radiosonde data
instead of the Global Data Assimilation System (GDAS) \citep{bib:GDAS},
the effects on \mxm{} and \sxm{} are found to be 5.9~g/cm$^2$ and
7.4~g/cm$^2$, respectively.

We use a fluorescence yield model which uses the absolute yield
measurement by Kakimoto et al. \citep{Kakimoto:1995pr} and the
fluorescence spectral measurement by the FLASH
experiment \citep{Abbasi:2007am}. A $\sim 5-6$~g/cm$^2$ effect is
expected if we use different fluorescence modeling. For example, we
found a $+5.6$~g/cm$^2$ shift in \mxm{} and a 3.7~g/cm$^2$ effect in
\sxm{} when we use the model based on the measurements by the AirFly
experiment \citep{Ave:2007xh,Ave:2012ifa} on the absolute yield, the
spectrum, and the atmospheric parameter dependencies.

A systematic effect in \xm{} also comes from the reconstruction
program used in the analysis. We have two reconstruction programs
independently developed in TA for the same data. The reconstruction
bias can be estimated by an event-by-event comparison of \xm{} values
calculated by these two separate reconstruction procedures, and this
is smaller than 4.1~g/cm$^2$ for events with $E > 10^{18.2}$~eV.

Some of these contributions are not fully independent. For example,
the uncertainties evaluated from the BR-LR difference and the
comparison of different analysis programs could be correlated. In the
calculation of the total systematic uncertainty, we use a linear sum
of these two sources of uncertainty (14.1~g/cm$^2$) as a conservative
estimate. Other sources are added in quadrature, and we find the
total systematic uncertainty in \mxm{} to be 17.4~g/cm$^2$. The results
are summarized in Table~\ref{tab:SysUncInMeanXmax}.

The systematic uncertainties of \sxm{} from the sources discussed
above are also evaluated, and given in
Table~\ref{tab:SysUncInSigmaXmax}. Adding in quadrature, we obtain
21.1~g/cm$^2$.

\begin{table}
    \begin{tabular}{ccp{0.5\columnwidth}}
        \centering Items & $\Delta$\mxm & Notes
        \\ \hline \hline \multicolumn{3}{c}{Independent sources}
        \\ \hline Detector & 5.1~g/cm$^2$ & Relative timing between FD and
        SD (3.8~g/cm$^2$), pointing direction of the telescope
        (3.3~g/cm$^2$)\\ Atmosphere & 6.8~g/cm$^2$ & Aerosol (3.4~g/cm$^2$),
        atmospheric depth (5.9~g/cm$^2$) \\ Fluorescence yield & 5.6~g/cm$^2$
        & Difference in yield models \\ \hline Quadratic sum & 10.2~g/cm$^2$
        & \\ \hline \hline \multicolumn{3}{c}{Not fully
          independent sources} \\ \hline Detector & 10.0~g/cm$^2$ &
        Difference in two FD stations \\ Reconstruction & 4.1~g/cm$^2$ &
        Difference in reconstructions \\ \hline Linear
        sum & 14.1~g/cm$^2$ & \\ \hline \hline Total & 17.4~g/cm$^2$ & \\
    \end{tabular}
    \caption{The systematic uncertainties in \mxm{} of TA hybrid
      BR/LR reconstruction.}
    \label{tab:SysUncInMeanXmax}
\end{table}

\begin{table}
    \begin{tabular}{ccp{0.5\columnwidth}}
        \centering Items & $\Delta$\sxm & Notes
        \\ \hline Detector & 4.3~g/cm$^2$ & Relative timing between FD and
        SD (1.7~g/cm$^2$), pointing direction of the telescope (4.0~g/cm$^2$)
        \\ Atmosphere & 20.3~g/cm$^2$ & Aerosol (18.9~g/cm$^2$),
        atmospheric depth (7.4~g/cm$^2$) \\ Fluorescence yield & 3.7~g/cm$^2$
        & Difference in yield models \\ \hline Quadratic sum & 21.1~g/cm$^2$ &
    \end{tabular}
    \caption{The systematic uncertainties in \sxm{} of TA hybrid
      BR/LR reconstruction.}
    \label{tab:SysUncInSigmaXmax}
\end{table}

\begin{figure}
  \centering
  \includegraphics[clip,width=\textwidth]{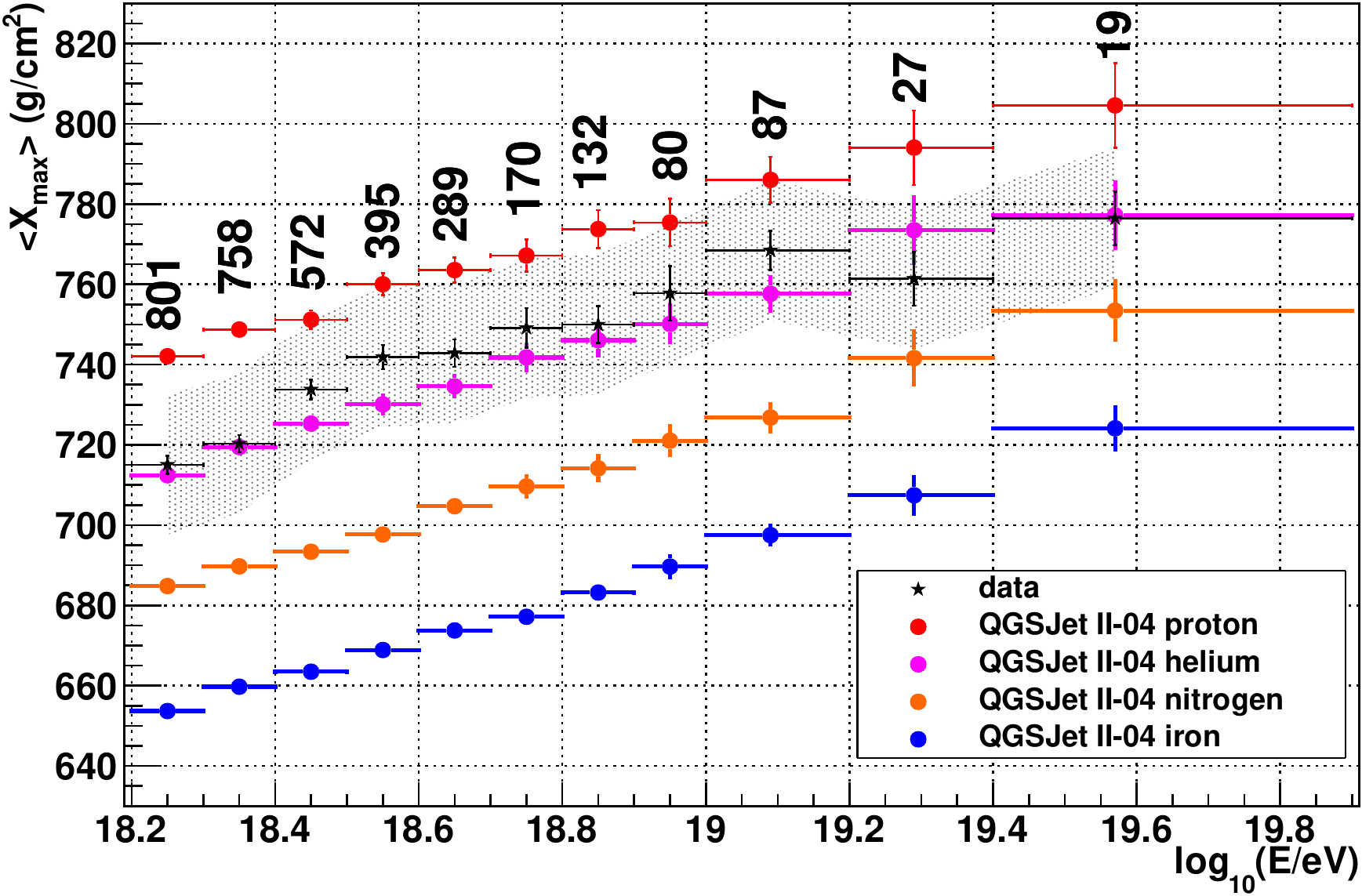}
  \caption{Mean \xm{} as a function of energy as observed by Telescope
    Array in BR/LR hybrid mode over 8.5 years of data collection. The
    numbers above the data points indicate the number of events
    observed. The gray band is the systematic uncertainty of this
    analysis. Reconstructed Monte Carlo of four different primary
    species generated using the QGSJet~II-04 hadronic model are shown
    for comparison.}
  \label{fig:xmax_elongation}
\end{figure}

As seen in Figure~\ref{fig:xmax_elongation}, within systematic
uncertainties, \mxm{} of the data is in agreement with QGSJet~II-04
protons and helium for nearly all energy bins. There is clear
separation between the region of systematic uncertainty and heavier
elements such as nitrogen and iron. In the last two energy bins there
is some overlap between the systematic uncertainty region of the data
and the nitrogen, but statistics in the data there are very poor. Care
must be taken in interpreting Figure~\ref{fig:xmax_elongation}, since
\mxm{} by itself is not a robust enough measure to fully draw
conclusions about UHECR composition. When comparing \mxm{} of data to
Monte Carlo, in addition to detector resolution and systematic
uncertainties in the data which may hinder resolving the between
different elements with relatively similar masses, the issue of
systematic uncertainties in the hadronic model used to generate the
Monte Carlo must also be recognized. This will be discussed in
Section~\ref{sec:hypothesis_tests}. Referring back to
Figures~\ref{fig:mxm_sxm_01} and \ref{fig:mxm_sxm_02}, we can see that
though the \mxm{} of the data in Figure~\ref{fig:xmax_elongation}, lies
close to QGSJet~II-04 helium, the \sxm{} of the data is larger than
the helium model allows for energy bins with good data statistics. For
this reason, we will test the agreement of data and Monte Carlo by
comparing not just \mxm{} and \sxm{}, but by using the entire
distributions. The elongation rate of the data shown in
Figure~\ref{fig:xmax_elongation} found by performing a $\chi^2$ fit to
the data is found to be $56.8 \pm 5.3$~g/cm$^2$/decade. The
$\chi^2$/DOF of this fit is 10.67/9. Table~\ref{tab:mean_xmax}
summarizes the observed first and second moments of TA's observed
\xm{} for all energy bins.


\begin{table}
  \centering
  \begin{tabular}{rrrrrr}
    \hline
    \hline
    $E_{\mathrm{low}}$ &$\left<E\right>$ &$E_{\mathrm{high}}$ &$N_{\mathrm{data}}$ &\mxm &\sxm\\
    \hline
    18.20 &18.25 &18.30 &801 &$715 \pm 2^{+17.4}_{-17.4}$ &$63 \pm 2^{+3}_{-4}$\\
    18.30 &18.35 &18.40 &758 &$720 \pm 2^{+17.4}_{-17.4}$ &$59 \pm 2^{+4}_{-4}$\\
    18.40 &18.45 &18.50 &572 &$734 \pm 2^{+17.4}_{-17.4}$ &$58 \pm 2^{+4}_{-4}$\\
    18.50 &18.55 &18.60 &395 &$742 \pm 3^{+17.4}_{-17.4}$ &$61 \pm 3^{+4}_{-4}$\\
    18.60 &18.65 &18.70 &289 &$743 \pm 3^{+17.4}_{-17.4}$ &$58 \pm 3^{+4}_{-4}$\\
    18.70 &18.75 &18.80 &170 &$749 \pm 5^{+17.4}_{-17.4}$ &$65 \pm 6^{+3}_{-4}$\\
    18.80 &18.85 &18.90 &132 &$750 \pm 5^{+17.4}_{-17.4}$ &$52 \pm 5^{+4}_{-4}$\\
    18.90 &18.95 &19.00 &80  &$758 \pm 7^{+17.4}_{-17.4}$ &$61 \pm 8^{+4}_{-4}$\\
    19.00 &19.09 &19.20 &87  &$769 \pm 5^{+17.4}_{-17.4}$ &$46 \pm 4^{+5}_{-5}$\\
    19.20 &19.29 &19.40 &27  &$761 \pm 7^{+17.4}_{-17.4}$ &$35 \pm 4^{+6}_{-7}$\\
    19.40 &19.57 &19.90 &19  &$777 \pm 7^{+17.4}_{-17.4}$ &$29 \pm 4^{+7}_{-9}$ \\
  \end{tabular}
  \caption{\mxm{} and \sxm{} observed over 8.5 years of data by
    Telescope Array in BR/LR hybrid collection mode. Energy is in
    units of $\log_{10}(E/\mathrm{eV})$ and \mxm{} and \sxm{} are in
    g/cm$^{2}$.}
  \label{tab:mean_xmax}
\end{table}

\section{\label{sec:hypothesis_tests}Statistical Hypothesis Tests}
\subsection{\label{sec:stat_test_method}Method}
If one wishes to draw conclusions about agreement between the data and
the models we should employ a test that measures the agreement of the
entire distributions instead of relying upon the first and second
moments of the \xm{} distributions. Comparisons of the means and
standard deviations of \xm{} distributions are difficult to fully
characterize agreement or disagreement because these distributions are
naturally skewed. The deep \xm{} is problematic for energy bins with
low exposure, which can lead to misinterpretation of the results if
care is not taken and only the first and second moments of the
distributions are considered. To test the compatibility of the data
and the Monte Carlo, we use an unbinned maximum likelihood test.

To perform these tests we fit the Monte Carlo \xm{} distributions to a
continuous function described by a convolution of a Gaussian with an
exponential function then uniformly shift the \xm{} distributions of
the data within $\pm 100$~g/cm$^2$ in 1~g/cm$^2$ steps, calculate the
log likelihood, then record which $\Delta X_{\mathrm{max}}$ shift
gives the best likelihood between data and Monte Carlo. We allow for
shifting of the data to account for possible systematic uncertainties
in our reconstruction and for uncertainties in the models that we are
testing against. An additional benefit of this method is that if the
required shift is significantly larger than the combined experimental
and theoretical uncertainties, that pure elemental composition is
strongly disallowed. However, in the current paper, we focus on the
shape comparisons exclusively.

\begin{figure}
  \centering
  \includegraphics[clip,width=\textwidth]{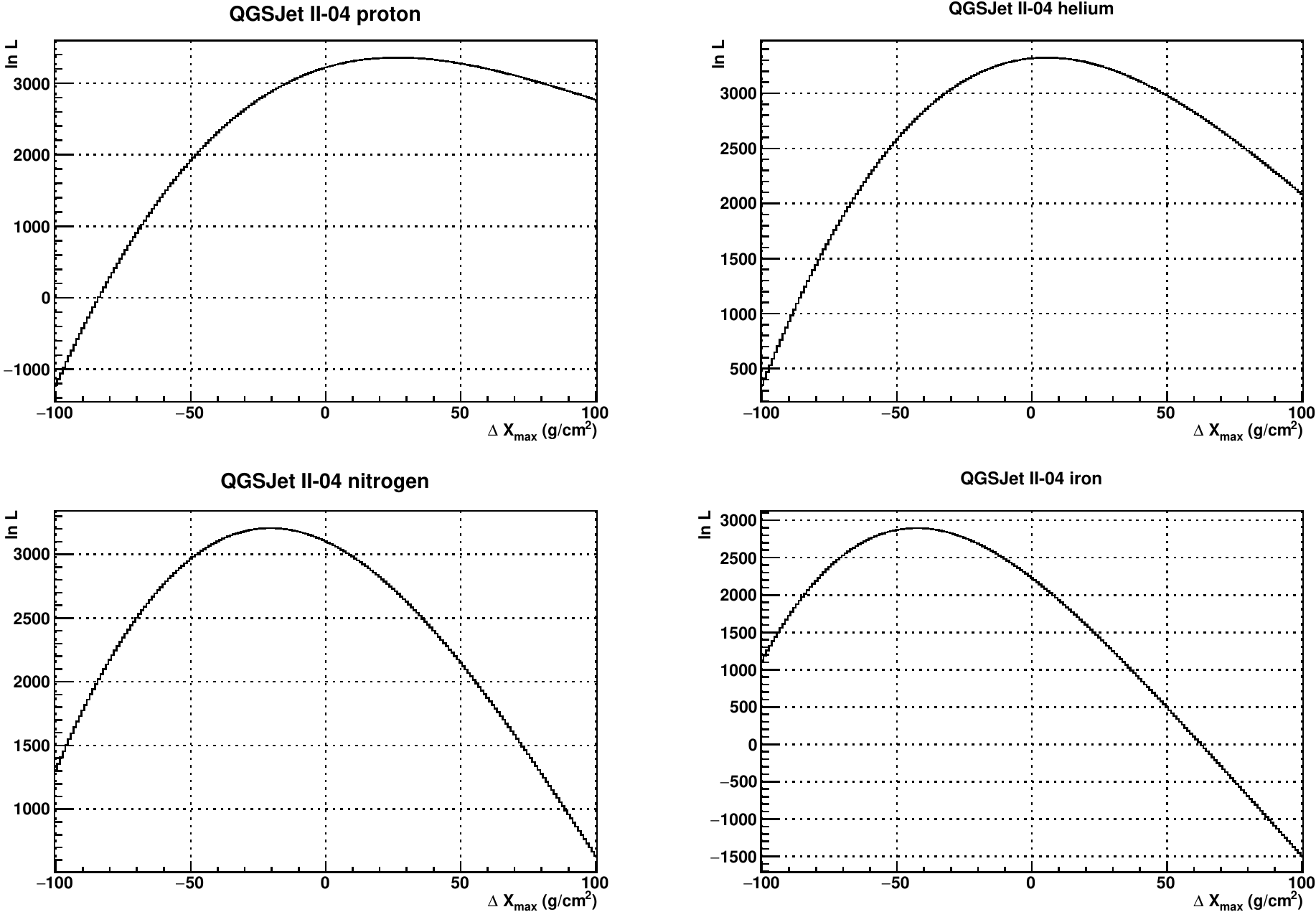}
  \caption{The log likelihood in the energy bin $18.2 \leq
    \log_{10}(E/\mathrm{eV}) < 18.3$ measured between data and Monte
    Carlo for the four chemical elements under test in this work. We
    systematically shift the data in each energy bin between $-100
    \leq \Delta X_{\mathrm{max}} \leq 100$~g/cm$^{2}$ and calculate
    the unbinned maximum likelihood. The $\Delta X_{\mathrm{max}}$
    which corresponds to the maximum measured likelihood is used to
    test the compatibility of the data and Monte Carlo \xm{}
    distributions.}
  \label{fig:ML_lnL}
\end{figure}

\begin{figure}
  \centering
  \includegraphics[clip,width=\textwidth]{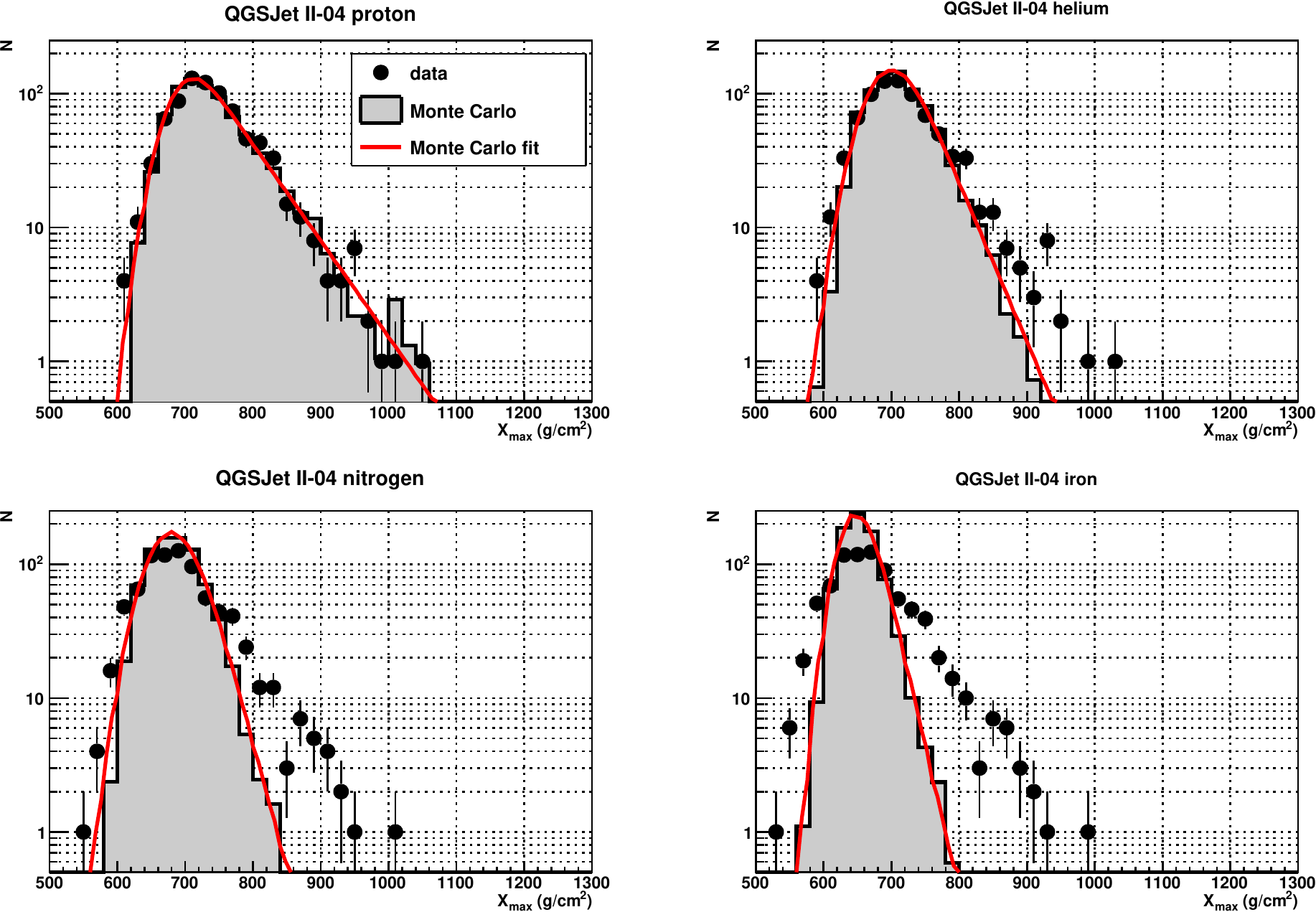}
  \caption{Data after shifting by the $\Delta X_{\mathrm{max}}$ which
    provides the maximum likelihood with the Monte Carlo in the $18.2
    \leq \log_{10}(E/\mathrm{eV}) < 18.3$ energy bin. The unbinned
    data is tested against the fit to the Monte Carlo (red line). For
    this energy bin the data and QGSJet~II-04 protons were found to be
    compatible, with a $p$-value of 0.32. All other chemical elements
    had $p$-values $< 0.05$ and found to be incompatible with the data.}
  \label{fig:ML_fit}
\end{figure}

For example, Figure~\ref{fig:ML_lnL} shows the log likelihood values
measured for the chemical elements tested against the data in the
$18.2 \leq \log_{10}(E/\mathrm{eV}) < 18.3$ energy bin.
Figure~\ref{fig:ML_fit} shows the data after shifting by the best
$\Delta X_{\mathrm{max}}$ and the Monte Carlo for each chemical
element for the same energy bin. Data and protons appear to agree well
over their entire distributions. Data and helium match well up until
about 850~g/cm$^2$, where the helium tail begins to fall off faster
than the data. Nitrogen and iron show less agreement in the tails as
well. Note that the same data is used for each subfigure shown in
Figure~\ref{fig:ML_fit}, but it is shifted by a different amount in
each one. The shifts applied to the data are +29, +7, -19, and
-41~g/cm$^2$ in the proton, helium, nitrogen, and iron subfigures
respectively. The slight variation of the shape of the data histogram in
each subfigure is due to the effect of systematic shifting of all data
points and then binning in the plot. However, the maximum likelihood
calculated for our tests use an unbinned method.

\begin{figure}
  \centering
  \includegraphics[clip,width=\textwidth]{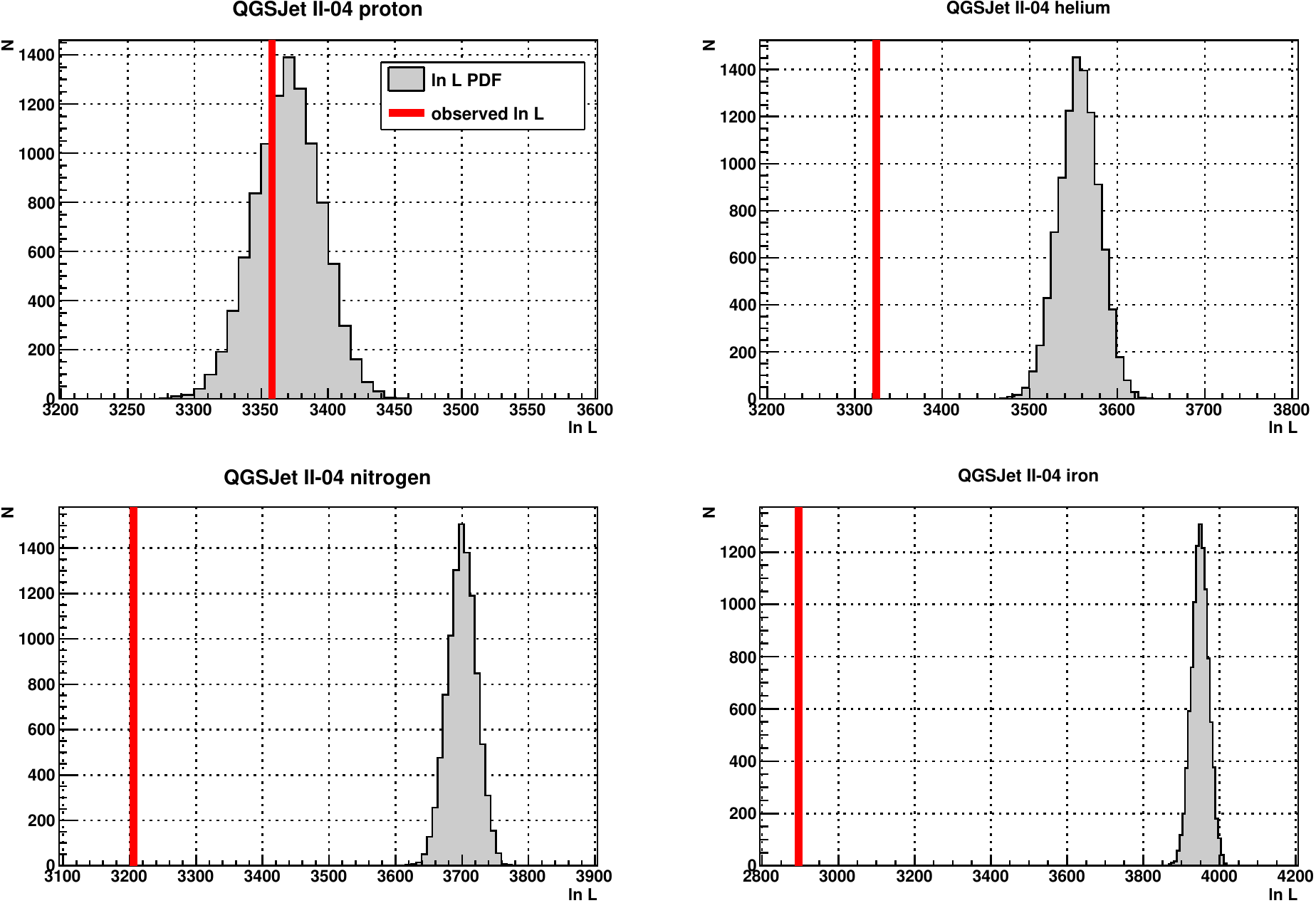}
  \caption{Calculation of $p$-value of the unbinned maximum likelihood
  test between data and Monte Carlo for the $18.2 \leq
  \log_{10}(E/\mathrm{eV}) < 18.3$ energy bin. The data is shifted by
  the $\Delta X_{\mathrm{max}}$ corresponding to the maximum
  likelihood, the value of which is indicated by the red line. The
  Monte Carlo is sampled according to the data statistics to measure
  the distribution of log likelihood expected. The $p$-value is the
  probability of measuring log likelihood less than or equal to that
  observed in the shifted data. The calculated $p$-values for this
  energy bin are 0.32 for proton, $10^{-25}$ for helium, $10^{-93}$
  for nitrogen, and $< 10^{-324}$ for iron.}
  \label{fig:ML_pdf}
\end{figure}

We then calculate the probability ($p$-value) of measuring a log
likelihood in the Monte Carlo equal to or more extreme than the one
measured in the data shifted by the best $\Delta X_{\mathrm{max}}$.
Figure~\ref{fig:ML_pdf} shows the distributions of log likelihood
calculated for the $18.2 \leq \log_{10}(E/\mathrm{eV}) < 18.3$ energy
bin, as well as the likelihood measured for the data when shifted by
the best $\Delta X_{\mathrm{max}}$. The null hypothesis being tested
is that the data after shifting and the Monte Carlo are drawn from the
same continuous distribution. If the $p$-value we measure from this
test statistic is 0.05 or less, we reject the null hypothesis at the
95\% confidence level, and we say the two distributions are not
compatible. If the $p$-value is greater than 0.05 we fail to reject
the systematically shifted data and Monte Carlo as being
compatible.

Hadronic models in the UHECR energy regime are based upon measurements
made in accelerators. Cross section, multiplicity, and elasticity of
the primary particle are fundamental parameters used by these models
that are particularly sensitive for UHECR \xm{}. These parameters are
measured at relatively low energies ($\sqrt{s} = 14$~TeV corresponds
to about $10^{17}$~eV in the lab frame), which need to be extrapolated
up to $10^{20}$~eV to fully describe the physics up to the highest
energy cosmic rays observed. Abbasi and Thomson have examined the
uncertainty in \mxm{} in several different popular hadronic models
introduced by extrapolating these parameters. The estimated lower
limits on the uncertainty in \mxm{} from the extrapolation was found
to $\sim 6$~g/cm$^2$ at $E_{\mathrm{lab}} = 10^{17}$~eV and $\sim
35$~g/cm$^2$ at $E_{\mathrm{lab}} = 10^{19.5}$~eV
\citep{Abbasi:2016sfu}. This uncertainty in \mxm{} at $10^{19.5}$~eV
is about the same as the difference in \mxm{} predicted among the
deepest model (EPOS LHC) and the shallowest model (QGSJet01c). The
shapes of the \xm{} distributions have a much smaller dependence on
hadronic model assumptions. Because of these large uncertainties in
the models that we compare our observed \xm{} to, we simultaneously
systematically shift the data and test the shapes of the distributions
to measure compatibility between the data and model.

\subsection{\label{sec:stat_test_results}Results}
\begin{figure}
  \centering
  \includegraphics[clip,width=\textwidth]{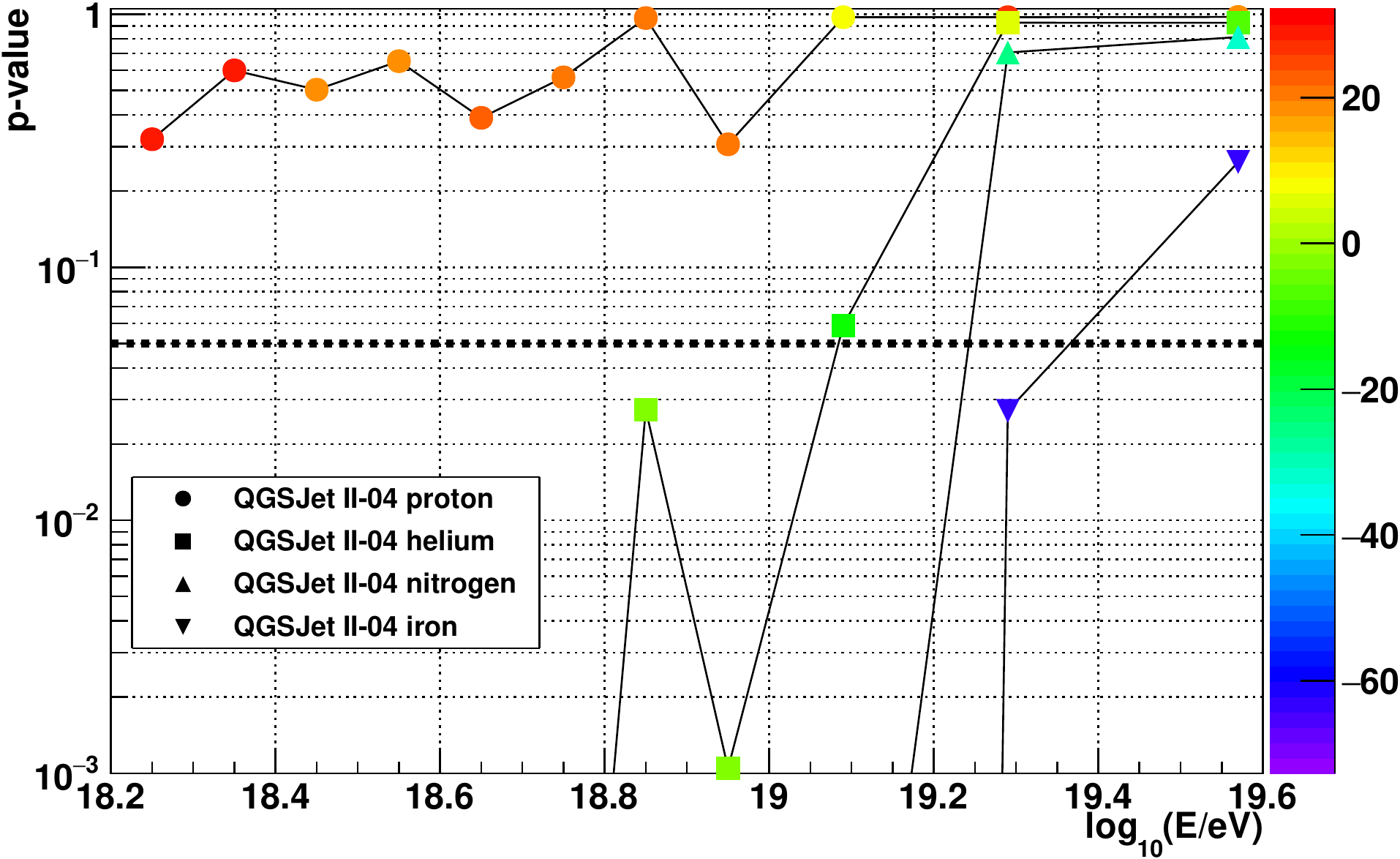}
  \caption{Unbinned maximum likelihood test on observed and simulated
    QGSJet~II-04 \xm{} distributions after systematic shifting of the
    data to find the best log likelihood. Each point represents the
    probability of measuring a log likelihood more extreme than that
    observed in the data after it is shifted by the best $\Delta
    X_{\mathrm{max}}$. The color of the point indicates the $\Delta
    X_{\mathrm{max}}$ measured in g/cm$^{2}$ required to find the
    maximum log likelihood value. The dashed line at $p$-value = 0.05
    indicates the threshold below which the data is deemed
    incompatible with the Monte Carlo at the 95\% confidence level.}
    \label{fig:dataMCMLTest}
\end{figure}

Table~\ref{tab:dataMCMLTest} shows the results of these tests. For
each QGSJet~II-04 model tested against the data, the $\Delta
X_{\mathrm{max}}$ which gave the best log likelihood is shown, as well
as the $p$-value for that shift. For QGSJet~II-04 protons, in most
energy bins the $\Delta X_{\mathrm{max}}$ shifts are about the size of
or slightly larger than the systematic uncertainty of \mxm{} for this
analysis. The $p$-values, which measure the agreement of the shapes of
the entire \xm{} distributions after shifting, all have values $>
0.05$, therefore we fail to reject protons as being compatible with
the data for all energy bins using this test. QGSJet~II-04 helium has
$\Delta X_{\mathrm{max}}$ shifts smaller than protons and within our
quoted systematic uncertainties of \mxm{}, but the $p$-values indicate
that once shifting is performed the shapes of the data and Monte Carlo
do not agree for $\log_{10}(E/\mathrm{eV}) < 19.0$. For those energy
bins, the test rejects QGSJet~II-04 helium as being compatible with
data after systematic shifting. Above $\log_{10}(E/\mathrm{eV}) =
19.0$, the $p$-values are $> 0.05$ and we fail to reject helium as
being compatible with the data. QGSJet~II-04 nitrogen requires $\Delta
X_{\mathrm{max}}$ shifts slightly larger than our systematic
uncertainty, but the $p$-values of the tests reject nitrogen for
$\log_{10}(E/\mathrm{eV}) < 19.2$. Iron requires $\Delta
X_{\mathrm{max}}$ shifts larger than our systematic uncertainty, and
is rejected as being compatible with the data for
$\log_{10}(E/\mathrm{eV}) < 19.4$. Figure~\ref{fig:dataMCMLTest}
visually summarizes the results of the tests and the data in
Table~\ref{tab:dataMCMLTest}.

\begin{table}
  \centering
\begin{tabular}{l rr rr rr rr}
  \hline
  \hline
  &\multicolumn{2}{c}{proton} &\multicolumn{2}{c}{helium}
  &\multicolumn{2}{c}{nitrogen} &\multicolumn{2}{c}{iron} \\
  \cline{2-9}
  energy &$\Delta X_{\mathrm{max}}$ &$p$-val &$\Delta
  X_{\mathrm{max}}$ &$p$-val
  &$\Delta X_{\mathrm{max}}$ &$p$-val &$\Delta X_{\mathrm{max}}$ &$p$-val\\
  \hline
  18.2-18.3 &$29\pm 2$ &0.32  &$7\pm 2$  &---
  &$-19\pm 1$ &---    &$-41\pm 1$ &---\\
  18.3-18.4 &$30\pm 2$ &0.59  &$6\pm 2$  &$2 \times 10^{-18}$
  &$-19\pm 1$ &---    &$-43\pm 1$ &---\\
  18.4-18.5 &$19\pm 2$ &0.50  &$-2\pm 2$ &$9 \times 10^{-11}$
  &$-28\pm 2$ &---    &$-53\pm 1$ &---\\
  18.5-18.6 &$19\pm 2$ &0.65  &$-2\pm 2$ &$2 \times 10^{-11}$
  &$-33 \pm 2$ &---    &$-54\pm 2$ &---\\
  18.6-18.7 &$22\pm 3$ &0.38  &$-1\pm 3$ &$3 \times 10^{-7}$
  &$-25\pm 2$ &---    &$-52\pm 2$ &---\\
  18.7-18.8 &$20\pm 4$ &0.55  &$2\pm 3$  &$6 \times 10^{-6}$
  &$-24\pm 3$ &---    &$-53\pm 2$ &---\\
  18.8-18.9 &$20\pm 4$ &0.97  &$2\pm 3$ &0.027
  &$-27\pm 3$ &$3 \times 10^{-6}$    &$-51\pm 2$ &---\\
  18.9-19.0 &$21\pm 5$ &0.30  &$1\pm 5$  &0.0010
  &$-25\pm 4$ &$1 \times 10^{-14}$    &$-42\pm 3$ &---\\
  19.0-19.2 &$10\pm 5$ &0.98  &$-7\pm 4$ &0.059
  &$-34\pm 4$ &$1 \times 10^{-5}$     &$-57\pm 3$ &---\\
  19.2-19.4 &$26\pm 8$ &0.98  &$9\pm 8$  &0.93
  &$-18\pm 7$ &0.71   &$-50\pm 5$ &0.027\\
  19.4-19.9 &$19\pm 8$ &0.98  &$-3\pm 8$ &0.93
  &$-23\pm 7$ &0.81   &$-50\pm 6$ &0.26\\
  \hline
\end{tabular}
\caption{Results of unbinned maximum likelihood test of TA BR/LR
  hybrid \xm{} data against four pure QGSJet~II-04 chemical
  models. For each model the $\Delta X_{\mathrm{max}}$ shift
  required to find the maximum log likelihood is shown, as well as the
  $p$-value of the likelihood.  After systematic shifting of the data,
  the maximum likelihood $p$-values reject all species except
  QGSJet~II-04 protons at the 95\% confidence level for energies below
  $10^{19}$~eV. Above $10^{19}$~eV statistics are rapidly falling, and
  the likelihood test fails to reject at the 95\% confidence level
  even very heavy elements. Entries shown as ``---'' have a $p$-value
  $< 7.6 \times 10^{-24}$ (significance $>10 \sigma$). \xm{} shifts
  are measured in g/cm$^{2}$.}
\label{tab:dataMCMLTest}
\end{table}

We can understand why the likelihood test finds our data
simultaneously compatible with elements with very different masses
such as protons and nitrogen in the last two energy bins if we
consider Figures~\ref{fig:mxm_sxm_bin_09} and
\ref{fig:mxm_sxm_bin_10}. We see that because of poor detector
exposure leading to very few events collected in these energy bins,
the confidence intervals of \mxm{} and \sxm{} of the Monte Carlo
overlap among the different elements. Given our current exposure, we
should not expect to be able to distinguish the difference between
protons, helium, or nitrogen in these energy bins. Iron too is found
compatible with the data in the last energy, but a shift of
$\Delta X_{\mathrm{max}}$ larger than our systematic uncertainty is
required. The agreement of the maximum likelihood test then comes from
very few events in the data (19 events) and the lack of a tail in the
data deep \xm{} distribution ($X_{\mathrm{max}} > 850$~g/cm$^{2}$) as
seen in Figure~\ref{fig:dataMC_xmax_bin_10}. This lack of deep \xm{}
tail allows the shapes of the data and Monte Carlo to resemble each
after a sufficient amount of \xm{} shifting.

\xm{} analysis using five years of data analyzed by the Middle Drum FD
also found compatibility of the data with protons and incompatibility
with iron. In that analysis, the $\chi^2$ test was applied to QGSJet
II-03 protons and iron in three energy ranges: $10^{18.2}$ -
$10^{18.4}$, $10^{18.4}$ - $10^{18.6}$, and $10^{18.6}$ -
$10^{18.8}$~eV. For these energy bins, the $p$-values of the $\chi^2$
tests rejected iron, but not protons \citep{Abbasi:2014sfa}.

\section{\label{sec:conclusions}Conclusions}
Telescope Array has completed analyzing 8.5 years of data collected in
hybrid mode using events observed simultaneously by the surface
detector array and the Black Rock and Long Ridge fluorescence
detectors. This data provided 3330 events after reconstruction and
cuts are applied, and was used to analyze the depth of shower maxima
(\xm). Good operation of the detector was verified by using an
extensive Monte Carlo suite with showers pre-generated using
CORSIKA. This Monte Carlo allows us to verify that we understand the
detector with a high degree of confidence and also to compare the
observed \xm{} distributions with CORSIKA models of four different
single element primaries: protons, helium, nitrogen, and iron, all
generated using the QGSJet~II-04 hadronic model. The data can be
compared to the Monte Carlo by the traditional method, comparing the
first and second moments (\mxm{} and \sxm{}) of the observed \xm{}
distributions to the Monte Carlo. This method may be overly simplistic
and misleading especially for energy bins with low exposure, which can
change the shapes of the observed distributions. We have presented a
new way to visualize \mxm{} and \sxm{} by plotting their joint
distributions in the data as well as the confidence intervals expected
from Monte Carlo. We have extended the analysis of \xm{} by using
unbinned maximum likelihood, which allows us to measure the
compatibility of the data and Monte Carlo using the entire
distributions. This is especially important for statistical
distributions that potentially exhibit a high degree of skew, such as
those of light elements with \xm{} distributions with deeply
penetrating tails. Using this test we can empirically reject certain
chemical elements at a given confidence level as being compatible with
our data.

After allowing for systematic shifting of the data \xm{} distributions
and performing the likelihood test on the data and Monte Carlo
distributions of four pure chemical species, we find that we fail to
reject QGSJet~II-04 protons as being compatible with the data for all
energy bins at the 95\% confidence level. QGSJet~II-04 helium is
rejected as being compatible with the data for $\log_{10}(E) <
19.0$. QGSJet~II-04 nitrogen is rejected for $\log_{10}(E) < 19.2$ and
iron is rejected for $\log_{10}(E) < 19.4$. We've demonstrated that
for $\log_{10}(E) \geq 19.0$, TA has insufficient exposure to
accurately distinguish the difference between different individual
elements. Energy bins in this energy range have poor statistics due to
low exposure and agreement among several of the models is found with
the data. However, this agreement is physically unrealistic for the
case of iron because of the large shifts required, in excess of our
systematic uncertainty.

\section{Acknowledgements}
The Telescope Array experiment is supported by the Japan Society for
the Promotion of Science through Grants-in-Aid for Scientific Research
on Specially Promoted Research (21000002) ``Extreme Phenomena in the
Universe Explored by Highest Energy Cosmic Rays'' and for Scientific
Research (19104006), and the Inter-University Research Program of the
Institute for Cosmic Ray Research; by the U.S. National Science
Foundation awards PHY-0601915,
PHY-1404495, PHY-1404502, and PHY-1607727; 
by the National Research Foundation of Korea \linebreak
(2015R1A2A1A01006870, 2015R1A2A1A15055344, 2016R1A5A1013277,
\linebreak 2007-0093860, 2016R1A2B4014967); by the Russian Academy of
Sciences, RFBR grant 16-02-00962a (INR), IISN project No. 4.4502.13,
and Belgian Science Policy under IUAP VII/37 (ULB). The foundations of
Dr. Ezekiel R. and Edna Wattis Dumke, Willard L. Eccles, and George
S. and Dolores Dor\'e Eccles all helped with generous donations. The
State of Utah supported the project through its Economic Development
Board, and the University of Utah through the Office of the Vice
President for Research. The experimental site became available through
the cooperation of the Utah School and Institutional Trust Lands
Administration (SITLA), U.S. Bureau of Land Management (BLM), and the
U.S. Air Force. We appreciate the assistance of the State of Utah and
Fillmore offices of the BLM in crafting the Plan of Development for
the site.  Patrick Shea assisted the collaboration with valuable advice 
on a variety of topics. The people and the officials of Millard County, 
Utah have been a source of
steadfast and warm support for our work which we greatly appreciate. 
We are indebted to the Millard County Road Department for their efforts 
to maintain and clear the roads which get us to our sites. 
We gratefully acknowledge the contribution from the technical staffs of
our home institutions. An allocation of computer time from the Center
for High Performance Computing at the University of Utah is gratefully
acknowledged.

\bibliography{taHybridXmaxAPJ2018}

\end{document}